\documentclass[11pt]{article}

\usepackage{epsf,graphics,epic,eepic}
\usepackage[normalsize]{caption}

\def\barr{\begin{array}}
\def\earr{\end{array}}

\begin{document}

\title{Virtual Compton Scattering off the nucleon}
\author{ P.A.M. Guichon and M. Vanderhaeghen}
\date{}
\maketitle

\vspace{-1cm}
\center{\it CEA-Saclay, DAPNIA/SPhN, F-91191 Gif-sur-Yvette, France}

%
\vspace{2cm}
\abstract{
We review the recent developments of virtual Compton scattering. 
We focus on the 
kinematical regimes which look the most promising. The threshold
regime gives access to the generalized polarizabilities of the
proton. The hard scattering 
regime allows tests of perturbative QCD predictions and of the valence
quark wave function. 
The Bjorken regime is closely related to deep inelastic scattering. 
It gives access to the off-forward parton distributions which 
generalize the ordinary parton distributions and may shed a new light 
on the spin problem. 
For each regime we discuss the experimental perspectives and the 
role of the Bethe-Heitler background.
}
%
%
%


%
%

\vspace{6cm}
{\it to appear in Prog.Part.Nucl.Phys., Vol.41 (1998)}

\newpage

\tableofcontents

\newpage

\section{INTRODUCTION}

Virtual Compton Scattering generally refers to any process where two
photons are involved and where at least one of them is virtual. In
the following we shall use Virtual Compton Scattering (VCS) in a more
restricted sense, that is the reaction where a \em space-like \em virtual
photon is absorbed by a hadronic target which returns to its initial
state by emitting one real photon. The reaction can be accessed through
electroproduction of photons, \( (e,e'\gamma ) \). In practice the target will be the
proton but one could extend our considerations to other targets provided
the experiments, which are very difficult, become feasible. The experimental
difficulty has two origins. Firstly, the cross section is suppressed
by a factor \( \alpha \sim 1/137 \) with respect to the purely elastic case. 
Secondly, the
emission of a neutral pion which decays into two photons creates a
physical background which may prevent the extraction of the VCS signal.
When the target is a proton one can use kinematical conditions where
its recoil momentum is large enough to be measured in a high resolution
spectrometer, thus allowing to separate the VCS events from the \( \pi ^{0} \)events.
Even in this favourable case the experiments are still delicate and
have become possible only due to the advent of the new generation
of electron accelerators. This is why, even though VCS has been put
forward many times by theorists, the experiments are just beginning.

VCS was first proposed as a possible test of Quantum Electrodynamics
(QED) at small distance \cite{Bjo58} but it is amusing to note that
the first calculation \cite{Ber58} of the VCS cross section was motivated
by the experimental program of the Mark-III accelerator in Stanford,
where the electroproduction of photons was considered as an embarassing
background of the electroproduction of pions, the physically interesting
reaction. Later the VCS process was recognized as interesting by itself
and was studied by the same authors \cite{Ber61} in the framework of
dispersion relations. The interest for this reaction then disappeared
for a while due to the lack of experimental data. Actually VCS is
a small background!

After the discovery of approximate scaling in deep inelastic scattering,
VCS was proposed \cite{Geo71} to get further insight in the light cone
singularity of current commutators and later the process was analysed
using the operator product expansion \cite{Wat82}, providing tests
of Quantum Chromodynamics (QCD) and of the models of hadrons. In the
framework of the hard scattering picture \cite{Bro80}, the first, and
till now unique, calculation \cite{Far90a} of the VCS amplitudes in
the framework of Perturbative QCD (PQCD) was stimulated by physicists
who where planning the first experiment of VCS \cite{vBi89} 
using the high luminosity 
apparatus of the (abandonned) PEGASYS project. 

The VCS program came to live thanks to the advent of the CEBAF accelerator.
The high luminosity and high resolution of the latter allowed to propose
a convincing program of VCS \cite{Ber93}. Due to the relatively low
energy, the best physics case here turned out to be the threshold region
where the so called generalized polarizabilities \cite{Gui95} of the
proton could be measured. It was then soon realised that a part of
this threshold physics was in fact immediately feasible \cite{d'Ho95}
at the MAMI accelerator. The data taking took place in 1996 and at
the time of writing are still under analysis. In the meanwhile the
first few events of VCS were observed \cite{vdB95}.

The last important development is the recent proposal \cite{Ji97a}\cite{Rad96a}
that VCS in the Bjorken regime could be used to extract new structure
functions which reduce to the ordinary parton distributions in the
forward limit. The first moments of these functions are related
to the hadronic form factors and the second moment gives access to the
total spin carried by the quarks in the nucleon. This is an unexpected
development which illustrates the interplay between the existence
of a performant accelerator and the apparitions of new and promising
ideas.
\begin{figure}[h]
\epsfxsize=13.cm
\epsfysize=4.cm
\centerline{\epsffile{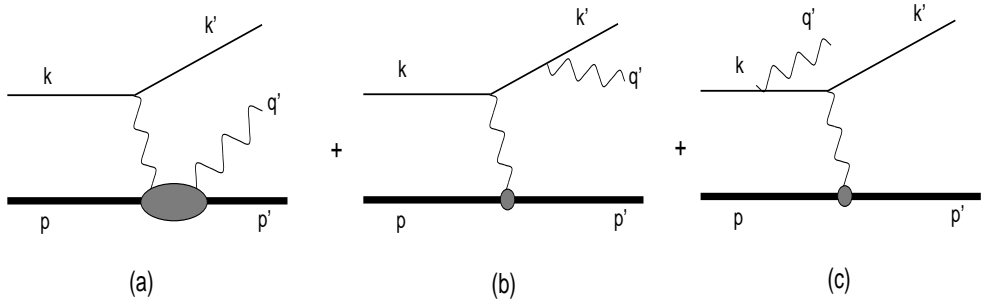}}
\caption{\small (a) FVCS amplitude, (b,c) BH amplitude.}
\label{fig:bhvcs}
\end{figure}


In the \( (e,e'\gamma ) \) reaction on the proton

\[
e+p\rightarrow e'+p'+\gamma \;, \]
the final photon can be emitted either by the proton, giving access
to the VCS process, or by the electrons, which is referred to as the
Bethe-Heitler (BH) process. 
If we call \( T^{ee'\gamma } \) the amplitude of the \( (e,e'\gamma ) \) 
reaction, we have
\[
T^{ee'\gamma }=T^{FVCS}+T^{BH} \;,\]
where \( T^{FVCS} \) and \( T^{BH} \) corresponds respectively to graphs (a) and
(b,c) of Fig.(\ref{fig:bhvcs}). The meaning of the superscript FVCS is that the
corresponding amplitude contains the leptonic current while
VCS refers only to the subprocess 
\[
p+\gamma ^{*}\rightarrow p'+\gamma \;.\]
The BH amplitude can be calculated
exactly in QED, if one knows the elastic form factors of the proton.
Therefore it contains no new information on the structure. Unfortunately
light particles such as electrons radiate much more than the heavy
proton. So the BH process generally dominates or interferes strongly
with the VCS process, and this may jeopardize the interpretation of
the \( (e,e'\gamma ) \) reaction, independently of the experimental difficulty. The
only way out of this problem is either to find kinematical regions
where the BH process is suppressed or to have a very good theoretical
control over the interference between the BH and the FVCS amplitudes.

Assuming that this problem is fixed, another difficulty of the theoretical
analysis is the complexity of the VCS amplitude itself. For a spin
$S$ target, the number of possible helicity states is \( 2\times (2S+1)\times 3\times 2 \) since the virtual
and real photons have respectively 3 and 2 degrees of freedom. Parity
invariance reduces this number by a factor 1/2 so that for the proton
one needs 12 complex functions to specify completely the VCS amplitude.
These functions depend, for instance, on the virtuality of the initial
photon (\( Q^{2} \)), the scattering angle (\( \theta _{\gamma ^{*}\gamma } \)) and the energy in the proton-photon
center of mass (\( \sqrt{s} \)). Extracting the whole set of functions would involve
such a huge amount of very difficult 
experiments that it looks definitely out of reach.

In view of the two difficult points discussed above, the safe attitude
has been to restrict the use of the \( (e,e'\gamma ) \) reaction to cases where the
situation is simple enough to allow a fruitful interpretation. Until
now 3 cases have been identified:

\begin{itemize}

\item The low energy region, below the pion production threshold, where
the VCS amplitudes can be parametrised by 6 generalized polarizabilities
which are real functions of \( Q^{2} \).

\item The hard scattering region, where \( \sqrt{s} \) is large with respect to the typical
hadronic scale, and \( \theta _{\gamma ^{*}\gamma } \) is close to \( 90^{\rm o} \). The virtuality needs not to be
large. There one can test the prediction of PQCD in the framework
of the Hard Scattering Picture (HSP) \cite{Bro80}. In particular the
phase of the amplitude which is predicted to be large \cite{Far90a},\cite{Kro91}
 can be accessed through the interference between the BH and the VCS
processes.

\item The Bjorken region, where \( s \) and \( Q^{2} \) are large, \( Q^{2}/s \) finite and \( \theta _{\gamma ^{*}\gamma } \) close to
zero. This is often referred to as Deeply Virtual Compton Scattering
(DVCS). In this limit the amplitude depends only on 4 parton distributions
(called Off Forward Parton Distributions, OFPD's).

\end{itemize}

In Section 2 we give   our notations and conventions. We present  the
theoretical framework and collect the formulae that are used in the rest of the
paper. Sections 3 presents the threshold regime. We have tried to propose a
qualitative introduction to the physics of polarizabilities by considering the
real photon as an applied external field. The low energy parametrization of the amplitude
in terms of six generalized polarizabilities is then discussed in details and we
propose an experimental program to determine them. In Section 4 we explain the
interest of the hard scattering regime and we use the diquark model to delineate
the experimental possibilities. We also discuss  real Compton scattering
because the experimental prospects are better in this case. Section 5 deals with
the Bjorken regime. The DVCS amplitude is computed in the handbag 
approximation for
which the non perturbative input are the OFPD's. We use the same OFPD's to
compute the meson production amplitude. Using a simple ansatz for the OFPD's we
present some estimates of cross sections in kinematics corresponding to existing 
and future facilities. 
We conclude in Section 6 with a discussion of the open problems and perspectives.

\newpage

\section{PRELIMINARIES}

\def\rmqp{ {\rm q}'}

\subsection{Units and constants}

The system of units used in this paper is the (natural) Heaviside-Lorentz
one \cite{Jac75}, but occasionally we shall also write some formulae
in the Gauss system so as to fix some misunderstandings. Table (\ref{Table_2.0})
summarises our notations or values for the various constants of this
paper.\begin{table}
{\centering \begin{tabular}{|c|c|c|}
\hline 
&Heaviside-Lorentz&Gauss\\
\hline 
\hline 
Electron mass&\( m_{e} \sim 0 \)&\( m_{e} \sim 0 \)\\
\hline 
Proton mass&\( m=.938 \)GeV&\( m=.938 \)GeV\\
\hline 
Planck
constant&\( \hbar  \)=1&\( \hbar  \)=1\\
\hline 
Light velocity&\( c=1 \)&\( c=1 \)\\
\hline 
Proton charge&\( e \)&\( e_{G} \)\\
\hline 
QED fine structure
constant&\( \alpha _{QED}=e^{2}/4\pi =1/137 \)&\( \alpha _{QED}=e_{G}^{2}=1/137 \)\\
\hline 
\end{tabular}\par}

\caption{Constants used in the paper. \label{Table_2.0}}
\end{table}

\subsection{Kinematics}

Greek letters are used to note Lorentz indices. For the space
components we use latin letters. Summation over repeated indices is
implicit and the metric is \( g_{00}=1,\, g_{ii}=-1. \) We have collected 
in Table (\ref{Table_2.1}) our notations
for the 4-momenta and helicities or spin projections. We often neglect the
mass of the electron with respect to its energy, which is permitted
except when the angle between the electrons or between one electron
and the final photon is exactly zero. These situations are irrelevant
experimentally, so the approximation is safe. 
As a consequence of this approximation, the
electron helicity is conserved, which simplifies a lot the formalism.
For the proton the same would happen in the hard scattering regime
at very large values of \( s \), but in the threshold region the use of
the helicity quantum number is not so compelling. In this case we
shall use the (rest frame) spin projection because it leads to a simpler
angular analysis. \begin{table}
{\centering \begin{tabular}{|c|c|c|c|}
\hline 
&4-momentum&Helicity&Spin projection\\
\hline 
\hline 
Incident electron&\( k \)&\( h \)&\\
\hline 
Scatterred
electron&\( k' \)&\( h'=h \)&\\
\hline 
Target proton&\( p \)&\( h_{N} \)&\( \sigma  \)\\
\hline 
Recoil proton&\( p' \)&\( h_{N}' \)&\( \sigma ' \)\\
\hline 
Final photon&\( q' \)&\( \lambda ' \)&\\
\hline 
\end{tabular}\par}

\caption{Notation for the 4-momenta and polarizations. \label{Table_2.1}}
\end{table}

The virtual photon exchanged in the VCS process has 4-momentum \( q=k-k' \), 
while it is \( q-q' \) in the BH process. In the following we shall note
with a Roman letter the modulus of a 3-vector, e.g. 
\( \rm q=\left| \vec q \right| . \) A hat on a 3-vector will denote 
the direction of this vector. 
\newline
\indent
We shall often work in the C.M. frame defined by \( \vec p +\vec q =0, \) and \em by default
every non Lorentz covariant quantity will refer to this particular frame. \em The
unit vectors of this frame are noted 
[\( \vec e \, \left( 1\right) ,\, \vec e \, \left( 2\right) ,\, \vec e \,
 \left( 3\right)  \)] and are such that the components
of the 3-momenta are
\begin{eqnarray}
\vec q \, \,  & = & \rm q\left( 0,\, 0,\, 1\right) \;, \nonumber \\
\vec q \, ' & = & \rm q'\left( \sin \theta ,\, 0,\, \cos \theta \right) \;, \nonumber \\
\vec k \, \,  & = & \rm k\left( \sin \alpha \cos \phi ,\, 
\sin \alpha \sin \phi ,\, \cos \alpha \right) \;, \nonumber \\
\label{eq_2_1} \vec k \, ' & = & \rm k'\left( \sin \alpha '\cos \phi ,\, 
\sin \alpha '\sin \phi ,\, \cos \alpha '\right) \;,
\end{eqnarray}
 and the scattering angle of the electron is
\( \theta _{e}=\alpha '-\alpha . \) The invariant quantities which appear frequently in the following
are:
\begin{equation}
Q^{2} = -q^{2},\, t = (q-q')^{2},\, s = \left( p+q\right) ^{2} \;, 
\label{eq_2_2}  
\end{equation}
and we have the relations :
\begin{equation}
\label{eq_2_3}
{\rm q'}=\frac{s-m^{2}}{2\sqrt{s}},\, q_{0}=\frac{s-m^{2}-Q^{2}}{2\sqrt{s}},\, p_{0}=\frac{s+m^{2}+Q^{2}}{2\sqrt{s}},\, p_{0}'=\frac{s+m^{2}}{2\sqrt{s}}.
\end{equation}

\subsection{BH and FVCS amplitudes}

For definiteness we consider the case of an electron beam. For a positron
beam one must change the relative sign of the BH and FVCS amplitudes.
In the one photon exchange approximation and in Lorentz gauge, the
BH amplitude is 
\begin{equation}
\label{eq_2_4}
T^{BH}=\frac{-e^{3}}{t}\varepsilon _{\mu }'^{*}L^{\mu \nu }\, \overline{u}(p')\Gamma _{\nu }(p',p)u(p) \;,
\end{equation}
where the spinors are defined in App. \ref{app1}. 
The polarization of the final photon is \( \varepsilon '(q',\lambda ') \) and
the vertex \( \Gamma  \) is defined as
\begin{eqnarray}
\Gamma ^{\nu }(K',K) & = & F_{1}(X)\gamma ^{\nu }+iF_{2}(X)\sigma ^{\nu \rho }(K'-K)_{\rho }/2m \;,\nonumber \\
\label{eq_2_5} X & = & (K'-K)^{2},\, F_{1}(0)=1,\, F_{2}(0)=1.79 \;.
\end{eqnarray}
From QED one has the expression of the
non symmetric lepton tensor
\begin{equation}
\label{eq_2_6}
L^{\mu \nu }=\overline{u}(k')\left( \gamma ^{\mu }\frac{1}{\gamma .(k'+q')-m_{e}+i\epsilon }\gamma ^{\nu }
+\gamma ^{\nu }\frac{1}{\gamma .(k-q')-m_{e}+i\epsilon }\gamma ^{\mu }\right) u(k) \;,
\end{equation}
which explicitely satisfies the gauge
invariance conditions
\begin{equation}
\label{eq_2_7}
q_{\mu }'L^{\mu \nu }=L^{\mu \nu }\left( q_{\nu }-q_{\nu }'\right) =0.
\end{equation}

The FVCS amplitude has a similar expression
\begin{equation}
\label{eq_2_8}
T^{FVCS}=\frac{e^{3}}{q^{2}}\varepsilon _{\mu }'^{*}H^{\mu \nu }\, \overline{u}(k')\gamma _{\nu }u(k).
\end{equation}
Note that in Eqs.(\ref{eq_2_4}, \ref{eq_2_8})
we have extracted a factor \( e^{3} \) for convenience. The hadronic
tensor \( H^{\mu \nu } \) depends on the proton structure. At this point we do not
want to make hypothesis which are specific of a model. 
We only assume that it is described by a strong interaction
Hamiltonian \( H_{S} \) which depends on some generic charged field and its
momentum (\( \psi ,\pi ). \) Under a gauge transformation they transform as
\begin{eqnarray}
\label{eq_2_9} \psi \rightarrow \exp (-ie\alpha (x))\psi ,\, \pi \rightarrow 
\exp (ie\alpha (x))\pi ,
\end{eqnarray}
with analogous
transformation for \( (\psi ^{*},\, \pi ^{*}) \). The distinction between Fermi and Bose fields
is irrelevant here. For definiteness we assume canonical commutation
rules
\begin{equation}
\label{eq_2_10}
\left[ \pi (\vec x ,\, t),\psi (\vec x \, ',\, t)\right] =-i\delta (\vec x -\vec x \, ').
\end{equation}
The explicit form of \( H_{S} \) is not necessary for the moment. We simply
assume the following form
\begin{equation}
\label{eq_2_11}
H_{S}=\int d\vec x \, {\cal H_{S}}\left( \pi ,\, \psi ,\, \partial _{i}\psi ,\, \pi ^{*},\, \psi ^{*},\, \partial _{i}\psi ^{*}\right) ,
\end{equation}
and that \( H_{S} \) is invariant under the transformations
(\ref{eq_2_9}) when \( \alpha  \) is constant. In general \( \cal H_{S} \) does not depends on the gradient
of the momenta but including such a dependence would not lead to any
difficulty, apart the fact that the formulae would be longer. Also
one could include non localities in \( {\cal H_{S}} \) which are stronger than the
first derivative. 

To introduce the coupling to the electromagnetic field one imposes
that the Hamiltonian equations of motion 
\begin{eqnarray}
\frac{\partial \psi }{\partial t} & = & i\left[ H_{S},\, \psi \right] \;,\nonumber \\
\label{eq_2_12} \frac{\partial \pi }{\partial t} & = & i\left[ H_{S},\, \pi \right] \;,
\end{eqnarray}
be invariant under the gauge
transformations (\ref{eq_2_9}). This is achieved by the minimal substitution
\begin{eqnarray}
\partial _{i}\psi  & \rightarrow  & \partial _{i}\psi +i\, e\, A_{i}\,\psi ,\nonumber \\
\partial _{i}\psi ^{*} & \rightarrow  & \partial _{i}\psi ^{*}-i\, e\, A_{i}\,\psi ^{*},\nonumber \\
\label{eq_2_13} {\cal H}_{S} & \rightarrow  & {\cal H}_{S}+ie\left( \psi ^{*}\pi ^{*}-\psi \pi \right) A_{0},
\end{eqnarray}
where
the gauge transformation for the electromagnetic field \( A_{\mu } \) is
\begin{equation}
\label{eq_2_14}
A_{\mu }\rightarrow A_{\mu }+\partial _{\mu }\alpha .
\end{equation}

We need to compute \( T^{FVCS} \)only to lowest significant order in \( e \). Therefore
when we perform the substitution (\ref{eq_2_13}) in Eq.(\ref{eq_2_11}) we only need to expand
the Hamiltonian to order \( e^{2}. \) This leads to the effective Hamiltonian
\begin{equation}
\label{eq_2_15}
H_{eff}=H_{S}+V,\, \, \, \, \, \, \rm with:\, \it V=\int d\vec x \, e\, j_{\mu }A^{\mu }
+\frac{e^{2}}{2}\, S_{\mu \nu }A^{\mu }A^{\nu } \;,
\end{equation}
where
we have defined \scshape 
\begin{eqnarray}
j^{0} & = & i\psi ^{*}\pi ^{*}+{\rm h.c.},\, \, \, 
j^{i}=-i\psi ^{*}\frac{\partial \cal H_{S}\it }
{\partial \partial _{i}\psi ^{*}}+{\rm h.c.},\nonumber \nonumber \\
\label{eq_2_16} S^{ij} & = & \psi \psi ^{*}
\frac{\partial ^{2}{\cal H}_{S}}
{\partial \partial _{i}\psi \, \partial \partial _{j}\psi ^{*}}
-\psi ^{2}\frac{\partial ^{2}{\cal H}_{S}}{\partial \partial _{i}\psi \, \partial \partial _{j}\psi }
+{\rm h.c.},\nonumber \nonumber \\
 S^{\mu 0} & = & S^{0\mu }=0 \;,
\end{eqnarray}
\upshape and h.c. denotes the hermitian
conjugate. If we note the field strength 
\( F^{\mu \nu }=\partial ^{\mu }A^{\nu }-\partial ^{\nu }A^{\mu } \), 
Eq.(\ref{eq_2_15}) implies the field equations
\begin{equation}
\label{eq_2_16a}
\partial _{\mu }F^{\mu \nu }=ej^{\nu }+e^{2}S^{\nu \mu }A_{\mu } \;.
\end{equation}

Note that the contact (or seagull ) term \( \mathcal{S}^{\mu \nu } \) has only space components.
This is characteristic of the Hamiltonian formulation where the variables
are the fields and their momenta. In the Lagrangian formalism the
time components \( S^{0\mu },\, S^{\mu 0} \) of the contact term are not zero but they disappear
when one builds the Hamiltonian. Here the use of the Hamiltonian
formalism is compulsory because we must use perturbation theory
without too specific hypotheses about the structure. So even if we
had started from the Lagrangian we would have been forced to go to
the Hamiltonian at some point.

By computing the effect of \( V \) in Eq.(\ref{eq_2_15}) 
in perturbation theory to lowest significant order
in \( e \) one gets the following expression for \( H^{\mu \nu } \)

\begin{eqnarray}
H^{\mu \nu } & = & \int d\vec x \, {\rm e}^{-i\vec q \, '.\vec x }<\vec p \, '|j^{\mu }(\vec x ,0)\frac{1}{p_{0}'+\rmqp -H_{S}+i\epsilon }j^{\nu }(0)\nonumber \\
&& \hspace{2.5cm} + j^{\nu }(0)\frac{1}{p_{0}-\rmqp -H_{S}+i\epsilon }j^{\mu }(\vec x ,0)|\vec p >+H_{seagull}^{\mu \nu },\nonumber \\
\label{eq_2_17} H_{seagull}^{\mu \nu } & = & < \vec p \, '|S^{\mu \nu }(0)| \vec p >.
\end{eqnarray}
This can also be written
\begin{equation}
\label{eq_2_18}
H^{\mu \nu }=-i\int d^{4}x\, {\rm e}^{-iq.x}
< p \, '|T\left[ j^{\nu }(x),j^{\mu }(0)\right] | p >
+ H^{\mu \nu }_{seagull} \;,
\end{equation}
with \( T \) the time ordering symbol, and since
the equations of motion are by construction gauge invariant, \( H^{\mu \nu } \) satisfies
the gauge invariance conditions

\[
q_{\mu }'H^{\mu \nu }=H^{\mu \nu }q_{\nu }=0.\]
In the C.M. frame one has \( \vec p =-\vec q  \) and \( \vec p \, '=-\vec q \, ' \). The hadronic tensor, which depends
on \( (\vec p ,\, \, \vec p \,' \)) and the spin labels
 \( (\sigma ,\, \, \sigma ' \)) or helicity labels \( (h ,\, \, h'\)) 
 will be noted
  \( H^{\mu \nu }(\vec q \, ',\sigma ',\vec q ,\sigma ) \) or 
\( H^{\mu \nu }(\vec q \, ',h',\vec q ,h) \)   when necessary .

\subsection{The VCS amplitude}

It is convenient to expand the lepton current which appears in Eq.(\ref{eq_2_8})
on the basis of the polarization vector of the virtual photon which
are defined in App. \ref{app1}. 
Using the expressions for the helicity spinors (App. \ref{app1}) one gets
\begin{eqnarray}
 &  & \overline{u}(k',h)\gamma ^{\nu }u(k,h)=\sum _{\lambda =0,\pm 1}
 \Omega (h,\lambda )\varepsilon ^{\nu }(q,\lambda ),\nonumber \\
 &  & \Omega (h,\lambda )=\left( -\lambda {\rm e}^{-i\lambda \phi }\alpha (\lambda h)
 +\delta (\lambda ,0)\frac{q_{0}}{Q}\sqrt{2\varepsilon }\right) 
 \frac{Q}{\sqrt{1-\varepsilon }},\nonumber \\
&  & \alpha (z)=\frac{\sqrt{1+\varepsilon }
+2z\sqrt{1-\varepsilon }}{\sqrt{2}},\nonumber\\ 
\label{eq_2_20}  & &\varepsilon ={\rm \frac{(k+k')^{2}-q^{2}}{\left( k+k'\right) ^{2}+q^{2}}} 
\;=\;\left[ 1 + 2 {{\rm q^2} \over {Q^2}} \tan^2 \theta_e/2 \right]^{-1}\;,
\end{eqnarray}
where \( \varepsilon  \) is the usual linear polarization parameter.
It is equal to 1 when the virtual photon is linearly polarized in
the lepton scattering plane. We \em define \em the VCS amplitude by 
\begin{equation}
\label{eq_2_21}
T^{VCS}(\lambda ',\lambda )=\varepsilon _{\mu }'^{*}(\lambda ')H^{\mu \nu }\varepsilon _{\nu }(\lambda ),
\end{equation}
which
allows us to write 
\begin{equation}
\label{eq_2_21a}
T^{FVCS}(\lambda ')=\frac{e^{3}}{-Q^{2}}\sum _{\lambda }\Omega (h,\lambda )T^{VCS}(\lambda ',\lambda ).
\end{equation}
Thanks to the decoupling of the lepton current
in Eq.(\ref{eq_2_21a}) one can take the limit \( Q^{2}\rightarrow 0 \) in \( T^{VCS}. \) In this limit the transverse
part \( T^{VCS}(\lambda =\pm 1) \) coincides (up to a factor \( e^{2} \)) with the amplitude for real Compton
scattering.

\subsection{Tensor expansion of the VCS amplitude}

\def\pol{\vec\varepsilon} \def\polp{\vec\varepsilon\,'^*}\def\Ncal{\cal
 N\it}

It is often useful to have a general expansion of the VCS amplitude
defined by Eq.(\ref{eq_2_21a}). A convenient, though not unique, tensor expansion
in the C.M. frame and in Lorentz gauge has been proposed in \cite{Gui95}.
For the longitudinal part \( (\lambda =0) \) it has the form

\begin{eqnarray}
\Ncal ^{-1}T^{VCS} & = & a^{l}\polp .\hat q \nonumber \\
\label{eq_2_22} & + & i[b^{l}_{1}\polp .(\hat q \times \hat q \, ') \vec\sigma .\vec e (1) 
+ b^{l}_{2}\polp .\hat q \vec\sigma .\vec e (2) 
+ b^{l}_{3}\polp .(\hat q \times \hat q \, ')\vec\sigma .\vec e (3)] \;,
\end{eqnarray}
and for the transverse one \( (\lambda =\pm 1) \)
\begin{eqnarray}
\Ncal ^{-1}T^{VCS} & = & a^{t}\polp .\pol + a '^{t}\polp .\hat q \pol .\hat q \, 
'\nonumber \\
 & + & i[b^{t}_{1}\polp .\hat q \pol .(\hat q \times \hat q \, ') + 
 b_{1}'^{t}\polp .(\hat q \times \hat q \, ')\pol .
 \hat q \, ']\vec\sigma .\vec e (1)\nonumber \\
 & + & i[b^{t}_{2}\polp .\pol 
 + b_{2}'^{t}\polp .\hat q \pol .\hat q \, ']\vec\sigma .\vec e (2)
 \nonumber \\
\label{eq_2_23}  & + & i[b_{3}^{t}\polp .\hat q \pol .(\hat q \times \hat q \, ')
+b_{3}'^{t}\polp .(\hat q \times \hat q \, ')\pol .
\hat q \, ']\vec\sigma .\vec e (3) \;,
\end{eqnarray}
where the normalisation factor \( {\cal N}=2\sqrt{p_{0}p_{0}'} \) is
for future convenience and \( \vec\sigma  \) are the Pauli matrices acting on the
nucleon spin. The helicity labels \( \lambda ,\, \lambda ' \) 
have been omitted to simplify the equations.

The 12 scattering coefficients \( (a^{l},\, b_{1}^{l},...) \) 
are complex functions of \( (\rm q,\, q',\, \cos \theta ) \) which become
real below the pion threshold due to time reversal invariance \cite{Gui95}.

\subsection{The ( $e, e'\,\gamma$ ) cross sections}

In this section, we are giving the observables 
of the (\(e,\ e'\,\gamma\) ) reaction. 
Due to the small cross sections the experiments will
require such high luminosities that the use of polarized hydrogen targets
seems excluded. Moreover detecting the polarization of the outgoing electron 
or photon looks impossible as this can be done only through 
another electromagnetic process, thereby decreasing by a factor $\sim 1/137$ 
the already small counting rates. Therefore it looks
reasonable to restrict our considerations to the case of a polarized beam and
to the detection of the recoil proton polarization.

Let (\(\xi_e^\mu,\,\xi_p^\mu\)) the 4-vectors which describe the polarization
state of the initial electron and final proton. By definition \(\xi^\mu\) is
 orthogonal to the particle momentum (that is \( \xi_e.k=\xi_p.p'=0\)),  and,
  {\em in the particle rest frame}, \(\vec{\xi}\) 
 is twice the expectation value of the particle 
 spin . We define the following density matrices 
\begin{eqnarray}
\rho^e_{h\overline{h}}(\xi^e)&=&\frac{1}{2m_e}
\overline{u}(k,h)\frac{(1+\gamma^5\gamma.\xi^e)}{2}u(k,\overline{h}),\nonumber\\
\rho^p_{\sigma'\overline{\sigma}'}(\xi^p)&=&\frac{1}{2m}
\overline{u}(p',\sigma')\frac{(1+\gamma^5\gamma.\xi^p)}{2}u(p',\overline{\sigma}').
\label{eq_2_220}
\end{eqnarray}
and the Lorentz invariant quantity

\begin{equation}
\label{eq_2_22a}
{\cal M}(\xi^e ,\xi^p)=\frac{1}{2}
\sum _{\sigma \sigma '\lambda 'h h'}
T^{ee'\gamma}(\lambda '\sigma 'h',\sigma h)\rho^e_{h\overline{h}}(\xi^e)
T^{ee'\gamma\dagger}(\lambda '\overline{\sigma}'h',\sigma \overline{h})
\rho^p_{\overline{\sigma}',\sigma '}(\xi^p),\label{eq_2_23b}
\end{equation}
in terms of which the invariant cross section is 
\begin{equation}
\label{eq_2_23a}
d\sigma(\xi^e,\xi^p)=\frac{\left( 2\pi \right) ^{-5}}{32\sqrt{(p.k)^{2}-m^{2}m^{2}_{e}}}
\frac{d\vec k \, '\, d\vec p \, '\, d\vec q \, '}{k_{0}'\, p_{0}'\, q_{0}'}
\delta ^{4}(p+k-p'-k'-q'){\cal M}(\xi^e ,\xi^p).
\end{equation}
Note that in the above expression, the states are supposed to be normalized as 
\begin{equation}
\label{eq_2_24}
<\vec p |\vec p \, '>=(2\pi )^{3}2p_{0}\delta (\vec p -\vec p \, ').
\end{equation}
After integration over
the energy of the final photon and defining
\begin{equation}
{\cal M}={\cal M}(0,\xi^p)+{\cal M}(0,-\xi^p) =2{\cal M}(0,0), 
\label{eq_2_24a}
\end{equation}
one gets
for the unpolarized cross section

\begin{equation}
\label{eq_2_25}
\frac{d\sigma }{dQ^{2}dsdtd\phi d\phi _{e}}
= \frac{(2\pi )^{-5}}{128 m^{2} k_{0,lab}^{2}\Lambda(s,-Q^2,m^2)}{\cal M} \;,
\end{equation}
where the azimuthal angle of the electron \( \phi _{e} \) is arbitrary and 
\begin{equation}
\Lambda (x,y,z)=\sqrt{x^2+y^2+z^2-2xy-2xz-2yz}
\label{eq_2_25a}.
\end{equation}

The experiment is often performed by detecting the electron in coincidence
with the proton and the \( (e,\ e'\,\gamma ) \) event is tagged by a zero missing mass,
\( (k+p-k'-p')^{2}=0. \) The corresponding laboratory cross section is then
\begin{equation}
\label{eq_2_26}
\frac{d\sigma }{d{\rm k}'_{lab}d\hat k \, '_{lab}d\hat p '_{cm}}=\frac{(2\pi )^{-5}}{64m}\frac{{\rm k'}_{lab}}{{\rm k}_{lab}}\frac{s-m^{2}}{s}{\cal M}.
\end{equation}

The next interesting observable is the single (electron) spin asymmetry (SSA) because
it 
should be measurable with the existing highly polarized beams. We define it as

\begin{equation}
{\cal A}=\frac{ {\cal M}(\xi^e,0)-{\cal M}(-\xi^e,0)}
{ {\cal M}(\xi^e,0)+{\cal M}(-\xi^e,0)}.
\label{eq_2_27}
\end{equation}
In practice \(\xi^e\) is proportional to $k$ which corresponds to a 
longitudinal polarization of the electron. 
Then, because of time reversal invariance and up to corrections of 
order \(\alpha_{QED}\), one has \({\cal A}=0\) below the
pion production threshold (see discussion of Section \ref{sec4}). 
Also   \({\cal A}\)  vanishes by reflection symmetry
if the electron and hadron planes  coincide.

Finally for a fixed electron polarisation \(\xi^e\) one can try to measure the average
polarization \({\cal P}\) of the recoiling proton . The components along
\(\xi^p\) of this double polarization observable are
\begin{equation}
-{\cal P}.\xi^p=
\frac{ {\cal M}(\xi^e,\xi^p  )-{\cal M}(\xi^e,-\xi^p )}
{ {\cal M}(\xi^e,\xi^p  )+{\cal M}(\xi^e,-\xi^p ) } \;.
\label{eq_2_28}
\end{equation}

\subsection{Virtual photon cross sections}

The ($e,\ e'\gamma$) reaction is somewhat unique among the electroproduction
reactions due to the presence of the BH process. However, there may be  situations
where this process  can be neglected in which case it is interesting to present
the cross sections in the usual manner that is in terms of virtual photon cross
sections. To this aim we define the following helicity amplitudes \cite{Kro96}

\begin{eqnarray}
&&  \Phi_{1}=e^2 T^{VCS}(\ \ 1\ \  1/2,\ \ 1\ \ 1/2) \;,\nonumber\\
&&  \Phi_{2}=e^2 T^{VCS}(-1-1/2,\ \ 1\ \  1/2) \;,\nonumber\\ 
&&  \Phi_{3}=e^2 T^{VCS}(-1\ \  1/2,\ \ 1\ \  1/2) \;,\nonumber\\
&&  \Phi_{4}=e^2 T^{VCS}(\ \ 1-1/2,\ \ 1\ \  1/2) \;,\nonumber\\
 \nonumber\\
&&\Phi_{5}=e^2 T^{VCS}(\ \ 1-1/2,\ \ 1-1/2) \;,\nonumber\\
&&\Phi_{6}=e^2 T^{VCS}(-1\  \ 1/2,\ \ 1-1/2) \;,\nonumber\\
&&\Phi_{7}=e^2 T^{VCS}(-1-1/2,\ \ 1-1/2) \;,\nonumber\\
&&\Phi_{8}=e^2 T^{VCS}(\ \ 1\ \ 1/2,\ \ 1-1/2) \;,\nonumber\\
 \nonumber\\
&&\Phi_{9}=e^2 T^{VCS}(\ \ 1\ \ 1/2,\ \ 0\ \  1/2)\,q_0/Q \;,\nonumber\\
&&\Phi_{10}=e^2 T^{VCS}(-1-1/2,\ \ 0\ \ 1/2)\,q_0/Q \;,\nonumber\\
&&\Phi_{11}=e^2 T^{VCS}(-1\ \  1/2,\ \ 0\  \ 1/2)\,q_0/Q \;,\nonumber\\
&&\Phi_{12}=e^2 T^{VCS}(\ \ 1-1/2,\ \ 0\ \  1/2)\,q_0/Q \;,
\label{eq_2_29}
\end{eqnarray}
where $T^{VCS}(\lambda ',h_N',\ \lambda ,h_N)$ is the amplitude defined in
Eq.(\ref{eq_2_21}). The factor $q_0/Q$ which multiplies the longitudinal
amplitudes ($\Phi_9$ to $\Phi_{12}$) is a matter of convention 
and is consistent with the polarization vector of App.\ref{app1}.
\newline
\begin{minipage}[b]{13.9cm}
\hspace{.45cm}
The unpolarized ($e,\ e'\gamma$) cross section where one retains only the VCS
contribution reads then 
\begin{eqnarray}
\label{eq_2_30}
\frac{d\sigma}{d{\mathrm k}_{lab}'\,d\hat{k}_{lab}'\, d\phi\, dt}=\frac{1}{2\pi}\,\Gamma_v\,
\left(\frac{d\sigma_T}{dt}+\varepsilon\,\frac{d\sigma_L}{dt} 
+\varepsilon\cos{2\phi}\,\frac{d\sigma_{TT}}{dt}
+\sqrt{2\varepsilon(1+\varepsilon)}\cos{\phi}\,\frac{d\sigma_{LT}}{dt}\right)\;,
\end{eqnarray}
where  the virtual photon flux is
\begin{equation}
\Gamma_v=\frac{\alpha_{QED}}{2\pi^2}
        \frac{k_{0,lab}'}{k_{0,lab}}
        \frac{s-m^2}{2 m \, Q^2 (1-\varepsilon)} \;,
\label{eq_2_31}
\end{equation}
and the various virtual photon cross sections \footnotemark[1] are defined as 
\footnotetext[1]{Note that only 
$\sigma_T$ and $\sigma_L$ can be considered as cross sections. The other ones
may be negative.} 
\end{minipage}

\noindent
i) The cross-section for transverse photons (which, at $Q^2=0$, reduces to the 
unpolarized cross section $d\sigma/dt$ for real Compton scattering)
\begin{equation}
\label{sigt}
\frac{d\sigma_T}{dt}=\frac{c}{2}\;\sum_{i=1}^{8}\; |\Phi_i|^2 \;.
\end{equation}
ii) The cross-section for longitudinal photons 
\begin{equation}
\label{sigl}
\frac{d\sigma_L}{dt}=c\;\sum_{i=9}^{12}\; |\Phi_i|^2 \;.
\end{equation}
iii) The transverse-transverse interference term
\begin{equation}
\label{sigtt}
\frac{d\sigma_{TT}}{dt}=
-\frac{c}{2}\;\Re e\left[\Phi_1^\ast\,\Phi_7-\Phi_2^\ast\,\Phi_8
                     +\Phi_3^\ast\,\Phi_5-\Phi_4^\ast\,\Phi_6\right] \;.
\end{equation}
iv) The longitudinal-transverse interference term
\begin{equation}
\label{siglt}
\frac{d\sigma_{LT}}{dt}=-\frac{c}{\sqrt{2}}
\;\Re e\left[\Phi_9^\ast\,(\Phi_1-\Phi_7)
      +\Phi_{10}^\ast\,(\Phi_2+\Phi_8)
      +\Phi_{11}^\ast\,(\Phi_3-\Phi_5)
      +\Phi_{12}^\ast\,(\Phi_4+\Phi_6)\right] \;.
\end{equation}
The two-body phase space factor $c$ is given by
\begin{equation}
c=\frac{1}{16\pi (s-m^2)^2 } \;.
\end{equation}

\newpage

\section{THE THRESHOLD REGIME AND THE POLARI\-ZA\-BILITIES}
\label{sec3}

\def\exp{\hbox{\rm e}}

\def\rmqp{{ \hbox{\rm q}'} }

\def\rmq{{\hbox{\rm q}}}

\def\qt0{\tilde{q}_0}

\def\gqt{\tilde{Q}}

\def\calm{{\cal M}}

\def\dcalm{\Delta{\cal M}}

\def\rmkt{ {\rm k}_T}

\subsection{Introduction}

The best way to understand the physics case of the VCS at threshold
is to imagine that the final photon plays the role of an external
applied field \( A^{ext} \). This may look strange but is possible provided the
applied field has the adequate asymptotic conditions. Near the threshold,
the energy of the final photon of VCS is small. So the applied field
and its electric and magnetic fields are almost constant in time and
space. So \em 
VCS at threshold can be interpreted as electron scattering by a target
which is in constant electric and magnetic fields. \em The physics
is exactly the same as if one were performing an \em elastic \em electron
scattering experiment on a target placed between the plates of a capacitor 
or between the poles of a magnet.
\newline
\indent
From elementary physics we all know what happens: under the influence
of the applied field, the charge \( J^{0} \) and the current density 
\( \vec J  \) inside
the target get modified. For a weak field this modification is linear
in the field and the coefficients of proportionality are the so called
polarizabilities. For a uniform isotropic medium only two are necessary:
the electric and magnetic ones. In the general case one needs
a tensor which moreover depends on the space point. If we note \( \delta J^{\mu }(x) \) the
modification of the current density we have, in the linear response
approximation
\begin{equation}
\label{eq_3_1}
\delta J^{\mu }(x)=\int d^{4}yP^{\mu \nu }(x,y)A^{ext}_{\nu }(y).
\end{equation}
The polarizability tensor \( P^{\mu \nu } \) actually quantifies how
the system adjusts its internal structure to the applied field. The
way to measure it is obviously to measure \( \delta J^{\mu }(x) \) and we do this by electron
scattering, exactly in the same way as we measure the \em equilibrium
\em current \( J^{\mu }(x) \) of a free target. In the latter case we know that the
experiment can be interpreted in term of the Fourier transform 
of \( J^{\mu }(x) \), given by
\begin{equation}
\label{eq_3_2}
J^{\mu }(q)=\int d^{4}x\exp^{iq.x} J^{\mu }(x).
\end{equation}
Similarly
a VCS experiment measures the Fourier transform of \( \delta J^{\mu }(x) \), the current
induced by the applied field.
\newline
\indent
At this point we can see how the VCS generalizes the real Compton
scattering. Though we always think of the zero energy limit, in practice
this energy is of course never exactly zero. The corresponding non
zero frequency of \( A^{ext} \) induces a non uniform time dependence in \( \delta J^{\mu }(x) \) which
then radiates an electromagnetic field. The latter, 
having a non uniform time dependence, decreases only as \( 1/r \). 
So it can be detected at large distance. 
This is the final real photon of the Compton scattering. What we get is again
the Fourier transform of \( \delta J^{\mu }(x) \) but now computed along the real photon
line, that is for \( q^{0}=\left| \vec q \right| \simeq 0 \) by energy conservation. This is to be contrasted
with the case of VCS for which \( q^{0} \) and \( \vec q  \) are independent. 
Moreover since the photon is transverse in the real Compton scattering,
only the transverse component of \( \delta J^{\mu }(x) \) come into play. 
\newline
\indent
Now we come to the difficult point of the problem, at least for the
experimentalists. In the above reasoning we have tacitely assumed
that \( \delta J^{\mu }(x) \) contains only the response of the internal degrees of freedom
of the target. The reason is that it is the interesting part. However
if the target has a global charge and/or a global magnetic moment,
\( \delta J^{\mu }(x) \) also contains a trivial part due to the global 
motion under the influence of the applied field. 
If we put a proton in an electric
field the first effect that we observe is that it moves as a whole.
Similarly the magnetic field produces a precession of the magnetic
moment. Since in a weak field the acceleration is not large, if we
scatter an electron on this object we will learn no more than what
we learn with a {\it fixed} target! This problem is absent when one studies
 the polarizability of a macroscopic sample because it can be
fixed in space by appropriate means, which is not possible for the proton.
This absence of restoring force explains why the trivial response
due to the motion as a whole of the position and magnetic moment dominates
over the response of the internal degrees of freedom. This is the
physical origin of the low energy theorem \cite{Low58} for VCS. 
\newline
\indent
According to the above development, it is clear that to extract a
physically interesting information from a VCS experiment one will
have to subtract the trivial part which is due to the global motion.
This part must be defined carefully and this will be done in a precise
way through the low energy theorem. As stated above, this theorem originates
from the fact that at low energy the response is dominated by the
global motion of the proton in the external field. So one can ignore
the internal structure to calculate this part of the response. All
what we need are the parameters which control the motion, that is
the mass, the charge, and the magnetic moment. Once the motion is
known, we have to compute the amplitude for scattering an electron
on this moving proton. For this we only need the elastic form factors
and they are known. The calculation in this way is rather involved,
mainly due to relativistic effects. This has been done only in the
case of real Compton scattering \cite{Gel54}. In the VCS case, one prefers
to go directly to the quantum derivation where the use of gauge invariance
and perturbation theory greatly simplifies the work. Nevertheless
we think it is useful to keep in mind that the LET has a well identified
classical origin.

\subsection{Electron scattering in an external field}

\begin{minipage}[b]{13.9cm}
Let us put the above considerations on a more quantitative level.
In the presence of the external field the total \footnotemark[2]  
\footnotetext[2]{that is the one that scatters the electron} current is, from 
Eq.(\ref{eq_2_16a})
 
\begin{equation}
\label{eq_3_3}
J^{\mu }=ej^{\mu }+e^{2}S^{\mu \nu }A_{\nu }^{ext}.
\end{equation}
Note that we have set \( A=A^{ext} \) in the seagull term as the other part of the
field would only give a contribution of order \( e^{3} \) or more.

\hspace{0.45cm}
To get the full current we need to evaluate the effect of \( A^{ext} \) on the
motion. In quantum mechanics this amounts to compute the modification
of the nucleon state by \( A^{ext} \) . The perturbation is given by Eq.(\ref{eq_2_15})
but clearly the seagull term would again produce an effect of order
\( e^{3} \). So we can write the relevant Schroedinger equation in the form
\begin{equation}
\label{eq_3_4}
i\frac{\partial }{\partial t}|t>=(H_{S}+V)|t>,
\end{equation}
with 
\begin{equation}
\label{eq_3_5}
V=e\int d\vec x j^{\mu }A^{ext}_{\mu }=e \hat{V}.
\end{equation}
Note that we work in Schroedinger representation so 
that \( j^{\mu } \) does not depend on time. 
\vspace{0.2cm} 
\end{minipage}

\noindent
To specify the boundary conditions we assume that the electron 
scattering happens at time~\( t \).  
The scattering takes place on a hadronic
state \( |t,\, in> \) whose time evolution under the perturbation due to \( A^{ext} \) is such
that
\begin{equation}
\label{eq_3_6}
|t,\, in>\rightarrow |N_{i}>  {\rm when}\, t\rightarrow -\infty .
\end{equation}
Similarly the scattering leads to a state \( |t,\, out> \) such that
\begin{equation}
\label{eq_3_7}
|t,\, out>\rightarrow |N_{f}>  {\rm when}\, t\rightarrow +\infty .
\end{equation}
To make
the boundary conditions consistent with the dynamics, we assume an adiabatic
switching of the interaction by multiplying \( V \) by \( \exp ^{\epsilon t} \) 
when computing
\( |t,\, in> \)and by \( \exp ^{-\epsilon t} \) when computing \( |t,\, out>. \)

Using standard time dependent perturbation theory we get to lowest
order

\begin{eqnarray}
|t,\, in> & = & \exp ^{-iH_{S}t}\left( 1-i\int ^{t}_{-\infty }dt'\exp ^{iH_{S}t'}V\exp ^{\epsilon t'}\exp ^{-iH_{S}t'}\right) |N_{i}>\nonumber \\
\label{eq_3_8} |t,\, out> & = & \exp ^{-iH_{S}t}\left( 1-i\int _{t}^{\infty }dt'\exp ^{iH_{S}t'}V\exp ^{\epsilon t'}\exp ^{-iH_{S}t'}\right) |N_{f}>.
\end{eqnarray}
At low energy we know that the target feels essentially a constant
electric and magnetic fields \( (\vec E ,\, \vec B ). \) In this limiting situation the perturbation
is time independent and, combining Eqs.(\ref{eq_3_3}, \ref{eq_3_8}) we get 
\begin{eqnarray}
<t,\, out|J^{\mu }(\vec r )|t,\, in> & = & \exp ^{i(E_{f}-E_{i})t}
\left\{<N_{f}|ej^{\mu }(\vec r )+e^{2}S^{\mu \nu }(\vec r )A_{\nu }^{ext}(\vec r
)\right.\nonumber \\
\label{eq_3_9}   & + & \left.  e^{2}(\hat{V}\frac{1}{E_{f}-H_{S}+i\epsilon }j^{\mu }(\vec r )
+j^{\mu }(\vec r )\frac{1}{E_{i}-H_{S}+i\epsilon }\hat{V})|N_{i}>\right\},
\end{eqnarray}
and we can identify
the induced current as the part which goes like \( e^{2}. \)

In Eq.(\ref{eq_3_9}) we see that if we insert a complete set of intermediate
states between \( V \) and \( j^{\mu } \) the state corresponding to the nucleon itself
is singular. This pathology is due to the fact that we have taken
the strict limit of a constant field and is the quantum manifestation
of the global motion discussed previously. Since this part contains
no new information, we define the intrinsic induced current by excluding
the nucleon from the intermediate states. Since the next possible
state is a nucleon and a pion, the energy difference is then finite
and we can set \( \epsilon =0. \) Thus we get (at \( t=0 \) )
\begin{eqnarray}
\delta J_{int.}^{\mu }(\vec r ) & = & 
e^{2}\left\{<N_{f}|S^{\mu \nu }(\vec r )A_{\nu }^{ext}(\vec r )|N_{f}>\right. \nonumber \\
\label{eq_3_10} & + & \left. \sum _{n\neq N}\frac{<N_{f}|\hat{V}|n><n|j^{\mu }(\vec r )|N_{i>}}
{E_{f}-E_{n}}+\frac{<N_{f}|j^{\mu }(\vec r )|n><n|\hat{V}|N_{i>}}{E_{i}-E_{n}}\right\}.
\end{eqnarray}
 To simplify a little we set
\( E_{i}=E_{f}=m \) which amounts to neglect the recoil energy of the proton in the electron
scattering event. This is only justified at small \( Q^{2} \) so in the full
treatment later we shall not make this approximation. 

We now consider separately the effect of a constant electric or magnetic
field. The gauge potential corresponding to a constant electric field
is 
\begin{equation}
\label{eq_3_11}
A^{0}_{ext}=-\vec r .\vec E \, , \hspace{1cm} \vec A _{ext}=0.
\end{equation}
Using Eq.(\ref{eq_3_5}) and remembering that the seagull has no time component
we get
\begin{equation}
\label{eq_3_12}
\delta J_{int.,E}^{\mu }(\vec r )=e^{2}\sum _{n\neq N}\frac{<N_{f}|\vec d .\vec E |n><n|j^{\mu }(\vec r )|N_{i}>}{E_{n}-m}+{\rm c.c.},
\end{equation}
where we have defined the dipole moment operator
\begin{equation}
\label{eq_3_13}
\vec d =\int d\vec r \,\vec{ r} j^{0}(\vec r ).
\end{equation}

For a constant magnetic field the gauge field is
\begin{equation}
\label{eq_3_14}
A_{ext}^{0}=0, \hspace{0.5cm} \vec A _{ext}=-\frac{1}{2}\vec r \times \vec B .
\end{equation}
So we get
\begin{eqnarray}
\delta J_{int.,B}^{\mu }(\vec r ) & = & \frac{e^{2}}{2}<N_{f}|S^{\mu i}(\vec r )\varepsilon _{ijk}r^{j}|N_{i}>B^{k}\nonumber \\
\label{eq_3_15}  & + & e^{2}\sum _{n\neq N}\frac{<N_{f}|\vec\mu .\vec B |n><n|j^{\mu }(\vec r )|N_{i}>}{E_{n}-m}+{\rm c.c.},
\end{eqnarray}
where the
magnetic dipole operator is
\begin{equation}
\label{eq_3_16}
\vec\mu =\frac{1}{2}\int d\vec r \, \vec r \times \vec j (\vec r ).
\end{equation}
If we succeed in eliminating the trivial
part of the response, then a low energy VCS experiment will allow
to measure the Fourier transform of the induced currents of Eqs.(\ref{eq_3_12},
\ref{eq_3_15}).

We can compute from Eqs.(\ref{eq_3_12}, \ref{eq_3_15}) 
the induced dipole moments. After
averaging over the nucleon spin projection \( \sigma  \) we get 
\begin{eqnarray}
\delta \vec d  & = & \int d\vec r \, \vec r \delta J_{int.,E}^{0}=\alpha \vec E ,\nonumber \\
\label{eq_3_17} \alpha  & = & \frac{e^{2}}{3}\sum _{n\neq N,\sigma ,\sigma '}
\frac{|<N,\sigma |\vec d \, |n,\sigma '>|^{2}}{E_{n}-m} \;>\; 0.
\end{eqnarray}
For the magnetic dipole we assume
that the seagull term is as in a scalar field theory, that is \( S^{ij}(\vec r )=-2\delta (i,j)j^{0}(\vec r ), \) and
we get
\begin{eqnarray}
\delta \vec\mu  & = & \frac{1}{2}\int d\vec r \, \vec r \times \vec J =(\beta _{para}+\beta _{dia})\vec B ,\nonumber \\
\beta _{para} & = & \frac{e^{2}}{3}\sum _{n\neq N,\sigma ,\sigma '}
\frac{|<N,\sigma |\vec\mu |n,\sigma '>|^{2}}{E_{n}-m} \;>\;0, \nonumber \\
\label{eq_3_18} \beta _{dia} & = & -\frac{e^{2}}{6}
\sum _{\sigma }<N,\sigma |\int d\vec r r^{2}j^{0}(\vec r )|N,\sigma > \;<\;0.
\end{eqnarray}
Note that in the above formulae the polarizabilities \( \alpha  \) and
\( \beta  \) are in Heaviside-Lorentz units. To get them in Gauss units, one
must replace \( e \) by \( e_{G} \).

\subsection{The low energy theorem}

Here we abandon the semi-quantitative picture and go back to the true
reaction. As explained above we need to separate the trivial part
due to the global motion of the proton. However it is easy to see
that this separation is not gauge invariant. By itself this is not
a problem since only the full amplitude must be gauge invariant, but
it is much more comfortable from the theoretical point of view to
deal with two parts which are independently gauge invariant. 

To this aim we start from the expression of the full VCS amplitude,
Eq.(\ref{eq_2_17}), and instead of separating the on mass-shell nucleon
contribution we write
\begin{equation}
\label{eq_3_19}
H=H_{B}+H_{NB},
\end{equation}
with \( H_{B} \)\em , \em the so called Born term, \em defined
\em by
\begin{eqnarray}
H_{B}^{\mu \nu } & = & \overline{u}(p')\Gamma ^{\mu }(p',p'+q')\frac{\gamma .(p'+q')+m}{(p'+q')^{2}-m^{2}}\Gamma ^{\nu }(p'+q',p)u(p)\nonumber \\
\label{eq_3_20}  & + & \overline{u}(p')\Gamma ^{\nu }(p',p-q')\frac{\gamma .(p-q')+m}{(p-q')-m^{2}}\Gamma ^{\mu }(p-q',p)u(p).
\end{eqnarray}
Clearly \( H_{B} \) contains not only the nucleon contribution but also
the pair excitation term. Moreover the vertex \( \Gamma  \) is now evaluated for
off mass shell values of one of its arguments because neither \( (p'+q')^{2} \) nor
\( (p-q')^{2} \) are equal to \( m^{2}. \) The advantage is that, if one uses the vertex decomposition
of Eq.(\ref{eq_3_20}), it is a simple excercise to show that
\begin{equation}
\label{eq_3_21}
q_{\mu }'H_{B}^{\mu \nu }=H_{B}^{\mu \nu }q_{\nu }=0,
\end{equation}
which in turn implies
the gauge invariance of the non Born term \( H_{NB} \).

It has been shown in ref.\cite {Gui95} that \( H_{NB} \) is a regular function
of the 4-vector \( q'^{\mu } \), which amounts to say that \( H_{NB} \) 
has a polynomial expansion
of the form (this also holds for \( q^{\mu } \) but we do not need it for 
the moment)

\begin{equation}
\label{eq_3_22}
H_{NB}^{\mu \nu }=a^{\mu \nu }(q)+b_{\alpha }^{\mu \nu }(q)q'^{\alpha }+c_{\alpha \beta }^{\mu \nu }(q)q'^{\alpha }q'^{\beta }+\cdots 
\end{equation}
Since the final photon is real we have \( q'^{0}=\rmqp  \) and therefore 
we can write

\begin{equation}
\label{eq_3_23}
H_{NB}^{\mu \nu }=a^{\mu \nu }+b_{\alpha }^{\mu \nu }q'^{\alpha }+O(\rmqp ^{2}).
\end{equation}
The gauge invariance of \( H_{NB} \) implies
\begin{equation}
\label{eq_3_24}
0=q_{\mu }'a^{\mu \nu }+O(\rmqp ^{2})=\rmqp a^{0\nu }+q_{i}'a^{i\nu }
+O(\rmqp ^{2}).
\end{equation}
The coefficients \( a^{\mu \nu } \)are constant
with respect to \( \vec q \, ' \). So averaging 
Eq.(\ref{eq_3_24}) over the direction \( \hat q ' \) on the one hand 
or multiplying Eq.(\ref{eq_3_24}) by \( \hat q ' \) and averaging again 
on the other hand, we get 
\begin{equation}
\label{eq_3_25}
0=\rmqp a^{0\nu }=\rmqp a^{i\nu }\;,
\end{equation}
which implies that \( a^{\mu \nu }=0. \) This low energy
theorem (LET), which was first shown in ref.\cite {Low58}, tells us
that the expansion of \( H_{NB} \), the unknown part of the VCS amplitude, starts
at order \( q' \) . Since, for finite \( Q^{2} \) the Born part starts at order \( 1/q' \) we
conclude that the first two term of the VCS amplitude are known as
soon as the form factors which define the vertex \( \Gamma  \) are known.

\subsection{Multipole expansion of \protect\( H_{NB}\protect \)}

The LET defines in a precise way the separation between the trivial
and non trivial part of he VCS amplitude. The non trivial part begins
at order \( q' \) and is contained in \( H_{NB}. \) There is of course a contribution
of order \( q' \) in \( H_{B} \) but it is exactly known and 
therefore can be subtracted,
at least in principle. So what we need now is an adequate parametrisation
of \( H_{NB}. \) For this we use the multipole expansion so as to take advantage
of angular momentum and parity conservation. Such an expansion implies
the choice of a frame and the natural one is the C.M. frame.

Because \( H_{NB} \) is a regular function of \( q \) and \( q' \), we know from elementary
analysis that
\begin{equation}
\label{eq_3_26}
\int d\hat q \, 'H_{NB}^{\mu \nu }(\vec q \, ',\vec q )
Y_{m'}^{l'}(\hat q \, ')\sim \left( \rmqp \right)^{l'}\;,
\end{equation}
where \( \sim  \) means
``at least of order''. A similar property holds for an average over \( \hat q  \). Here
\( Y^{l'}_{m'} \) stands for the spherical harmonic of rank \( (l',m') \) . Since we are interested
only in the first power of \( \rm q' \), we can restrict our considerations to
the multipoles of rank \( l'=1 \), which amounts to say that we are 
keeping only the {\it dipole } contribution of the outgoing photon. 

To factorize the angular and spin dependence of \( H^{\mu \nu }(\vec q \, '\sigma ',\vec q \sigma ) \) we introduce the
reduced multipoles
\begin{eqnarray}
&&4\pi \cal N\it H^{(\rho 'L',\rho L)S}_{NB}(\rmqp ,\rmq )  \nonumber\\
&&= \frac{1}{2S+1}\sum _{\sigma \sigma 'M'M}(-)^{1/2+\sigma '+L+M}<\frac{1}{2}-\sigma ',\frac{1}{2}\sigma |Ss> <L'M',L-M|Ss> \nonumber\\
\label{eq_3_27}  && \hspace{1cm} \int d\hat q \, 'd\hat q \, V_{\mu }^{*}(\rho 'L'M',\, \hat q \, ')
H_{NB}^{\mu \nu }(\vec q \, '\sigma ',\vec q \sigma )V_{\nu }(\rho LM,\hat q ) \;,
\end{eqnarray}
where the basis vectors \( V^\mu(\rho LM,\hat q ) \) are defined in 
Appendix \ref{app2}.
The Clebsch-Gordan coefficients are the same as in Ref.\cite{Edm57}.
In Eq.(\ref{eq_3_27}), $L$ ($L^{'}$) represents the angular momentum of the 
initial (final) electromagnetic transition whereas $S$ differentiates 
between the spin-flip ($S = 1$) or non spin-flip ($S = 0$) transition 
at the nucleon side. 

The index \( \rho  \) can take a priori 4 values: \( \rho =0 \) (charge), \( \rho =1 \) (magnetic),
\( \rho =2 \) (electric), \( \rho =3 \) (longitudinal), but gauge invariance relates the charge
and longitudinal multipoles according to
\begin{eqnarray}
\rmqp H_{NB}^{(3L',\rho L)S}+q_{0}'H^{(0L',\rho L)S} & = & 0 \;,\nonumber \\
\label{eq_3_28} \rmq H_{NB}^{(\rho 'L',3L)S}+q_{0}H_{NB}^{(\rho 'L',0L)S} & = & 0.
\end{eqnarray}
So if we define the new
basis
\begin{eqnarray}
W^{\mu }(\rho LM,\hat q ) & = & V^{\mu }(\rho LM,\hat q ),\, \, \, \, \, \, \rho =1,2,\nonumber \\
\label{eq_3_29} W^{\mu }(0,LM,\hat q ) & = & V^{\mu }(0LM,\hat q )+\frac{q_{0}}{\rmq }V^{\mu }(3LM,\hat q ),
\end{eqnarray}
which satisfy \( q_{\mu }W^{\mu }(\rho LM)=0,\, \, \rho =0,1,2, \) we have the following manifestly gauge invariant
expansion of \( H_{NB}^{\mu \nu } \)

\begin{eqnarray}
&&H_{NB}^{\mu \nu }(\vec q \, '\sigma ',\vec q \sigma ) =  4\pi {\cal N}\sum _{\rho 'L'M'\rho LM}g_{\rho '\rho '}W^{\mu }(\rho 'L'M',\hat q \, ')g_{\rho \rho }W^{\nu *}(\rho LM,\hat q )\nonumber \\
\label{eq_3_30} && \hspace{-1cm} \sum _{Ss}(-)^{1/2+\sigma
  '+L+M}<\frac{1}{2}-\sigma ',\frac{1}{2}\sigma |Ss> <L'M',L-M|Ss> 
H_{NB}^{(\rho 'L',\rho L)S}(\rmqp ,\rmq ),
\end{eqnarray}
where the sum over \( (\rho ',\rho ) \) is now restricted to (0,1,2). The expansion (\ref{eq_3_30})
is valid even if the two photons are virtual but for the real final
photon only the multipole \( \rho '=1,2 \) can appear. This of course does not mean
that the charge multipole does not exist: it is the contraction of the 
basis vector \( W \) with the transverse polarization vector 
of the photon \(\varepsilon '\) which
vanishes.

The selections rules due to parity and angular momentum conservation
are
\begin{equation}
\label{eq_3_31}
S=0,\, 1,\, \, \, \, |L'-S|\leq L\leq L'+S,\, \, \, \, (-)^{\rho '+L'}=(-)^{\rho +L}\;,
\end{equation}
and in addition one knowns that the electric and magnetic multipoles
are zero for \( L=L'=0. \)

\subsection{Low energy behaviour and generalized polarizabilities}

Using Eq.(\ref{eq_3_26}) and the definition of the basis vectors 
\( V^{\mu} \) it is clear
that for small values of \( \rmqp  \) the multipoles behave as
\begin{equation}
\label{eq_3_32}
H_{NB}^{(\rho 'L',\rho L)S}(\rmqp \,, \rmq )\sim \rmqp ^{L'}\, \, \, \, \, \, \, \, \hbox {\rm for}\, \, \rho '=0,1,
\end{equation}
A similar relation holds for small values
of \( \rmq  \). This follows from the fact that in the charge and magnetic vectors
the spherical harmonic has rank \( l=L \). For the electric multipoles one
has the same behaviour, but this does not follow from such a simple
argument because in this case the spherical harmonic has rank \( l=L\pm 1 \). One
has to use the Siegert relation \cite{Sie37} (see also Ref.\cite{Gui95})
which is a consequence of gauge invariance and which states that 
\begin{equation}
\label{eq_3_33}
H_{NB}^{(2L',\rho L)S}(\rmqp ,\rmq )=-\sqrt{\frac{L'+1}{L'}}\frac{q'^{0}}{\rmqp }
H_{NB}^{(0L',\rho L)S}(\rmqp ,\rmq )+O(\rmqp ^{L'+1}) \;,
\end{equation}
with
an analogous relation for the virtual photon. Since \( q'^{0}=\rmqp  \) for the real photon,
and \( q^{0}<\rmq  \) for the (space-like) virtual photon, we see that at small \( (\rmqp ,\rmq ) \)
all the multipoles behave at least as \( \rmqp ^{L'}\rmq ^{L}. \) As we are looking for the
part of \( H_{NB} \) which goes like \( \rmqp  \) and using the
selections rules we find
that, \em a priori, \em this involves the following multipoles
\begin{eqnarray}
H_{NB}^{(11,00)1},\, H_{NB}^{(11,02)1},\, H_{NB}^{(11,22)1},\, H_{NB}^{(11,11)0},\,
 H_{NB}^{(11,11)1}, &  & \nonumber \\
\label{eq_3_34} H_{NB}^{(21,01)0},\, H_{NB}^{(21,01)1},\, H_{NB}^{(21,21)0},\,
 H_{NB}^{(21,21)1},\, H_{NB}^{(21,12)1}. &  & 
\end{eqnarray}
So the low energy (but arbitrary \( \rmq ) \) behaviour of the non Born 
VCS amplitude could be parametrized by 10 functions of \( \rmq  \) defined by
\begin{equation}
\label{eq_3_35}
\hbox {\rm Limit\, of}\, \, \, \, \frac{1}{\rmqp }\frac{1}{\rmq
^{L}}H_{NB}^{(\rho' 1,\rho L)S}(\rmqp ,\, \rmq )\, \, \, \, \hbox {\rm when}\, \, \rmqp \rightarrow 0.
\end{equation}
with ( \(\rho 'L',\rho L)S \) as in Eq.(\ref{eq_3_34}).

However this parametrization is not adequate. 
The first reason is that the \( (\rho '=2,\, L'=1) \) multipoles
actually behave as \( \rmqp  \) because of gauge invariance. It often happens
that in model calculations gauge invariance is violated by some approximation
which otherwise would be innocuous. In this case the definition (\ref{eq_3_35})
may lead to a divergence. The second reason is more serious. In a
VCS experiment the functions defined in Eq.(\ref{eq_3_35}) are measured by taking
the limit \( \rmqp \rightarrow 0 \) at fixed \( \rmq  \). When one
lets \( \rmq \rightarrow 0 \) one would like to recover the
polarizabilities measured in real Compton scattering, but in the latter
case the experiment realizes the limit \( (\rmqp =\rmq \rightarrow 0 )\) 
and there is no guarantee that
the two limits are the same. It is not difficult to show \cite{Gui95}
that for the charge and magnetic multipoles the two limits are actually
the same. This is again due to the fact that in those multipoles the
rank of the spherical harmonic is \( l=L \) or \( l'=L' \). By contrast, from 
Eq.(\ref{eq_3_33}) one
can understand that it will not be the case for the electric multipoles
due to the factor \( q^{0}/\rmq  \) which goes to 1 in the Compton case while it obviously
goes to 0 in the VCS case. This problem was overcome in Ref.\cite{Gui95}
by the introduction of mixed multipoles which were neither electric
nor longitudinal. However the findings of a recent work \cite{Dre98}
 allow for a much more elegant solution. 

First one introduces the generalized polarizabilities (GP's) for the multipoles
which are not problematic, that is 
\begin{eqnarray}
P^{(11,00)1}(\rmq ) & = & \left[ \frac{1}{\rmqp }H_{NB}^{(11,00)1}(\rmqp ,\rmq )
\right] _{\rmqp =0},\, \, P^{(11,02)1}(\rmq )=\left[ \frac{1}
{\rmqp \rmq ^{2}}H_{NB}^{(11,02)1}(\rmqp ,\rmq )\right] _{\rmqp =0} \;,\nonumber \\
\label{eq_3_36} P^{(11,11)S}(\rmq ) & = & \left[ \frac{1}{\rmqp \rmq }H_{NB}^{(11,11)S}
(\rmqp ,\rmq )\right] _{\rmqp =0}.
\end{eqnarray}
Next one considers the multipoles
where only the real photon is electric. In this case one can use Siegert
relation to relate them to the corresponding charge multipoles which
again are safe. So one defines 
\begin{equation}
\label{eq_3_37}
P^{(01,01)S}(\rmq )=\left[ \frac{1}{\rmqp \rmq }H_{NB}^{(01,01)S}(\rmqp ,\rmq )\right] _{\rmqp =0},\, 
\, P^{(01,12)1}(\rmq )=\left[ \frac{1}{\rmqp \rmq ^{2}}H_{NB}^{(01,12)}
(\rmqp ,\rmq )\right] _{\rmqp =0} \;,
\end{equation}
which, using Eq.(\ref{eq_3_33}), specifies the low energy behaviour of
\(H_{NB}^{(21,01)S}\) and \(H_{NB}^{(21,12)1}\). In Ref.\cite{Dre98} it has been
shown that \(P^{(11,00)1}\) is not independent and can be eliminated according to
\begin{equation}
\label{eq_3_39}
P^{(11,00)1}(\rmq)=\sqrt{3}\frac{\rmq^2}{\tilde{q}_0}P^{(01,01)1}-
\frac{1}{\sqrt{2}}\rmq^2 P^{(11,02)1} \;,
\end{equation}
with \( \tilde{q}_{0} \) the limit of \( q_{0} \) when \(
\rmqp \rightarrow 0, \) that is \( \tilde{q}_{0}=m-\sqrt{m^{2}+\rmq ^{2}} \)
.

The remaining problematic multipoles are  
\( H_{NB}^{(11,22)1},\, H^{(21,21)0},\, H^{(21,21)1} \). The great value of 
the work of Refs.\cite{Dre97}\cite{Dre98} is that their
authors have been able to express the low energy limit of these multipoles
in term of the generalized polarizabilities already defined by Eqs.(\ref{eq_3_36},
\ref{eq_3_37}). Using their results, the definitions
(\ref{eq_3_36},\ref{eq_3_37}) and the relations (\ref{eq_3_33},\ref{eq_3_39})
one gets the following low energy  parametrization of all the needed multipoles 

\begin{eqnarray}
H_{NB}^{(11,00)1}(\rmqp ,\rmq ) & = & \rmqp \left(
\sqrt{3}\frac{\rmq^2}{\tilde{q}_0}P^{(01,01)1}(\rmq )-\frac{1}{\sqrt{2}}\rmq^2
P^{(11,02)1}(\rmq)\right)+O(\rmqp^{2}),\nonumber \\
H_{NB}^{(11,02)1}(\rmqp ,\rmq ) & = & \rmqp \rmq^2 P^{(11,02)1}(\rmq )+O(\rmqp
^{2}),\nonumber \\
H_{NB}^{(11,11)S}(\rmqp ,\rmq ) & = & \rmqp \rmq P^{(11,11)S}(\rmq )+O(\rmqp
^{2}),\nonumber \\
H_{NB}^{(21,01)S}(\rmqp ,\rmq ) & = & -\rmqp \rmq \sqrt{2} P^{(01,01)S}(\rmq ) 
+ O(\rmqp^{2}),\nonumber \\
H_{NB}^{(21,12)1}(\rmqp ,\rmq ) & = & -\rmqp \rmq^2 \sqrt{2} P^{(01,12)1}(\rmq ) 
+ O(\rmqp^{2}),\nonumber \\
H_{NB}^{(11,22)1}(\rmqp ,\rmq ) & = & \rmqp \rmq P^{(11,11)1}(\rmq )+O(\rmqp
^{2}),\nonumber \\
H_{NB}^{(21,21)0}(\rmqp ,\rmq ) & = & -\rmqp \tilde{q}_{0}P^{(11,11)0}(\rmq )+O(\rmqp ^{2}),
\nonumber \\
H_{NB}^{(21,21)1}(\rmqp ,\rmq ) & = & -\rmqp\left(2\frac{\rmq ^{2}}
{\tilde{q}_{0}}P^{(11,11)1}(\rmq )-\sqrt{2} \rmq ^{2}P^{(01,12)1}(\rmq )\right)
+O(\rmqp ^{2}),\label{eq_3_40}
\end{eqnarray}
Thus only 6 GP's, that is
\begin{eqnarray}
P^{(01,01)S}(\rmq),\ P^{(11,11)S}(\rmq),\ P^{(11,02)1}(\rmq),\
P^{(01,12)1}(\rmq),\label{eq_3_41}
\end{eqnarray}
are necessary to give the low energy behaviour of \(H_{NB}\). Moreover it was
shown in Ref.\cite{Dre98} that one has the relations
\begin{eqnarray}
P^{(01,01)1}(0)=\ P^{(11,11)1}(0)=0.\label{eq_3_42}
\end{eqnarray}
The relations of Ref.\cite{Dre98} which allow to eliminate 4 of the GP's
introduced in Ref.\cite{Gui95} can be proved by combining the nucleon crossing
symmetry with the charge conjugation invariance. These symmetries hold in any
relativistic theory but to exploit them one needs a covariant parametrization of
the amplitude. The reason is that nucleon crossing relates the reaction 
\begin{eqnarray}
\gamma^* +N(p)\to \gamma +N(p'),\label{eq_3_43}
\end{eqnarray}
to the reaction
\begin{eqnarray}
\gamma^* +\overline{N}(-p')\to \gamma +\overline{N}(-p).\label{eq_3_44}
\end{eqnarray}
So to implement the symmetry one needs an expression where the 4-momentum
formally appears while in the partial wave expansion only the 3 momentum
appears. Though the
correctness of the relations makes no doubt, their physical origin is not yet
clear. In particular one sees on Eqs.(\ref{eq_3_40}) that a purely magnetic GP
parametrizes a purely electric multipole. Since Lorentz
transformations mix the electric and magnetic field, one may suspect that the
origin of the relation is relativistic but a physical interpretation is still
lacking.

\subsection{Generalized polarizabilities in models}

In a static model one neglects all the effects which go like the velocity of the
initial or final nucleon. This allows to derive simple expressions for the GP's.
Here we recall the expressions derived in Ref.\cite{Gui95} for the most
important ones

\begin{eqnarray}
P^{(01,01)0}(\rmq)&=&\sqrt{\frac{2}{3}}\sum_{n\ne N}\frac{1}{m-E_n}\\ \nonumber
                  & & \left[<N|d_z(0)|X><X|d_z(\rmq)|N>+
                   <N|d_z(\rmq)|X><X|d_z(0)|N>\right] \;, \\ \nonumber
P^{(11,11)0}(\rmq)&=&\frac{4}{\sqrt{6}}\sum_{n\ne N}\frac{1}{m-E_n}\\ \nonumber
                  & & \left[<N|\mu_z(0)|X><X|\mu_z(\rmq)|N>+
                   <N|\mu_z(\rmq)|X><X|\mu_z(0)|N>\right] \;,\\ \label{eq_3_45}
\end{eqnarray}
where the generalized dipole moments are defined as
\begin{eqnarray}
\vec{d}(\rmq)&=&\int d\vec{r}\,\frac{3 j_1(\rmq r)}{\rmq r}j^0(\vec{r})\vec{r} 
\;,\\ \nonumber
\vec{\mu}(\rmq)&=&\frac{1}{2}\int d\vec{r}\,\frac{3 j_1(\rmq r)}{\rmq r}
\vec{r}\times\vec{j}(\vec{r}) \;, \label{eq_3_46}
\end{eqnarray}
where \(j_1\) is the spherical Bessel function and we have omitted a possible
seagull term to simplify. 

In the limit \(\rmq\to 0\), if we compare with
 Eqs.(\ref{eq_3_17}, \ref{eq_3_18}) we find the relations
with the usual polarizabilities

\begin{eqnarray}
P^{(01,01)0}(0)&=&-\sqrt{\frac{2}{3}}\frac{\alpha}{e^2}
=-\sqrt{\frac{2}{3}}\frac{\alpha_G}{e_G^2} \;,
\\ \nonumber
P^{(11,11)0}(0)&=&-\sqrt{\frac{8}{3}}\frac{\beta}{e^2}
=-\sqrt{\frac{8}{3}}\frac{\beta_G}{e_G^2} \;.
\label{eq_3_47}
\end{eqnarray}
\newline
\indent
The generalized polarizabilities have been calculated in various more 
or less realistic nucleon structure models. 
A first estimate was proposed using the constituent quark model \cite{Gui95}. 
In Refs.\cite{Met97a}\cite{Met97b}, 
calculations have been given in the linear sigma model. Calculations 
and a discussion of the polarizabilities 
in Chiral Perturbation Theory were given in Refs.\cite{Hem97a} \cite{Hem97b}.

\subsection{Low energy expansion of the scattering coefficients}

The calculation of the observables is easier and more transparent in terms of
the tensor expansion defined in Eqs.(\ref{eq_2_22}, \ref{eq_2_23}). The relation between
the scattering coefficients (\(a^l,\,b^l,\dots\))  and the
multipoles has been derived \cite{Gui95} using a generalization of the
method used in pion electroproduction \cite{Ber67}. Combining the
results of Ref.\cite{Gui95} and the relations of Ref.\cite{Dre98} one gets

\begin{eqnarray}
a^l_{NB}&=&-\sqrt{\frac{3}{2}}\varepsilon_s P^{(01,01)0}(\rmq)\,\rmq\rmqp
+O(\rmqp^2) \;,\nonumber\\
b^l_{NB,1}=\frac{\rmq\cos\theta
-\qt0}{\rmq-\qt0\cos\theta}\, b^l_{NB,2} &=&\frac{3\varepsilon_s}{2\sin\theta}
\left(\frac{\rmq}{\tilde{q}_0} \cos\theta -1\right)P^{(01,01)1}(\rmq)\,\rmq\rmqp
+O(\rmqp^2) \;,\nonumber\\
b^l_{NB,3}&=&-\frac{3\varepsilon_s}{2}\left(\frac{\rmq}{\qt0}P^{01,01)1}(\rmq)
-q\sqrt{\frac{3}{2}}P^{(11,02)1}(\rmq)\right)\,\rmq\rmqp
+O(\rmqp^2) \;,\nonumber\\
a^t_{NB} = - \left(\cos\theta-\frac{\qt0}{\rmq}\right) a'^t_{NB}
&=&\sqrt{\frac{3}{8}}\left(\cos\theta-\frac{\qt0}{\rmq}\right)
P^{(11,11)0}(\rmq)\,\rmq\rmqp+O(\rmqp^2) \;,\nonumber\\
b^t_{NB,1}=-b'^t_{NB,2}&=&\frac{3}{2\sin\theta}
\left(\frac{\rmq^2}{\qt0}P^{(11,11)1}(\rmq)-\sqrt{2}\rmq^2P^{(01,12)1}(\rmq) 
\right)\rmqp
+O(\rmqp^2) \;,\nonumber\\
b'^t_{NB,1} \sim b^t_{NB,2}&=&O(\rmqp^2) \;,\nonumber\\
b^t_{NB,3}=\frac{\rmq\cos\theta
-\qt0}{\rmq-\qt0\cos\theta}\,b'^t_{NB,3}&=&-\frac{3}{2\sin^2\theta}
\left(1-\frac{\rmq}{\qt0} \cos\theta \right)P^{(11,11)1}(\rmq)\,\rmq\rmqp
+O(\rmqp^2) \;,
\label{eq_3_48}
\end{eqnarray}
where NB indicates that we consider only the non Born part of the scattering
coefficients and where $\varepsilon_s = Q^2 / (q_0 \rmq)$.

\subsection{Observables}

We now investigate how one can analyse the \( (e,\ e'\,\gamma)\) observables in
order to extract the GP's. We recall that the observables can be expressed in
term of the Lorentz invariant quantity \({\cal M}(\xi_e,\xi_p)\) 
defined in Section 2. We consider it in the C.M. frame and we choose as
independent variables \( (\rmqp,\rmq,\varepsilon,\theta,\phi) \). 
The tilde will denote
the value of a dependent variable in the limit \(\rmqp=0\).

For the polarization observables we consider an electron beam in a pure helicity
state \(h=\pm 1/2\), which amounts to take \(\xi_e= 2h k/m_e\) 
up to terms of order \( m_e\) that we can neglect. 
(this yields from Eq.(\ref{eq_2_220})
$\rho^e_{h \, h}(\xi^e) = 1/2m_e
\overline{u}(k,h) \, (1+\gamma^5 2 h)/2 \, u(k,h) $ ). 
Below the pion threshold only the recoil polarization \({\cal P}\),
Eq.(\ref{eq_2_28}), is interesting. Since \({\cal P}.p'=0\) one only needs the
space part of  \({\cal P}\) and we consider its components  along the basis [\(
\vec e \, \left( 1\right) ,\, \vec e \, \left( 2\right) ,\, \vec e \,
 \left( 3\right)  \)] defined in Section 2. We shall denote
 \begin{eqnarray}
 \Delta{\cal M}(h,i)&=&{\cal M}\left[\xi_e=2h k/m_e,\
 \vec{\xi}_p=\vec{e}(i)\right]-{\cal M}\left[
 \xi_e=2h k/m_e,\ \vec{\xi}_p=-\vec{e}(i)\right].
\label{eq_3_49}
\end{eqnarray}
We {\it assume} that, at fixed \((\rmq,\varepsilon,\theta,\phi) \), 
the experiment is able to determine 
\({\cal M}\) and \(\Delta{\cal M}\) in the form
\begin{eqnarray}
\calm^{\rm exp}&=&\frac{\calm^{\rm exp}_{-2}}{\rmqp^2}
+\frac{\calm^{\rm exp}_{-1}}{\rmqp}
+\calm^{\rm exp}_0+O(\rmqp),
\nonumber\\
\dcalm^{\rm exp}&=&\frac{\dcalm^{\rm exp}_{-2}}{\rmqp^2}
+\frac{\dcalm^{\rm exp}_{-1}}{\rmqp}+\dcalm^{\rm exp}_0+O(\rmqp) \;.
\label{eq_3_50}
\end{eqnarray}
Due to the low energy theorem, the threshold coefficients 
\(\calm_{-2},\ \calm_{-1},\ \dcalm_{-2},\
\dcalm_{-1} \) are known. If we define  

\begin{eqnarray}
\calm^{\rm BH+Born}&=&\frac{\calm^{\rm BH+Born}_{-2}}{\rmqp^2}
+\frac{\calm^{\rm BH+Born}_{-1}}{\rmqp}
+\calm^{\rm BH+Born}_0+O(\rmqp),
\nonumber\\
\dcalm^{\rm BH+Born}&=&\frac{\dcalm^{\rm BH+Born}_{-2}}{\rmqp^2}
+\frac{\dcalm^{\rm BH+Born}_{-1}}{\rmqp}+\dcalm^{\rm BH+Born}_0+O(\rmqp) \;,
\label{eq_3_51}
\end{eqnarray}
where the index BH+Born means  that the quantity is computed by keeping only 
 the BH and Born contributions in the \( (e,\ e'\,\gamma)\) amplitude, 
then we have the following constraints due to the LET :
\begin{eqnarray}
\calm^{\rm exp}_{-k}=\calm^{\rm BH+Born}_{-k},\ 
\dcalm^{\rm exp}_{-k}=\dcalm^{\rm BH+Born}_{-k},\ k=1,2 \;.
\label{eq_3_52}
\end{eqnarray}
The information on the GPs is contained in \( \calm^{\rm exp}_0,\) and
 \( \dcalm^{\rm exp}_0\). These coefficients contain a part
  which comes from the (BH+Born) part
of the amplitude and another one which is a linear combination of the GPs with
coefficients determined by the kinematics. After some algebra
\cite{Gui95},\cite{Vdh97a} one finds that the observables depend on the 6
following  structure functions

\begin{eqnarray}
&&P_{LL}(\rmq)= - 2\sqrt{6} m G_E
P^{\left( {01,01} \right)0}\left(\rmq\right) \;,\nonumber\\
&&P_{TT}(\rmq)=-3G_M\frac{\rmq^2}{\qt0}\left(
P^{(11,11)1}(\rmq)-\sqrt{2}\qt0 P^{(01,12)1}(\rmq)\right) \;,
\nonumber\\
&&P_{LT}(\rmq)=\sqrt{\frac{3}{2}}\frac{m\rmq}{\tilde Q}G_E P^{(11,11)0}(\rmq)
+\frac{3}{2}\frac{\tilde{Q}\rmq}{\qt0} G_M P^{(01,01)1}(\rmq) \;,\nonumber\\
&&P_{LT}^z(\rmq)=\frac{3 \tilde{Q}\rmq}{2\qt0}G_M P^{(01,01)1}(\rmq)
-\frac{3m\rmq}{\gqt}G_E P^{(11,11)1}(\rmq) \;,
\nonumber\\
&&P_{LT}^{'z}(\rmq)=-\frac{3}{2}\gqt G_M P^{(01,01)1}(\rmq)
+\frac{3 m \rmq^2}{\gqt\qt0}G_E P^{(11,11)1}(\rmq) \;,\nonumber\\
&&P_{LT}^{'\perp}(\rmq)=\frac{3\rmq\gqt}{2 \qt0}G_M\left(
P^{(01,01)1}(\rmq)-\sqrt{\frac{3}{2}}\qt0 P^{(11,02)1}(\rmq)\right) \;,
\label{eq_3_53}
\end{eqnarray}
where $G_E$ and $G_M$ stand for $G_E( \tilde Q^2)$ and $G_M( \tilde Q^2)$.
Clearly measuring the 6 structure functions \footnotemark[3]  
defined in Eq.(\ref{eq_3_53})
amounts to measure the 6 independent GP's. For convenience we also introduce 
the combinations

\begin{eqnarray}
&&\hspace{-0.9cm} P_{LT}^{\perp}=
{{RG_E} \over {2G_M}}P_{TT}
-{{G_M} \over {2RG_E}}P_{LL}
,\nonumber\\
&&\hspace{-0.9cm} P_{TT}^{\perp}={{G_M} \over {R G_E}}
\left( P_{LT}^{z}-P_{LT} \right)
= - {{\rmq} \over 2} G_M \left( 3 P^{(11,11)1} 
+ \sqrt{{3 \over 2}} P^{(11,11)0} \right), \nonumber\\
&&\hspace{-0.9cm} P_{TT}^{'\perp}={{G_M} \over {R G_E}}
\left( P_{LT}^{'z}+\frac{\qt0}{\rmq} P_{LT} \right) 
= {{\rmq} \over 2} G_M \left( 3 {{\rmq} \over {\qt0}} P^{(11,11)1} 
+ \sqrt{{3 \over 2}} {{\qt0} \over {\rmq}} P^{(11,11)0} \right), 
\label{eq_3_54}
\end{eqnarray}
where $R = 2 m/\gqt$. The observables \footnotemark[4] 
can then be written as
(using the notation \((x,y,z)\) for the directions (1,2,3) ) 
\cite{Gui95},\cite{Vdh97a}

\noindent
\begin{minipage}[b]{13.9cm}
\footnotetext[3]{The unpolarized cross section and the 
double polarization observables yield 6 independent structure functions 
and not 7 as mentionned originally in Ref.\cite{Vdh97a}. 
Afterwards \cite{Dre98}, 
four relations were found between the polarizabilities which yield 
then 6 independent observables to determine the 6 independent 
polarizabilities.} 
\footnotetext[4]{Note the misprint in Eq.(4) of Ref.\cite{Vdh97a}: 
\(h\varepsilon\) should be replaced by \(\varepsilon\).} 
\begin{eqnarray}
&&\calm^{\rm exp}_0 -  \calm^{\rm BH+Born}_0 \nonumber\\
&& = 2 K_2 \; \left\{ 
v_1 \left[ {\epsilon  P_{LL}(\rmq)  -  P_{TT}}(\rmq)\right]
\;+\; \left(v_2-\frac{\qt0}{\rmq}v_3\right)\sqrt {2\varepsilon \left( {1+\varepsilon }\right)}
P_{LT}(\rmq) \right\},\nonumber\\
&&\dcalm^{\rm exp}_0(h,z) - \dcalm^{\rm BH+Born}_0(h,z)  \nonumber\\
&& = 4 (2h) K_2 \left\{ {-v_1  \sqrt {1-\varepsilon ^2}P_{TT}(\rmq)
\;+\;  v_2  \sqrt {2\varepsilon \left( {1-\varepsilon } \right)}P_{LT}^z(\rmq)
\;+\; v_3  \sqrt {2\varepsilon \left( {1-\varepsilon } \right)}
P_{LT}^{'z}(\rmq) } \right\} ,\nonumber
\end{eqnarray}
\end{minipage}
\newpage

\begin{eqnarray}
&&\dcalm^{\rm exp}_0(h,x) - \dcalm^{\rm BH+Born}_0(h,x) \nonumber\\
&& = 4 (2h) K_2 \left\{  v_1^x 
\sqrt {2\varepsilon \left( {1-\varepsilon } \right)} \, P_{LT}^{\perp}(\rmq) 
\;+\; v_2^x  \sqrt {1-\varepsilon ^2} \, P_{TT}^{\perp}(\rmq) \right. \nonumber\\
&&\hspace{1.6cm} \left. +\; v_3^x  \sqrt {1-\varepsilon ^2} \, P_{TT}^{'\perp}(\rmq)
\;+\; v_4^x \sqrt {2\varepsilon \left( {1-\varepsilon } 
\right)} \, P_{LT}^{' \perp}(\rmq) \right\},\nonumber\\
\nonumber\\
&&\dcalm^{\rm exp}_0(h,y) - \dcalm^{\rm BH+Born}_0(h,y) \nonumber\\
&& = 4 (2h) K_2 \left\{ 
 v_1^y \sqrt {2\varepsilon \left( {1-\varepsilon }
\right)} \, P_{LT}^{\perp}(\rmq) 
\;+\; v_2^y \sqrt {1-\varepsilon ^2} \, P_{TT}^{\perp}(\rmq)  \right. \nonumber\\  
&&\hspace{1.6cm} \left. +\; v_3^y \sqrt {1-\varepsilon ^2} \, P_{TT}^{' \perp}(\rmq)
\;+\; v_4^y \sqrt {2\varepsilon \left( {1-\varepsilon } \right)} \, 
P_{LT}^{'\perp}(\rmq)  \right\},
\label{eq_3_55}
\end{eqnarray}
with 
\begin{equation}
K_2=e^6\,{\rmq \over {\gqt^2}}  
{{2m} \over {1-\varepsilon }} 
\sqrt {{{2E_q} \over {E_q+m}}},\ \ E_q=\sqrt{m^2+\rmq^2}.
\label{eq_3_56}
\end{equation}
The angular dependent functions ($v_1$, $v_2$,\dots)   are given by 
\begin{eqnarray}
&&v_1  =  \sin \theta \left( {\omega^{''}\sin \theta
-\rmkt\omega^{'}\,\cos \theta \cos \phi  } \right) , \nonumber\\
&&v_2  = - \left( {\omega^{''}\sin \theta \cos \phi - 
\rmkt\omega^{'}\,\cos \theta } \right) , \nonumber\\
&&v_3  = - \left( {\omega ^{''}\sin \theta \cos \theta  \cos \phi 
- \rmkt\omega^{'}\,\left( {1-\sin^2 \theta  \cos^2 \phi} \right) } 
\right) ,\nonumber\\
\nonumber\\
&&v_1^x  =  \sin \theta  \cos \phi 
\left( {\omega^{''}\sin \theta
-\rmkt\,\omega^{'}\,\cos \theta \cos \phi  } \right), \nonumber\\
&&v_2^x  = -  {\omega^{''}\sin \theta + 
\rmkt\omega^{'}\,\cos \theta \cos \phi  } , \nonumber\\
&&v_3^x  = - \cos \theta \left( {\omega^{''}\sin \theta -
\rmkt\omega^{'}\,\cos \theta \cos \phi } \right) , \nonumber\\
&&v_4^x  = \rmkt \omega^{'}\sin \theta \sin ^2\phi,\nonumber\\
\nonumber\\
&&v_1^y  =  \sin \theta  \sin \phi 
\left( {\omega^{''}\sin \theta
-\rmkt\,\omega^{'}\,\cos \theta \cos \phi  } \right), \nonumber\\
&&v_2^y  = \rmkt \omega^{'}\cos \theta \sin \phi,\nonumber\\
&&v_3^y  = \rmkt \omega^{'}\sin \phi,\nonumber\\
&&v_4^y  = - \rmkt \omega^{'}\sin \theta \sin \phi  \cos \phi \;, \label{eq_3_57}
\end{eqnarray}

\noindent
\begin{minipage}[b]{13.9cm}
and finally \footnotemark[5] 
\footnotetext[5]{In Ref.\cite{Gui95} there is a misprint in Eq.(119) for 
$\omega$, which should be corrected by replacing \( p'.q'\) by \( p.q'\).}
\begin{eqnarray}
&&\omega=
\left[-\rmqp\left(\frac{1}{p.q'}+\frac{1}{k.q'}\right)\right]_{\rmqp=0},\  
\ \omega'=
\left[\rmqp\left(\frac{1}{k'.q'}-\frac{1}{k.q'}\right)\right]_{\rmqp=0},
\nonumber\\
&&\omega''=\omega\rmq-\omega'\sqrt{\tilde{{\rm k}}'^2-{\rm k}^2_T},\
\ \rmkt=\gqt\sqrt{\frac{\varepsilon}{2(1-\varepsilon}}.
\label{eq_3_58}
\end{eqnarray}
\end{minipage}

\subsection{Results for ($e, e' \gamma$) observables in the threshold regime}

In the previous sections, we have outlined the observables of the 
$p(e, e' p)\gamma$ reaction below threshold and have detailed 
how the nucleon structure effect can be parametrized in terms of 
six independent polarizabilities. 
In this section, results will be shown for the $p(e, e' p)\gamma$ 
observables in the threshold region in kinematics where the 
experiments are performed at MAMI \cite{d'Ho95} and at CEBAF \cite{Ber93}. 

\begin{figure}[h]
\epsfxsize=10 cm
\epsfysize=11 cm
\centerline{\epsffile{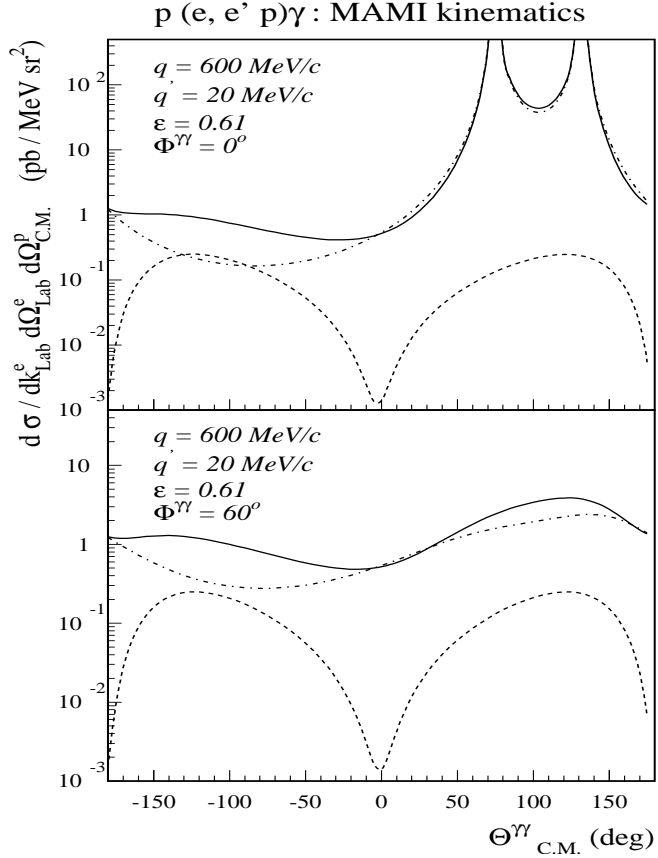}}
\caption{\small $p(e, e' p)\gamma$ differential cross section in MAMI kinematics :
BH (dashed-dotted lines), Born (dashed lines) and 
BH + Born contributions (full lines) in plane ($\phi = 0^0$) and 
out-of-plane ($\phi = 60^0$).}
\label{fig:mamicross}
\end{figure}

The experimental strategy of VCS in the threshold region consists of 
two steps. One measures the VCS cross section at several values of 
the outgoing photon energy. At low outgoing photon energies, the 
measurement of the VCS observables provides a test of the LET as 
can be seen from Eqs.(\ref{eq_3_50}, \ref{eq_3_52}). 
Once the LET is verified, the relative effect of the polarizabilities 
can be extracted using Eq.(\ref{eq_3_55}).

The predictions for the Bethe-Heitler (BH) and Born cross sections 
are shown in Fig.(\ref{fig:mamicross}) for an outgoing photon of 
low energy. The BH cross section has a characteristic angular shape 
and displays two "spikes". These "spikes" occur when the direction 
of the outgoing photon coincides with either the initial or final 
electron directions. In these regions, the cross section is 
completely dominated by the BH contributions. In order to measure 
the VCS contribution, one clearly has to detect the photon in the 
half-plane opposite to the electron directions where the BH 
contamination is the smallest. 

The first absolute measurement of the VCS cross section on the nucleon 
performed at MAMI \cite{d'Ho95} has shown that radiative corrections 
provide an important contribution to the $p(e, e' p)\gamma$ reaction. 
All diagrams that contribute to the cross section to order $\alpha_{em}^4$ 
have been calculated in Ref.\cite{Vdh98}.  
A preliminary comparison to the MAMI data at low outgoing photon energy 
shows that the VCS cross section can be understood by the 
radiatively corrected Bethe-Heitler (BH) + Born processes \cite{d'Ho97} 
which provides a test of the LET. 

\begin{figure}[ch]
\epsfxsize=11. cm
\epsfysize=7 cm
\centerline{\epsffile{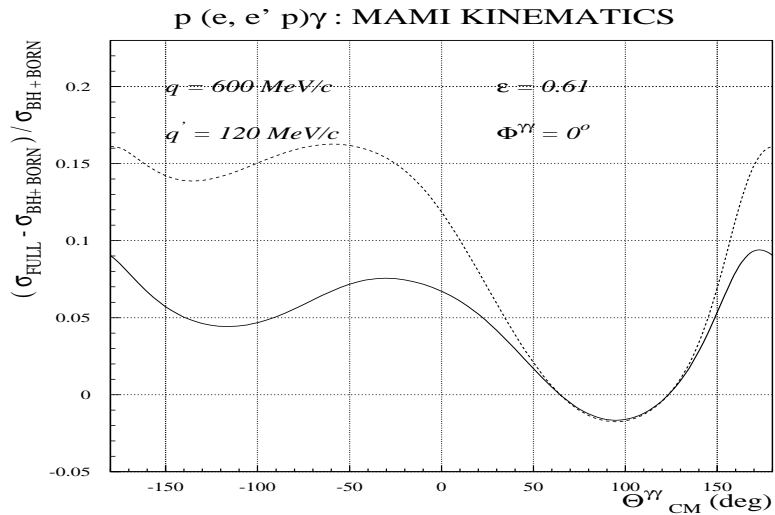}}
\caption{\small Relative nucleon structure effect in VCS in MAMI kinematics. 
$\sigma_{FULL}$ represents the $p(e, e' p)\gamma$ differential cross section 
in the full model described in the text and $\sigma_{BH + BORN}$
represents the BH + Born contribution. The ratio calculated with all
non-Born VCS diagrams is shown by the solid
curve, whereas the ratio calculated with only the $\Delta$ diagram
contribution to the non-Born VCS amplitude is shown by the dashed curve.}
\label{fig:mamiratio}
\end{figure}

For the sake of providing estimates of the nucleon structure effect accessible 
through VCS, the relativistic effective Lagrangian model of Ref.\cite{Vdh96} 
is used here as it takes into account the nucleon structure effects to all 
orders in the outgoing photon energy. 
In this model, the $p(e, e' p)\gamma$ reaction 
below pion production threshold is described in terms 
of BH, Born and non-Born diagrams. For the non-Born diagrams, 
a tree-level calculation was performed consisting of exchanges of the 
resonances in the $s$- and $u$-channel 
and the exchanges of the $\pi^0$ and $\sigma$ in the $t$-channel. 
For the resonances, the most important contributions come from the 
$\Delta$(1232) and $D_{13}$(1520) for which the radiative couplings were 
taken from pion photo/electro production \cite{Vdh95}.  
The model was tested against available data for real Compton scattering below 
pion threshold and yields for the proton polarizabilities at the photon point : 
$\alpha_G = 7.29 \cdot 10^{-4} fm^3$ and $\beta_G = 1.56 \cdot 10^{-4} fm^3$, 
which should be compared to the values extracted form experiment \cite{MGi95} :
$\alpha^{exp}_{G} = 12.1 \pm 0.8 \pm 0.5 \cdot 10^{-4} fm^3$ and 
$\beta^{exp}_{G} = 2.1 \mp 0.8 \mp 0.5 \cdot 10^{-4} fm^3$. 
The underestimation of $\alpha_G$ is probably due to the simplicity of the 
present tree-level effective Lagrangian model which does not include 
non-resonant $\pi$N intermediate states. 

\begin{figure}[h]
\epsfxsize=9. cm
\epsfysize=8 cm
\centerline{\epsffile{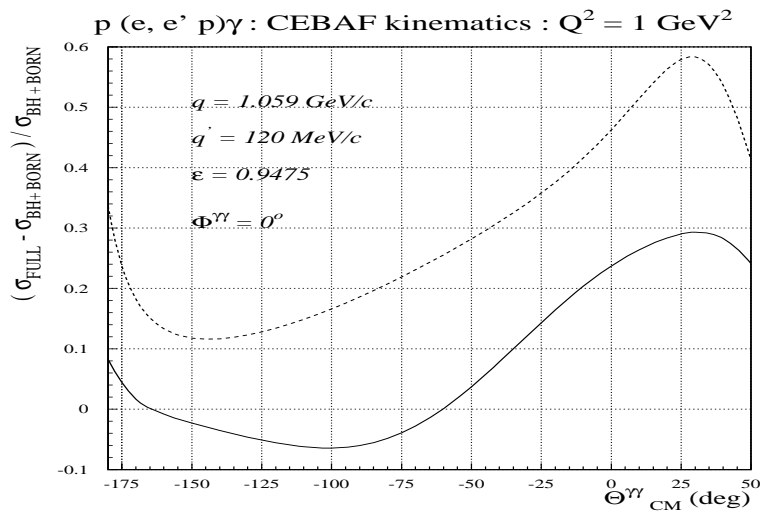}}
\vspace{-0.8cm}
\caption{\small Same ratio and curve conventions as in Fig.(\ref{fig:mamiratio}) 
but for CEBAF kinematics.}
\label{fig:cebafratio}
\end{figure}

To see the magnitude of the non-Born contributions, we plot in 
Figs.(\ref{fig:mamiratio}) and (\ref{fig:cebafratio}) the 
BH+B+NB differential cross section ($\sigma_{FULL}$) relative to 
the BH+B differential cross section ($\sigma_{BH+B}$)   
for MAMI kinematics and for CEBAF kinematics respectively. 
In Figs.(\ref{fig:mamiratio}, \ref{fig:cebafratio}), 
$\sigma$ stands for the fivefold differential $p(e, e' p)\gamma$ cross section 
${d \sigma}/{(d \Omega_e^{'})_{Lab} \, d |\bar{k}^{'}|_{Lab} \, 
(d \Omega_p^{'})_{CM}}$. 
The in-plane $p(e, e' p)\gamma$ differential cross section ratios are shown 
for a momentum of the outgoing real photon of $\rm q'$ = 120 MeV, 
which is close to but below the pion threshold. 
The MAMI experiment \cite{d'Ho95} was performed for a virtual photon 
momentum $\rm q$ = 600 MeV and a value of the virtual photon 
polarization around $\varepsilon = 0.61$. 
In the MAMI kinematics, the non-Born contribution 
is a 10-15 $\%$ effect to the 
$p(e, e' p)\gamma$ reaction at backward angles (where the photon is emitted 
opposite to the electron directions). 
\newline
\indent
The kinematics of the CEBAF experiment \cite{Ber93}, 
corresponds to a larger value of $\epsilon \approx 0.95$ due to the higher 
beam energy. The correspondingly larger virtual photon flux factor 
which goes like 1/(1 - $\varepsilon$) increases 
the VCS contribution relative to the BH cross section. 
Thus one expects a larger 
effect in CEBAF kinematics as is also seen from the model calculations 
on Fig.(\ref{fig:cebafratio}) which are performed at $\rm q \approx$ 1.06 GeV/c.
\newline
\indent
If one wants to extract the polarizabilities from experiment, 
we have seen that 
an unpolarized experiment is not sufficient as it gives access to only 
3 independent response functions. 
To separate the polarizabilities, one has to resort to 
double polarization observables. Experimentally, at existing high duty cycle 
electron facilities with a polarized electron beam such as at MAMI, 
MIT-Bates and CEBAF, double polarization VCS experiments can be 
performed by measuring the 
recoil polarization of the outgoing nucleon with a focal plane polarimeter. 
\newline
\indent
\begin{figure}[h]
\epsfxsize=11. cm
\epsfysize=8 cm
\centerline{\epsffile{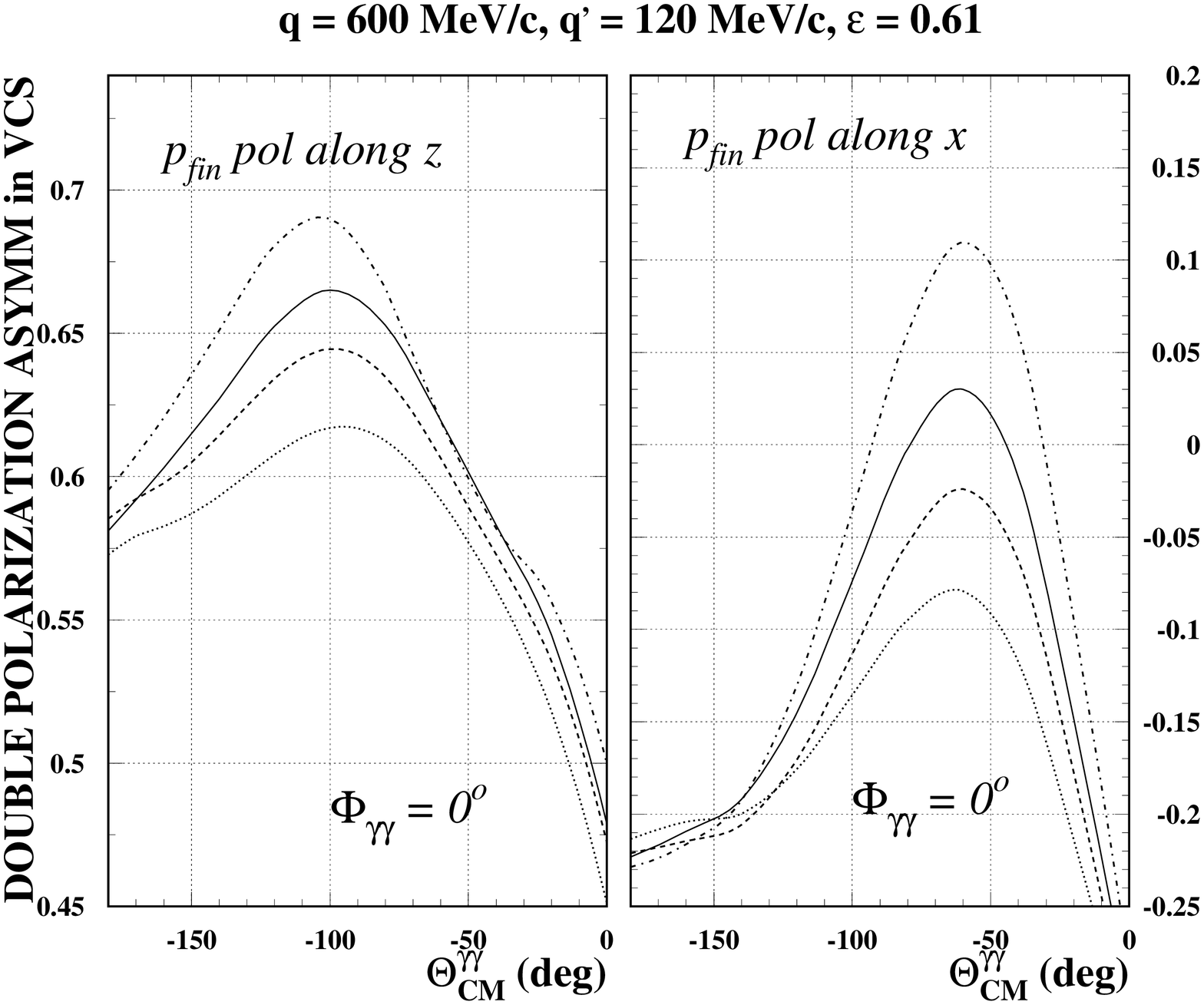}}
\caption{\small VCS double polarization asymmetry (polarized electron, 
recoil proton polarization along either the $x$- or $z$-directions) 
in MAMI kinematics ($Q^2 = 0.33 \; GeV^2$) as function of the 
CM angle between real and virtual photon. 
The BH + Born contribution is shown by the dashed lines.  
The result of the BH + Born + $\Delta$ contribution is shown by 
the dotted lines, whereas the BH + Born + $D_{13}$ + $\sigma$ contribution 
is shown by the dashed-dotted lines. 
The total effect in the model is shown by the full lines.}
\label{fig:polmami}
\end{figure}
In Fig.(\ref{fig:polmami}) the double polarization asymmetry of 
Eq.(\ref{eq_2_28}), where the recoil proton polarization is measured either 
along the $x$- or $z$- directions,  
is shown at $\rm q$ = 600 MeV/c for in-plane kinematics.
It is seen that the asymmetry where the 
final proton is polarized parallel to the virtual photon yields a 
large value (between 0.6 and 0.7). Remark that in Fig.(\ref{fig:polmami}), 
the electric and magnetic contributions in the present effective Lagrangian 
calculation yield large and opposite effects to the asymmetry.

\newpage

\section{THE HARD SCATTERING REGIME}
\label{sec4}
\subsection{Introduction}

Here we come to the hard part of this review. The threshold regime presented in
Section \ref{sec3} and the deep VCS which is the subject of Section \ref{sec5} are
going well. In both cases the theoretical activity is important and while in the
first case the experiments are  running, in the second case they are already at
the stage of experiment proposals. By an irony of history the threshold regime
and the deep VCS should be considered as the wealthy descendants of the VCS in
the hard scattering regime, a domain now somewhat sleeping. Though everybody
agrees that this is a splendid physics case for VCS, it suffers from a severe
drawback, namely the extreme  difficulty of the experiments. It was the conclusion of
the first feasibility study \cite{Arv93} that no existing accelerator is
adequate for such experiments. A new type of electron accelerator with a high
energy ($>30$ GeV), high duty factor ($\sim 100\%$), very good  energy resolution
(much better than the pion mass) and high intensity ($>100\mu {\mathrm A}$) is
actually necessary. Unfortunately no decision about the ELFE project 
\cite{Arv95}, which meets with these
requirements, is  expected in the coming years. The consequence is that, despite
its evident interest, the subject is in standby. In particular there is no motivation to
develop the very complicated perturbative QCD calculations for this process.
\newline
\indent
However we want to hope that one day the field will have the opportunity to take
off and the role of this review may be to prepare the way. We shall do it by first
introducing the physics of the hard scattering regime. Then we shall use the
diquark model to get realistic estimates of the cross sections. We shall see
that, even though the energy of CEBAF is too low to achieve a fruitful VCS
program in the hard scattering regime, some preliminary studies can be performed
with the polarized beam. Finally we shall
switch to the case of real Compton scattering. The first reason is that the
physics, though less rich, is closely related to the VCS because the relevant
energy scales are ($s,\,t,\ u$) and not $Q^2$, which can be zero. The second reason is
that the experimental prospects are better in this case \cite{d'Ho96}\cite{D'An97}.

\subsection{Incoherence between the short and long-distance physics}

\begin{figure}[h]
\epsfxsize=12.0cm
\centerline{\epsffile{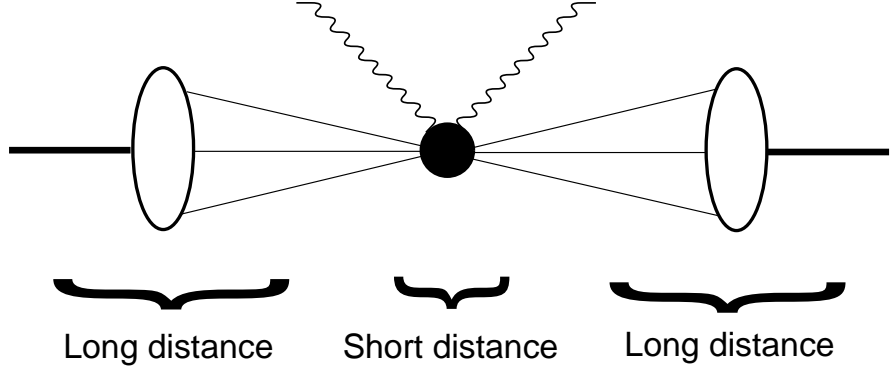}} 
\caption{\small Factorization of the scattering amplitude in the hard 
scattering regime.}
\label{fig:facto}
\end{figure}

This regime is defined  by requiring that all three variables ($s,\,t,\ u$) be
large with respect to a typical hadronic scale, say 1 GeV. In this case there
is a prejudice (actually proved in the case of elastic electron scattering
\cite{Ste98}) that the amplitude factorizes in a soft non-perturbative part, 
the distribution amplitude, and a
hard  scattering kernel  which is calculable from perturbative QCD. Because of
asymptotic freedom, 
the perturbative approach must be to some degree relevant to
the hard scattering regime. However, since the binding of the quarks and gluons in
the hadrons is a long distance, non-perturbative effect, the description of the
reaction requires a consistent analysis of both large and small scales.
Here we follow the nice argument of Ref.\cite{Ste98}. When the reaction is hard
enough the relative velocities of the participating particles are nearly
lightlike. Time dilatation increases the lifetime of the quantum configurations
which build the hadron. So the partonic content, as seen by the other
particles, is frozen. Moreover, due to the apparent contraction of the hadron size,
the time during which momentum can be exchanged is decreased. Therefore one
expects a lack of coherence between the long-distance confining effects and the
short distance reaction. This incoherence between the soft and hard physics is
the origin of the factorization which is illustrated on Fig.(\ref{fig:facto}).

The calculation of the amplitude can be done using the 
Brodsky-Lepage formalism \cite{Bro80} which leads to the factorized expression

 \begin{eqnarray}
\label{eq:fact}
T(\lambda ',h_N',\lambda,h_N) 
= \int dx_i dy_j \; \phi_N ^* (y_j) \;  
T_{H}(\lambda ',h_N',y_j,\,\lambda,h_N,x_i; s, t) \; \phi_N (x_i),
\end{eqnarray}
where ($x_i,\,y_i$) are respectively the momentum fractions of the quarks 
in the initial and final nucleon, $T_H$ is the hard scattering kernel and 
$\phi_N$ is the distribution amplitude (D.A.). 
The evaluation of Eq.(\ref{eq:fact}) requires a four-fold 
convolution integral since there are 
two constraint equations $(x_1 + x_2 + x_3 = 1$ and $y_1 + y_2 + y_3 = 1)$.
In  Eq.(\ref{eq:fact})  a sufficiently large 
momentum transfer is  assumed so as to neglect the transverse momentum dependence in the 
hard scattering amplitude $T_H$. In this limit, the integration over the 
transverse momenta $\vec{k}_{\perp i}$  
(where $\sum _i \vec{k}_{\perp i} = \mathbf 0$) 
acts only on the valence wavefunction  
\begin{equation}
\Psi_V ( x_1, x_2, x_3; \vec{k}_{\perp 1}, \vec{k}_{\perp 2},
\vec{k}_{\perp 3} )\;, 
\end{equation}
which is the amplitude of the 
three quark state in the Fock expansion of the proton:
\begin{equation}
\mid P > = \Psi_V \, \mid q q q >
 + \Psi_{q \overline q } \, \mid q q q, q {\overline q } >
 + \Psi_g \, \mid q q q, g >
 +...
\end{equation}
This valence wavefunction $\Psi_V$ integrated up to a scale $\mu$ 
(which separates the soft and hard parts of the wavefunction) defines the 
D.A. which appears in Eq.(\ref{eq:fact}) : 
\begin{equation}
\label{eq:wavefunction}
  \phi_N( x_i,\mu ) = \int ^\mu d^2 \vec{k}_{\perp i} \;
                                 \Psi_V ( x_i; \vec{k}_{\perp i} )\;.
\end{equation}

For $ \mu $ much larger than the average value of the transverse momentum 
in the proton, this function $ \phi_N $ depends only weakly on $ \mu $ 
\cite{Bro80} and this dependence can be neglected.

\begin{figure}[h]
\epsfysize=13cm
\centerline{\epsffile{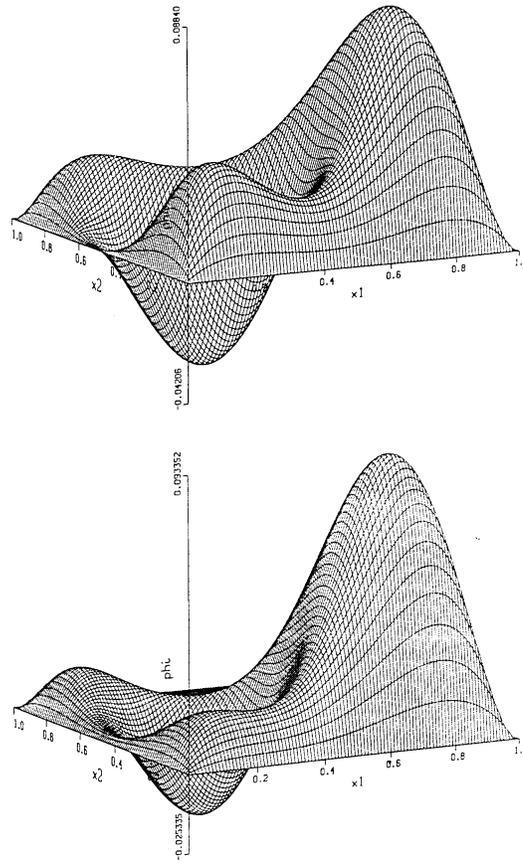} }
\vspace{ -0.5 cm}
\caption{\small Model distribution amplitudes for the nucleon : 
KS (upper figure) and COZ (lower figure), as function of the 
valence quark momentum fractions $x_1$ and $x_2$ ($x_1 + x_2 + x_3$ = 1).}
\label{fig:PQCD}
\end{figure}

The interest of the formalism is that the distribution amplitude is universal,
that is  independent of the particular reaction considered. 
Several  distribution amplitudes have 
been modeled  using QCD sum rules \cite{Che84}\cite{Che89}\cite{Kin87}.
They have a characteristic shape and 
predict that in a proton, the $u$-quark with helicity along the 
proton helicity carries about 2/3 of its longitudinal momentum (see
Fig.(\ref{fig:PQCD}) ).

\begin{figure}[h]
\epsfxsize=10.0cm
\centerline{\epsffile{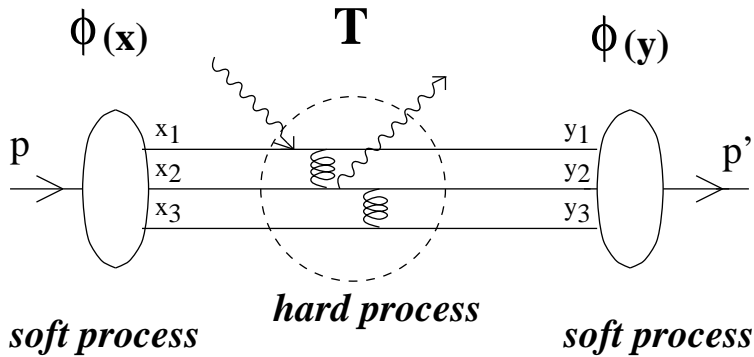}} 
\caption{\small The hard scattering kernel in leading order PQCD.}
\label{fig:PQCDdiag}
\end{figure}

For the computation of the hard scattering amplitude $T_H$, 
the leading order PQCD contribution corresponds 
to the exchange of the minimum number of 
gluons (in the present case two) between the three quarks as shown on
Fig.(\ref{fig:PQCDdiag}). The number of diagrams grows rapidly with the 
number of elementary particles involved in the reaction 
(42 diagrams for the nucleon form factor, 
336 diagrams in the case of real or virtual Compton scattering). 
Despite the large number of diagrams, the calculation of $T_H$ 
is a parameter free calculation once the scale $\Lambda_{QCD} \approx 200 ~MeV$ 
in $\alpha_s (Q^2)$ is given. Note that configurations with more than three
valence quarks are a priori allowed but since this implies to exchange more hard
gluons the corresponding contribution is suppressed by powers of $1/t$.

There are two characteristic features of the Brodsky-Lepage model which are
almost direct consequences of QCD: 
the dimensional counting rules \cite{Bro73} 
and the conservation of hadronic helicities \cite{Bro81}. 
The latter feature implies that any helicity flip amplitude is zero and, hence,
any single spin asymmetry too. The helicity sum rule is a consequence of 
utilizing the collinear approximation and of dealing with (almost) massless 
quarks which conserve their helicities when interacting with gluons. Whereas 
the dimensional counting rules are in reasonable agreement with experiment, 
the helicity sum rule seems to be violated even at moderately large 
momentum transfers. The prevailing opinion is that these phenomena cannot 
be explained in terms of perturbative QCD (see, for example, 
Ref.~\cite{Siv89}), rather they are generated by an interplay of 
perturbative and non-perturbative physics.

An interesting aspect of real and virtual Compton 
scattering is that they are  the  simplest processes in which 
the integrals over the longitudinal momentum fractions yield imaginary parts.
The reason is that, as in any scattering process, there are kinematical regions where internal quarks and 
gluons can go on their mass shell. 
The appearance of imaginary parts to leading order in $\alpha_s$ is a 
non-trivial prediction of PQCD which should be tested experimentally.
As explained below, the ($e, e' \gamma $) reaction  with polarized incoming
electrons seems to be a good candidate for this investigation.

\subsection{The diquark model}

The only calculation \cite{Far90a} of VCS in the Brodsky-Lepage formalism is unfortunately not
reliable due to a treatment of the singularities which leads to inaccurate
results. This is discussed in details in Section 4.4. To get realistic
predictions and to delineate what would be the most interesting VCS experiments
we rely on the diquark model calculation of Ref.\cite{Kro96}. Due to the
(relative) simplicity of this model it has been possible to analytically
integrate the singularities, which eliminates the problems encountered in Ref.
\cite{Far90a}.

In the diquark model \cite{Ans87} baryons are viewed as made up of
quarks and diquarks, the latter 
being treated as quasi-elementary constituents which partly survive medium 
hard collisions. The composite nature of the diquarks is taken into account by 
diquark form factors. Diquarks are an effective description of correlations in 
the wave functions and constitute a particular model for non-perturbative 
effects. The diquark model has been applied to a variety of processes and 
successfully confronted to data. Among these applications is a study of 
the nucleon's electromagnetic form factors \cite{Jak93}. In fact, predictions
are achieved for both the magnetic and the electric form factors. For the 
latter quantity no result is obtained in the pure quark hard scattering model 
because it requires helicity flips of the nucleon and, in so far, the electric 
form factor also represents a polarization 
effect. In the diquark model helicity flips are generated through spin 1 
diquarks. The diquark model is designed in such a way that it turns into the 
theoretically well established pure quark picture asymptotically. So 
the pure quark picture of Brodsky-Lepage and the diquark model do not 
oppose each other, they are not alternatives but rather complements.

\begin{figure}
\begin{center}
\epsfxsize = 12cm
\centerline{\epsffile{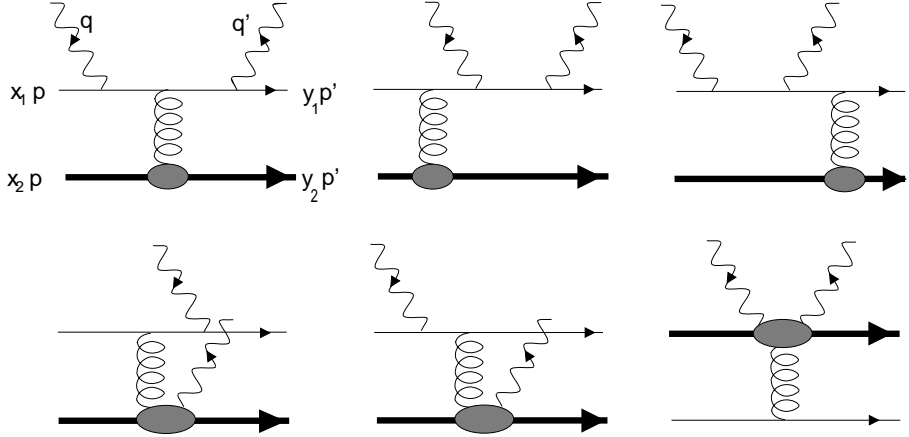} }
\end{center}
\caption{\small Feynman graphs contributing to the VCS process 
$\gamma^\ast\,p\rightarrow\gamma\,p$. Graphs with the two 
photons interchanged are not shown, and the thick line represents a diquark.}
\label{graph-diquark}
\end{figure}

Representative Feynman graphs contributing to the gauge invariant hard-scattering 
amplitudes are displayed in Fig.(\ref{graph-diquark}). 
The blobs appearing at the $gD$, $\gamma gD$ and 
$\gamma\gamma D$ vertices symbolize three-, four- and five-point 
functions. These n-point functions are evaluated for point-like diquarks 
and the results are multiplied with phenomenological vertex 
 form factors which take into account the 
composite nature of the diquarks.  The perturbative part of the model, i.e.~the 
coupling of gluons (and photons) to diquarks follows standard prescriptions. The
results for VCS shown below have been obtained with the same diquark parameters 
taken from previous applications of the diquark model \cite{Kro96}.

%
%
\begin{figure}[h]
\begin{center}
\epsfysize = 6cm
\centerline{\epsffile{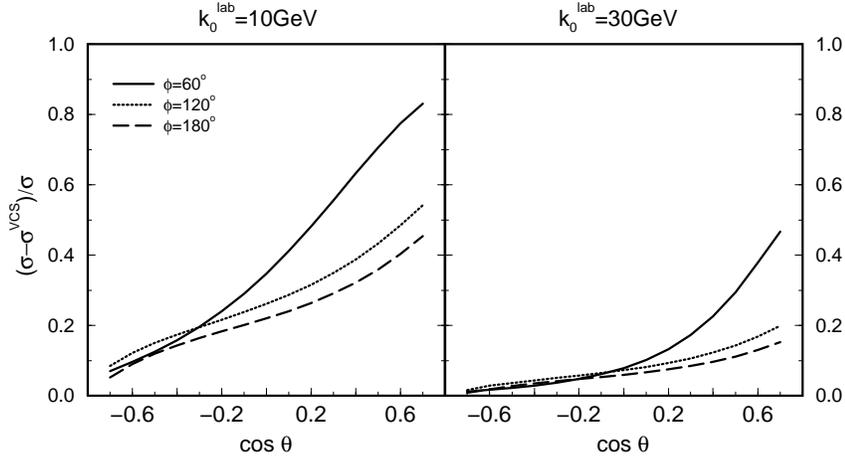} }
\end{center}
\vspace{-1cm}
\caption{\small The difference between the full photon electroproduction cross 
section and the VCS contribution to it over the full cross section vs.~$\cos 
\theta$ for $Q^2$ = 1 GeV$^2$, $s = 10$ GeV$^2$, 
two values of the beam energy $k_{0L}$, 
and for several values of the azimuthal angle $\phi$.  }
\label{VCS-BH}
\end{figure}

The first point to consider is the relative strength of the VCS and BH
processes. A strong interference between them, though not uninteresting by
itself, would obscure the interpretation of the results because it would prevent
the familiar analysis in terms of the virtual photon cross sections defined in
Eq.(\ref{eq_2_30}). For that purpose we plot in Fig.(\ref{VCS-BH})
the difference between the full $ep\to ep\gamma$ cross section and the
VCS contribution to it divided by the full cross section.   
This represents the BH contamination.  
Clearly the dominance of the VCS contribution requires a
high energy and requires to detect the outgoing photon in the half-plane 
opposite to the electron directions ($\phi = 180^0$).

\begin {figure}
\begin{center}
\epsfxsize = 12cm
\centerline{\epsffile{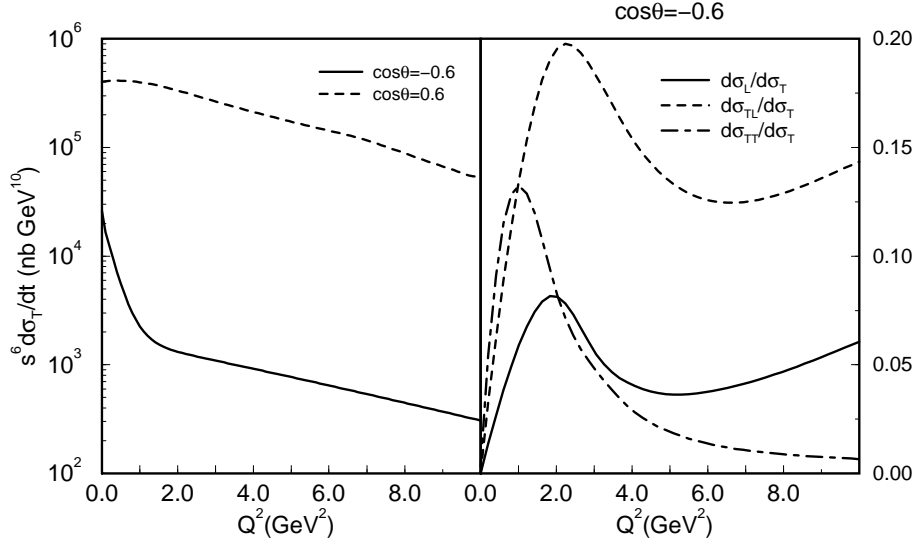} }
\end{center}
\vspace{-1cm}
\caption{\small The cross section for virtual Compton scattering vs.~$Q^2$ at 
$s=10$ GeV$^2$.}
\label{Q2-depend}
\end{figure}

The $Q^2$ dependence of the cross sections is obviously the new information
we can get from VCS by comparison with the real Compton scattering. 
In Fig.(\ref{Q2-depend}) the $Q^2$ dependence of the VCS cross sections is  shown
in a hard scattering situation ($s=10\;{\mathrm GeV}^2$, $\;\cos\theta= \pm 0.6$).  
The transition from real
Compton scattering to large $Q^2$ as predicted by the diquark 
model, is obviously non-trivial. 
In particular one observes that in the forward direction ($\cos \theta$ = 0.6), 
the variation is smooth, compatible with the diquark form factor. In the 
backward region ($\cos \theta$ = -0.6) there is a rapid variation 
at small $Q^2$ and the slope seems to be related to the scale of the diquark mass. 
So by studying the $Q^2$ variation as a function of the angle, one 
can observe the interplay of the two characteristic scales of the model. 

The ($L,\,LT,\,TT$) cross sections have relative sizes with respect 
to the dominant transverse cross section 
which are not negligible and their $Q^2$ dependence
seems to merit a thorough investigation. However one must keep an eye on 
the very small value of the cross sections. 
Without a dedicated accelerator nobody will have 
the opportunity to contemplate the actual curves.
Remark that at the real photon point, the diquark calculations predict a 
very small (non-zero) value for the photon asymmetry ($TT/T$) over most 
of the angular range which should be compared with the PQCD prediction 
shown in the next section.

By contrast to the above comments where the BH is considered  as a nuisance, 
the regions of strong BH contaminations offer an interesting possibility 
to measure the phases of the VCS amplitudes. These phases, due to the
possible poles of the  quarks, 
diquarks or gluon propagators, induce in the $(e,\
e'\gamma)$ amplitudes  non-trivial phases beyond the phases due to
 the azimuthal angle dependence 
(see Eq.~(\ref{eq_2_20})). In 
other words, the perturbative phases manifest themselves in the fact that 
the $T$ matrix is not self-adjoint. For the BH process on the other hand, 
$T=T^\dagger$ obviously holds. As shown in \cite{Kro96}, information 
on the absorptive part $T\,-\,T^\dagger$ can be obtained from the electron
single spin asymmetry.
Since at CEBAF the electron beam is polarized a measurement 
of the asymmetry seems feasible. 
It is important to note that this possibility of measuring the 
absorptive part of the $T$ matrix by the electron asymmetry follows from parity and
time reversal invariance and does not rely on any model hypothesis.

As is well known, a one-particle helicity
state transforms under the combined parity and time reversal operation as
\begin{equation}
   |{\vec  k},\lambda\rangle \longrightarrow \eta(\lambda)|{\vec  k},-\lambda\rangle \;,
\end{equation}
where $\eta(\lambda)$ is $\pm 1$ depending on $\lambda$, the spin of the 
particle and on its internal parity. So the combined
parity and time reversal operations transform a given helicity state into
a state with the same momentum but with reversed helicity. If the interaction
is invariant under the parity and time reversal operations, the $T$-matrix
elements for, say, a $2\to3$ process satisfy the relation
\begin{eqnarray}
\label{tpt}
&&\langle {\vec  k}_1^\prime,\lambda_1^\prime;{\vec  k}_2^\prime,\lambda_2^\prime; 
      {\vec  k}_3^\prime,\lambda_3^\prime|T|{\vec  k}_1,\lambda_1;
      {\vec  k}_2,\lambda_2\rangle \hspace{6.0cm}\nonumber\\
&&=\left(\prod_i\eta_i\right) 
\langle {\vec  k}_1^\prime,-\lambda_1^\prime;
       {\vec  k}_2^\prime,-\lambda_2^\prime;{\vec  k}_3^\prime,-\lambda_3^\prime
       |T^\dagger|{\vec  k}_1,-\lambda_1;{\vec  k}_2,-\lambda_2\rangle^\ast \;.
\end{eqnarray}
 Let us now assume that particle 1
is a spin $1/2$ one, say an electron and note $\sigma(\pm)$ is the differential
cross section for photon electroproduction 
with specified helicity of the incoming electron. Then, ignoring all kinematical variables,
the cross section with  particle 1  in a definite helicity state is
\begin{equation}
\sigma(\pm)=\sum_{\{\lambda_i,\lambda_i^\prime\}}|
     T^{e,e'\gamma}
     (\lambda_1^\prime,\lambda_2^\prime,\lambda_3^\prime ;\pm,\lambda_2)|^2.
\end{equation}
The difference of these cross sections $\Delta\sigma=\sigma(+)-\sigma(-)$ may be written as
\begin{eqnarray}
\Delta\sigma=\Re e\sum_{\{\lambda_i,\lambda_i^\prime\}}
         [T^{e,e'\gamma}
         (\lambda_1^\prime,\lambda_2^\prime,\lambda_3^\prime ;+,\lambda_2)
        +\prod_i\eta_i T^{e,e'\gamma}
        (-\lambda_1^\prime,-\lambda_2^\prime-\lambda_3^\prime;
         -,-\lambda_2)^\ast] \nonumber\\
\times  [T^{e,e'\gamma}(\lambda_1^\prime,\lambda_2^\prime,\lambda_3^\prime;
               +,\lambda_2)^\ast
     -\prod_i\eta_i T^{e,e'\gamma}(-\lambda_1^\prime,-\lambda_2^\prime,-\lambda_3^\prime;
        -,-\lambda_2)].
\end{eqnarray}
If there would be no absorptive part of $T$, i.~e., $T=T^\dagger$, then, 
according to Eq.(\ref{tpt}), the difference of the two helicity cross sections
and hence the electron asymmetry would be zero. Therefore, the asymmetry
measures the non-trivial phase as stated above.
\begin{figure}
\begin{center}
\epsfysize = 7cm
\centerline{\epsffile{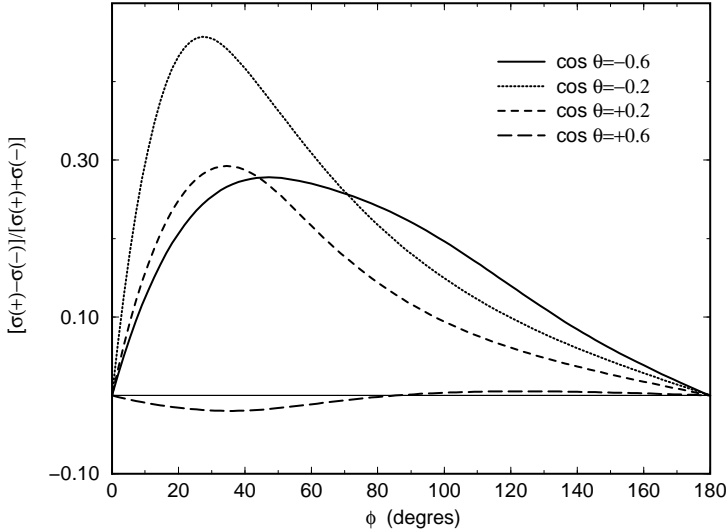} }
\end{center}
\vspace{-1cm}
\caption{\small The electron asymmetry at CEBAF (6 GeV) as function of $\phi$ 
for several values of $\cos \theta$ for $s$ = 5 GeV$^2$ and 
$Q^2$ = 1 GeV$^2$.}
\label{diquarkcebaf}
\end{figure}
If we consider only the VCS contribution, we find (dropping irrelevant 
kinematical factors)

\begin{eqnarray}
\Delta\sigma\simeq \sin \phi
\Im m\left[\Phi_9^\ast\,(\Phi_1-\Phi_7)
      +\Phi_{10}^\ast\,(\Phi_2+\Phi_8)
      +\Phi_{11}^\ast\,(\Phi_3-\Phi_5)
      +\Phi_{12}^\ast\,(\Phi_4+\Phi_6)\right]
\end{eqnarray}
Thus, we see that in this case $\Delta\sigma$ measures the imaginary part of the 
longitudinal-transverse interference term whereas $\sigma_{LT}$ 
 measures its real part. In other words, $\Delta\sigma$
measures the relative phase between the longitudinal and transverse VCS helicity
amplitudes. 

It turns out that the diquark model predicts
very small values for $\Delta\sigma$  in the pure VCS case. 
The ultimate reason is that the 
longitudinal helicity amplitudes are much smaller than the transverse ones
for VCS as illustrated on Fig.(\ref{Q2-depend}). 

However $\Delta\sigma $ can be strongly 
enhanced by the interference with the BH process.  
In the regions of strong BH contaminations
it then essentially measures the relative phase between the
transverse VCS amplitudes and the BH amplitudes. This 
enhancement can be seen on 
Fig.(\ref{diquarkcebaf}) where we show the electron asymmetry 
for a kinematical situation accessible in the future at CEBAF 
($s=5\;{\mathrm GeV}^2$, 
$\;k_{0L}=6\;{\mathrm GeV}$, $\;Q^2=1\;{\mathrm GeV}^2$). As can be seen from
that figure, the asymmetry is large in the diquark model for small values of
$|\cos\theta|\;$ ($\simeq 0.2$) and values of the azimuthal angle around
$30^\circ$. The magnitude of the effect is of course sensitive to details of 
the model and, therefore, should not be taken literally. However, this suggests
that preliminary tests of the theoretical predictions can be made with the low energy
polarized beam of CEBAF even though the cross section is in this case 
dominated by the BH process.

\subsection{Compton scattering in PQCD}

The VCS estimates in the diquark model make clear that a measurement 
of VCS at high energy and large angle will be very difficult 
and will require a dedicated accelerator even for  
moderate values of $Q^2$ ( $\sim$ 1 GeV$^2$).
At these rather low values of $Q^2$, the cross section is dominated 
by its transverse part. Therefore, it is natural to consider real 
Compton scattering at high energy and large angle to study the hard 
scattering. From the experimental point of view, there are much better 
hopes to realize a real Compton scattering experiment. 
Firstly, the count rates are much higher than for electroproduction. 
Also, the required high intensity, 
high energy ($E_\gamma \sim$ 10 - 15 GeV) real photon beam might be feasible 
at the existing HERA ring \cite{d'Ho96}\cite{D'An97}. 
This motivated us \cite{Vdh97b} to perform a leading order PQCD calculation 
for real Compton scattering. 
\newline
\indent
In a first stage, we simplified the calculations 
by approximating the $x, y$ dependence in the gluon virtuality $Q^2$ in 
$\alpha_s (Q^2)$ by their average values for a given distribution amplitude. 
In this way we also avoid large contributions in the end-point region 
($x_i \approx 0, x_i \approx 1$) where the asymptotic formula 
for $\alpha_s (Q^2)$ is no longer valid.

Once the hard scattering amplitude $T_H$ is evaluated, the four-fold 
convolution integral of Eq.(\ref{eq:fact}) has to be performed to obtain 
the Compton helicity amplitudes. The numerical integration 
requires some care because the quark and/or gluon propagators can go 
on-shell which leads to (integrable) singularities. 
The different numerical implementations of these singularities are  
probably at the origin of the different results obtained in 
two previous calculations \cite{Far90a}\cite{Kro91}.
In Refs.\cite{Far90a}\cite{Far90b}, the propagator singularities
were integrated by 
taking a finite value for the imaginary part +i$\epsilon$ 
of the propagator. Then the behaviour of the result was 
studied by decreasing the value of $\epsilon$. 
To obtain convergence with a practical number of samples 
in the Monte Carlo integration performed in Refs.\cite{Far90a} 
\cite{Far90b}, the smallest feasible  value for $\epsilon$ was 
$\epsilon \approx 0.005$. 
In Ref.\cite{Kro91} the propagator singularities were integrated by 
decomposing the propagators into a principal value (off-shell) part and 
an on-shell part. 
To compare these methods, we implemented both of them and found for 
the +i$\epsilon$ method differences of the order of 10\% for every diagram
as compared with the result of our final method. It is not surprising
that, 
when summing hundreds of diagrams, an error of 10\% on every diagram 
can easily be amplified due to the interference between the diagrams. 
To have confidence in the evaluation of the convolution of Eq.(\ref{eq:fact}), 
we compared the principal value integration method with a third 
independent method. 
This third method starts from the observation that the diagrams can 
be classified into four categories depending upon the number of 
propagators which can develop singularities : in the present case this number 
is 0, 1, 2 or 3. Besides the trivial case of zero singularities which can
be integrated immediately, the diagrams with one or two propagator 
singularities can be integrated by performing a contour integration in the 
complex plane for one of the four integrations. For the most difficult case 
of three propagator singularities, we found it possible to evaluate it by 
performing two contour integrations in the complex plane. In doing so, one 
achieves quite a fast convergence because the integrations along the real axis
are 
replaced by integrations along semi-circles in the complex plane which 
are far from the propagator poles.
We checked this method by also implementing the principal value integration 
method and found the same result up to 0.1\% for each type of singularity. 
The principal value method was found to converge much slower and is more 
complicated to implement especially for the case with three singularities 
due to the fact that the three principal value integrals are coupled.

Having exposed the PQCD calculational framework for Compton scattering, 
we now come to the calculations which are performed with several 
model distribution amplitudes denoted 
as CZ \cite{Che84}, COZ \cite{Che89} and KS \cite{Kin87}.

\begin{figure}[h]
\epsfxsize=10.0truecm
\centerline{\epsffile{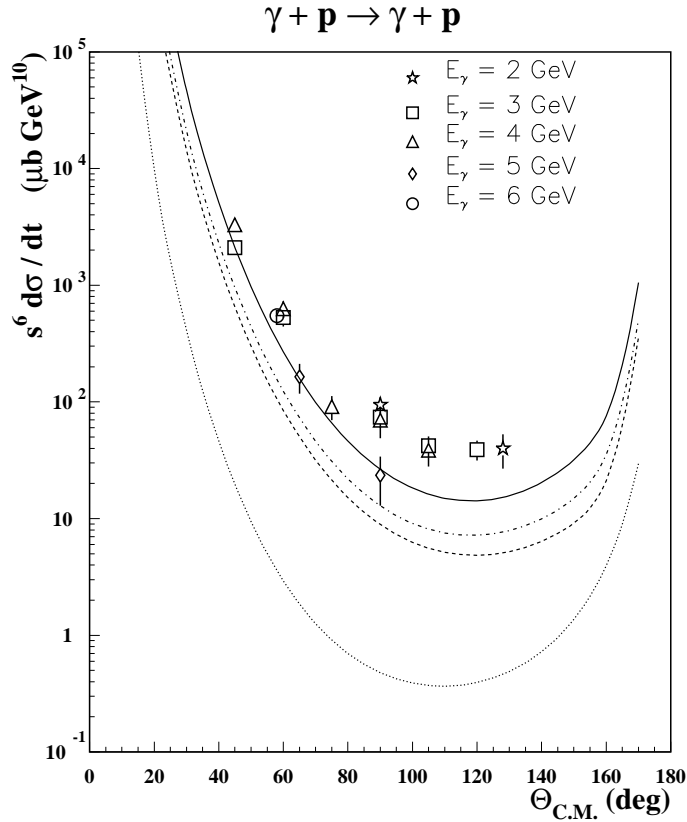} }
\caption{\small Unpolarized Compton cross section on the proton for different
nucleon D.A. : KS (full line), COZ (dashed-dotted line), CZ (dashed line) 
and asymptotic D.A. (dotted line). 
Data are from Refs.\protect\cite{Deu73}\protect\cite{Dud83}\protect\cite{Shu79}.}
\label{fig:unpol}
\end{figure}

The highest energy data which exist for real Compton scattering 
were taken around 5 GeV and are shown in Fig.(\ref{fig:unpol}).
Although the energy at which these experiments were performed 
is probably too low to justify a PQCD calculation, 
we nevertheless show the comparison with these existing data for 
illustrative purposes. 

Let us first mention that the hard scattering amplitude 
has the $s$-dependence ($ T \sim s^{-2}$) 
which leads to the QCD scaling laws \cite{Bro73}, that is 
${{d \sigma} \over {d t}} \sim s^{-6}$ for Compton scattering or VCS. 

The unpolarized real Compton differential cross section 
(multiplied by the scaling factor $s^6$) is shown 
as function of the photon C.M. angle. We first remark that the result 
with the asymptotic D.A. ($\sim 120 \, x_1 x_2 x_3$) is more than one decade below the 
results obtained with the QCD sum rules motivated amplitudes KS, COZ, CZ. 
The results with KS, COZ and CZ show a similar caracteristic angular 
dependence which is asymmetric around 90$^0$. Note that in the forward and 
backward directions, which are dominated by diffractive mechanisms, 
a PQCD calculation is not reliable. 
Comparing the results obtained with KS, COZ and CZ, one notices that 
although these distribution amplitudes have nearly the same lowest moments, 
they lead to differences of a factor of two in the Compton 
scattering cross section. Consequently, this observable is sensitive 
enough to distinguish between various distribution amplitudes.

\begin{figure}[h]
\epsfxsize=10.0cm
\centerline{\epsffile{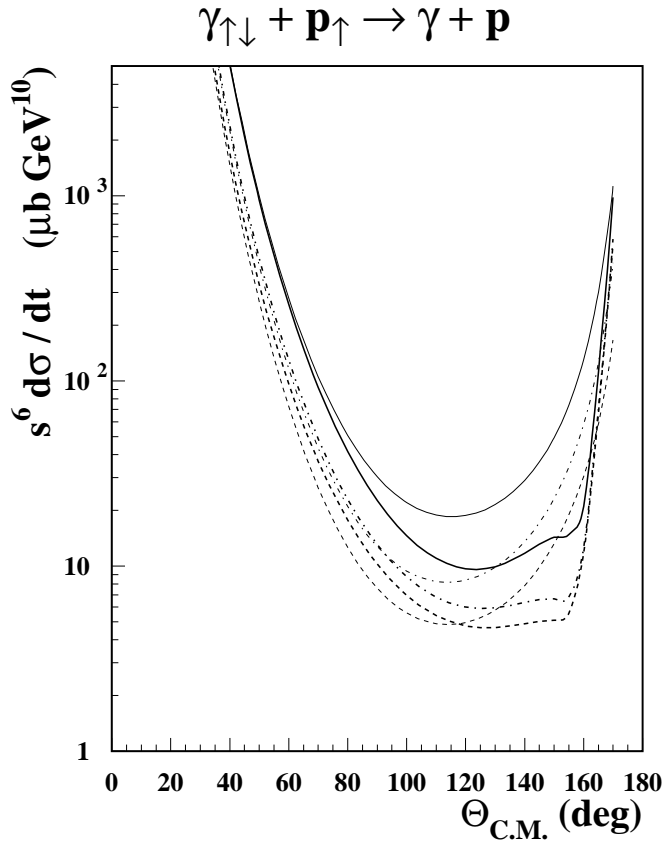} }
\caption{\small Calculations for the polarized Compton cross section 
for two helicity states of the photon : $\lambda = +1$ (thick lines), 
$\lambda' = -1$ (thin lines). Results are shown with 
KS (full lines), COZ (dashed-dotted lines), CZ (dashed lines).}
\label{fig:pol}
\end{figure}

\begin{figure}[h]
\epsfxsize=10.0cm
\centerline{\epsffile{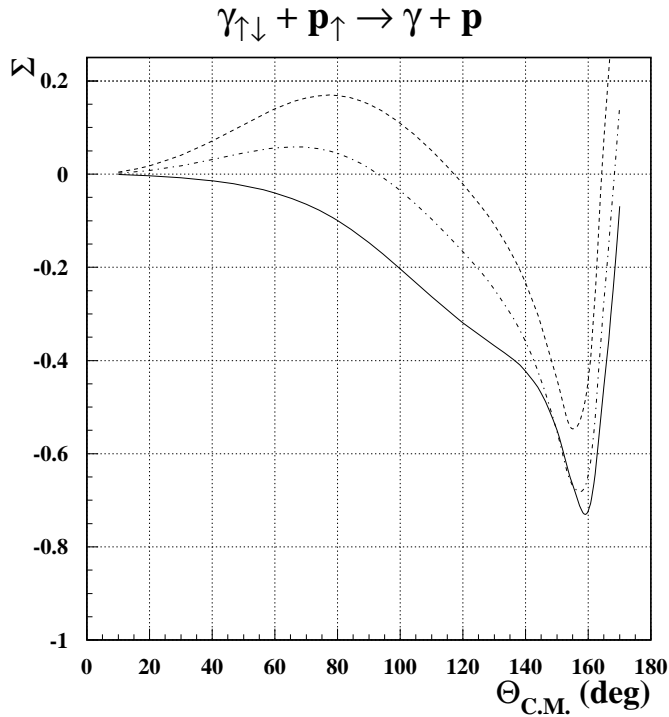} }
\caption{\small Calculations of the photon asymmetry for Compton scattering. 
Results are shown with 
KS (full lines), COZ (dashed-dotted lines) and CZ (dashed lines) D.A.'s.}
\label{fig:asymm}
\end{figure}

In Fig.(\ref{fig:pol}), we show the polarized Compton cross sections for 
the two helicity states of the photon and for a target 
proton with positive helicity. We remark that for all three D.A. 
there is a marked difference both in magnitude and 
angular dependence between the cross sections for the two photon helicities. 
Consequently, the resulting photon asymmetry $ \Sigma $ defined as
\begin{equation}
 \Sigma_\uparrow = \frac{\frac{d\sigma}{dt}(\uparrow,\lambda=1)
                 -\frac{d\sigma}{dt} (\uparrow,\lambda=-1)}
                 {\frac{d\sigma}{dt}(\uparrow,\lambda=1)
                 +\frac{d\sigma}{dt} (\uparrow,\lambda=-1)},
\end{equation}
where $ \lambda $ is the helicity of the incoming photon and where
 $ \uparrow $ means a positive helicity for the hadron, 
changes sign for different values of $\theta_{CM}$ 
as shown in Fig.(\ref{fig:asymm}). This suggests that the 
photon asymmetry might be a 
useful observable to distinguish between nucleon distribution amplitudes.

The predicted sensitivity of the asymmetry to the nucleon D.A. can be used in 
the extraction of a D.A. from Compton scattering data in the scaling region. 
In Ref.\cite{Vdh97b} we have outlined a procedure to extract a D.A. from 
Compton data in a model independent way by expanding the D.A. 
first in a set of basis functions and then using the angular informations 
of the cross sections to fit the expansion coefficients. It was seen that 
the precision on these coefficients is greatly improved when 
one measures both unpolarized cross sections and photon asymmetries. 
Such an experiment might be feasible at the HERA ring in the 
foreseeable future and might open 
up prospects to study the nucleon valence wave function in a direct way.

\newpage
\section{DEEPLY VIRTUAL COMPTON SCATTERING}
\label{sec5}

\subsection{Introduction}
\label{section-5.1}

\begin{minipage}[b]{13.9cm}
Deep-inelastic lepton scattering (DIS) in the Bjorken regime 
($Q^2 \rightarrow \infty$ and $x_B$ = $Q^2$/$2 p \cdot q$ finite) 
is a powerful tool for the study  of nucleon structure \cite{Ste95}. 
In particular, polarized DIS experiments have revealed that only about 
23$\%$ of the nucleon spin is carried by the quarks \cite{Sti96}. 
This has stimulated new investigations to understand the  nucleon spin.
Exclusive  virtual Compton scattering  
in the Bjorken regime (DVCS) has been proposed recently  
\cite{Ji97a}\cite{Ji97b}\cite{Rad96a} to access a new type of parton 
distributions, referred to as {\it off-forward parton distributions} (OFPD's)  
which  are generalizations of the parton distributions measured in DIS. 
Their first moments link them to the nucleon elastic form factors  
and it has been shown \cite{Ji97a} that their second moment 
gives access to the contribution  of the quark spin and quark orbital 
angular momentum to the nucleon spin. 

\hspace{0.45cm}
The off-forward or off-diagonal correlations of quark operators in the 
proton have a long history 
(see e.g. \cite{Dit88}\cite{Bal88}\cite{Jai93} and references therein). 
They appear in the recent literature also under the name of 
{\it nonforward parton distributions} or {\it double parton distributions} 
\cite{Rad97}. Despite their different names and definitions, 
the generalized parton distributions have in common that they 
include the additional degrees of freedom brought in by their 
non forward nature.

\hspace{0.45cm}
It has been proposed that these OFPD's can also be accessed 
through the hard exclusive electroproduction of mesons ($\pi^0$, $\rho^0$,...) 
for which a QCD factorization proof was given recently \cite{Col97}. 
This factorization is illustrated in 
Fig.(\ref{fig:diagrams}b) and is valid for the leading power in 
$Q$ and all logarithms. Furthermore, it is 
valid for all $x_B$ and thus generalizes previous results 
\cite{Rys93}\cite{Bro94}\cite{Rad96b}. According to Ref.\cite{Col97}, 
the factorization between hard and soft processes applies when the 
virtual photon is {\it longitudinally} polarized because in this case the 
end-point contributions in the meson wavefunction are power suppressed.  

\hspace{0.45cm}
In two very recent independent works \cite{Col98}\cite{Ji98}, 
it has been shown that factorization also holds for the DVCS amplitude in
QCD, up to power suppressed terms, to all orders in perturbation
theory. It has furthermore been shown in these works that 
the factorization remains valid independent of the virtuality of 
the final photon, so that it also applies to the production of a real
photon.  

\hspace{0.45cm}
In Ref.\cite{Vdh97c}, the leading order 
of the $\gamma$, $\pi^0$ and $\rho^0_L$ \footnotemark[6] 
leptoproduction amplitudes were calculated 
and first estimates for the cross sections were given 
using an educated guess for the OFPD's. 
These three reactions are highly complementary because 
apart from isospin factors, they depend on the same OFPD's.  
In the following, we present in more detail the 
$\gamma$, $\pi^0$ and $\rho^0_L$ amplitudes at the leading order in PQCD. 
We then explore the experimental opportunities which may allow to determine 
the OFPD's.
\footnotetext[6]{the index $L$ means longitudinally polarized $\rho^0$}
\vspace{3cm}
\end{minipage}

\newpage

\subsection{Light cone dominance of the amplitude in the Bjorken limit 
and definition of off-forward parton distributions}
\label{section-5.2}

\begin{figure}[h]
\epsfxsize=12 cm
\epsfysize=15 cm
\centerline{\epsffile{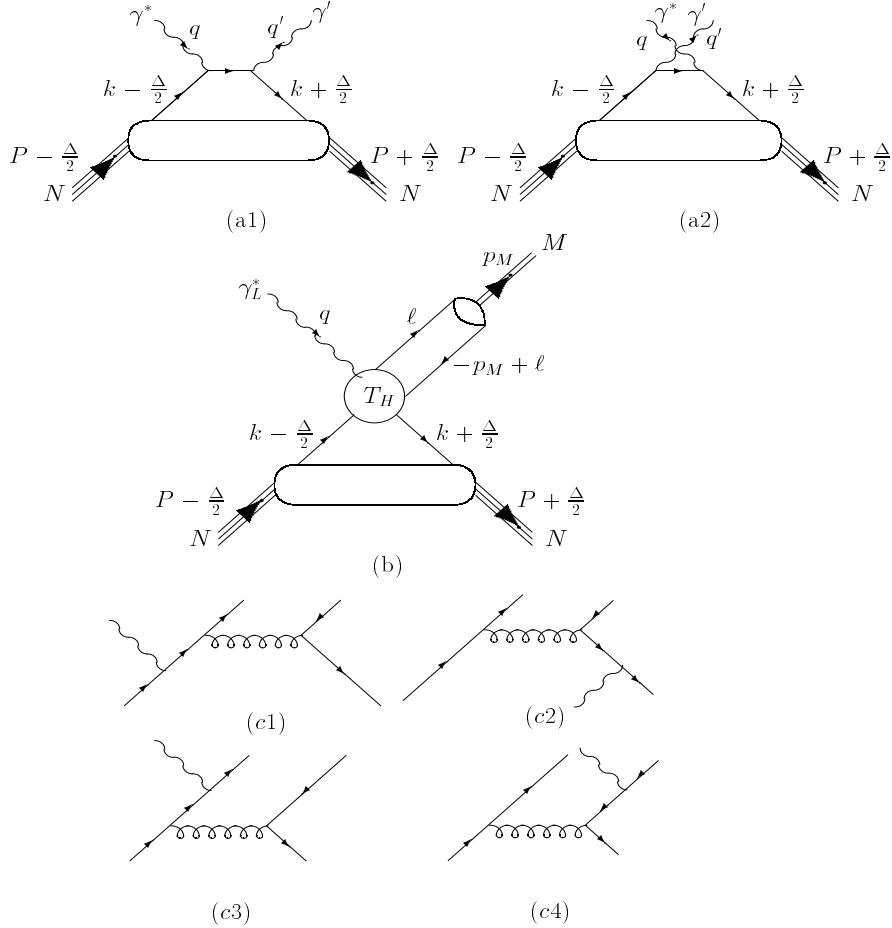}}
\vspace{-2 cm}
\caption[]{\small Direct (a1) and crossed (a2) handbag diagrams for DVCS; 
(b) diagram for the factorized meson electroproduction amplitude; 
(c) leading order diagrams for the hard scattering part $T_H$ 
of the meson electroproduction amplitude.}
\label{fig:diagrams}
\end{figure}
\indent
The leading twist contribution to the DVCS amplitude in the forward direction 
 is given \cite{Ji97b} by the handbag diagrams  
shown in Figs.(\ref{fig:diagrams})(a1, a2). 
To calculate the corresponding  amplitudes, it is convenient 
to use a frame where the virtual photon 
momentum $q^\mu$ and the average nucleon momentum $P^\mu$ 
(see Fig.(\ref{fig:diagrams}) for the kinematics) are collinear and 
along the $z$-axis. 
Furthermore, in the Bjorken regime it is natural to express the 
momenta in terms of the lightlike vectors  
\begin{equation}
\tilde p^\mu = {{P^+} \over {\sqrt{2}}} (1,0,0,1)\;, \hspace{1cm}
n^\mu = {1 \over {P^+\sqrt{2}}} (1,0,0,-1) \;,
\end{equation}
where the light-cone components $a^\pm$ are defined by  
$a^{\pm} \equiv 1/\sqrt{2} (a^0 \pm a^3)$.
The physical momenta have the following decomposition :
\begin{eqnarray}
&&P^\mu = \tilde p^\mu + {{\bar m^2} \over 2} \, n^\mu \;, 
\hspace{1.cm} 
q^\mu = - 2 \xi^{'} \, \tilde p^\mu + {{Q^2} \over {4 \xi^{'}}} \, n^\mu \;, \\
&&\label{eq:dvcsxi} \Delta^{\mu} \equiv p^{'\mu} - p^{\mu} = 
- 2 \xi \, \tilde p^\mu + \xi \bar m^2 \, n^\mu + \Delta^\mu_\perp \;, 
\end{eqnarray}
where the variables $\bar m^2$, $\xi^{'}$ and $\xi$ are given by
\begin{eqnarray}
&&\bar m^2 = m^2 - {{\Delta^2} \over 4} \;,\nonumber\\
&&2 \xi^{'} = {{P \cdot q} \over {\bar m^2}} \,
\left[ -1 + \sqrt{1 + {{Q^2 \bar m^2} \over {(P \cdot q)^2}}} \right]
\;\stackrel{Bj}{\longrightarrow}\; \frac{x_B}{1 - \frac{x_B}{2}} \;,\nonumber\\
&&2 \xi = 2 \xi^{'} \, {{Q^2 - \Delta^2} \over {Q^2 + \bar m^2 (2 \xi^{'})^2}} 
\;\stackrel{Bj}{\longrightarrow}\;  \frac{x_B}{1 - \frac{x_B}{2}} \;.
\label{eq:dvcskin}
\end{eqnarray}
Remark that for a fast moving proton along the positive z-axis 
(such as e.g. is the case in an 
infinite momentum frame) the $P^+$ component will be very large whereas 
the $P^-$ component will be negligible. 
For DVCS, the outgoing photon momentum is
\begin{equation}
q^{'\mu} \equiv q^{\mu} - \Delta^{\mu} = 
- 2 \left( \xi^{'} - \xi \right) \, \tilde p^\mu 
+ \left( {{Q^2} \over {4 \xi^{'}}} - \xi \bar m^2 \right) \, n^\mu 
- \Delta^\mu_\perp \;.
\label{eq:dvcsqp}
\end{equation}
\newline
\indent
In the Bjorken limit, the $q^-$ component of the virtual photon 
momentum is of order $Q^2$ whereas the component $q^+$ is of 
order 1. Therefore, the operator product in the VCS tensor of 
Eq.(\ref{eq_2_18}) is  dominated, as in DIS, by free quark currents 
separated by a lightlike distance. This is due on the one hand to the dominance of the 4-fold
integral in Eq.(\ref{eq_2_18}) by the region $y^+ \sim 1/|q^-|$ and on the other
hand to causality which forces $y^2>0$ \cite{Ell77}.  
So,  in the Bjorken regime Eq.(\ref{eq_2_18}) 
reduces to a one dimensional integral along a lightlike line. 
DVCS and hard electroproduction of mesons in the Bjorken regime  
will therefore select the leading twist part of the 
matrix element of the bilocal quark operator  
represented by the lower blob in Fig.(\ref{fig:diagrams})  
  
\begin{eqnarray}
&&\hspace{-0.5cm}{{P^+} \over {2 \pi}}\, \int d y^- e^{i x P^+ y^-} 
\langle p^{'} | \bar \psi_\beta (- {y \over 2}) \psi_\alpha ({y \over 2}) 
| p \rangle  {\Bigg |}_{y^+ = \vec y_{\perp} = 0} \nonumber\\
&&\hspace{-0.5cm}= {1 \over 4} \left\{ ({\gamma^-})_{\alpha \beta} 
\left[ {H^q(x,\xi,t) \; \bar u(p^{'}) \gamma^+ u(p) 
\;+\; E^q(x,\xi,t) \; \bar u(p^{'}) i \sigma^{+ \kappa} 
{{\Delta_{\kappa}} \over {2 m}} u(p)} \right] \right. \nonumber\\ 
&&\hspace{.2cm}\left. + ({\gamma_5 \gamma^-})_{\alpha \beta} 
\left[ {\tilde H^q(x,\xi,t) \; \bar u(p^{'}) \gamma^+ \gamma_5 u(p) 
\;+\; \tilde E^q(x,\xi,t) \; \bar u(p^{'}) \gamma_5 
{{\Delta^+} \over{2 m}} u(p) } \right] \right\}\;,
\label{eq:qsplitting}
\end{eqnarray}
where $\psi$ is the quark field and $u$ the nucleon spinor. 
In writing down the matrix element Eq.(\ref{eq:qsplitting}) we are 
implicitly working in the light-cone gauge $n \cdot A$ = 0 for the gluon 
field $A$. If one works in a covariant gauge, one has to add a gauge link 
between the quark fields. 

The leading twist matrix element of Eq.(\ref{eq:qsplitting}) is parametrized 
in terms of four OFPD's $H^q, E^q, \tilde H^q, \tilde E^q$. These OFPD's 
are defined for each quark flavor ($q = u, d$ and $s$) 
and depend upon the variables $x, \xi$ and $t$. 
The light-cone momentum fraction $x$ is defined by $k^+ = x P^+$ 
and $t = \Delta^2$.  
The support in $x$ of the OFPD's is $\left[ -1, 1\right]$ 
and a negative momentum fraction corresponds to the antiquark contribution. 
Note that in writing down Eq.(\ref{eq:qsplitting}), we have taken 
$\xi^{'} = \xi$ to simplify the presentation. As shown in 
Eqs.(\ref{eq:dvcskin}), these two variables have the 
same value in the Bjorken limit. 
From Eq.(\ref{eq:dvcsxi}) one realizes that  $\xi$ is  the longitudinal 
momentum transfer. 
As the longitudinal momentum fractions of the nucleons cannot be negative, 
$\xi$ is bounded by 
\begin{equation}
0 < \xi < {{\sqrt{- \Delta^2}/2} \over {\bar m}} < 1 \;.
\end{equation}
\newline
\indent
A glance at Figs.(\ref{fig:diagrams}a, b) shows that the active quark with 
momentum $k - \Delta/2$ has longitudinal (+ component) momentum fraction 
$ x + \xi$, whereas the one with momentum $k + \Delta/2$ 
has longitudinal momentum fraction $ x - \xi$. As negative 
momentum fractions correspond to antiquarks, one can identify two 
regions according to whether $|x| > \xi$ or $|x| < \xi$. 
When $x > \xi$, both quark propagators  represent 
quarks, whereas for $x < - \xi$ both represent antiquarks. In these regions, 
the OFPD's are the generalizations of the usual parton distributions from 
DIS. Actually, in the forward direction, the OFPD's $H$ and $\tilde H$ 
respectively  
reduce to the quark density distribution $q(x)$ and quark helicity 
distribution $\Delta q(x)$:
\begin{eqnarray}
\label{eq:qdistr}
&&H^q(x,\xi = 0,t = 0) = q(x) \;, \\
&&\tilde H^q(x,\xi = 0,t = 0) = \Delta q(x) \;,
\label{eq:delqdistr}
\end{eqnarray}
where $q$ and $\Delta q$ are defined as \cite{Jaf96}
\begin{eqnarray}
\label{eq:dis1}
q(x) \,&=&\, {{p^+} \over {2 \pi}}\, \int d y^{-} e^{i x p^{+} y^{-}} 
\langle p | \bar \psi (0) \gamma . n \psi (y) 
| p \rangle  {\Bigg |}_{y^+ = \vec y_{\perp} = 0} \;, \\
\Delta q(x) \,&=&\, {{p^+} \over {2 \pi}}\, \int d y^{-} e^{i x p^{+} y^{-}} 
\langle p S_\Vert | \bar \psi (0) \gamma . n \gamma_5 \psi (y) 
| p S_\Vert \rangle  {\Bigg |}_{y^+ = \vec y_{\perp} = 0} \;.
\label{eq:dis}
\end{eqnarray}
In Eqs.(\ref{eq:dis1}, \ref{eq:dis}), $p$ represents the 
initial nucleon momentum 
in the DIS process and $S_\Vert$ is the longitudinal nucleon spin projection. 
From Eqs.(\ref{eq:dis1}, \ref{eq:dis}) the quark distributions for negative 
momentum fractions are related to the antiquark distributions as 
$q(-x) = - \, \bar q(x)$ and $\Delta q(-x) = +\, \Delta \bar q(x)$. 
\newline
\indent
In the region $ -\xi < x < \xi$, one quark propagator represents a 
quark and  the other one an antiquark. In this region, the OFPD's 
behave like a meson distribution amplitude.

\subsection{Sum rules for off-forward parton distributions}
\label{section-5.3}

By integrating  Eq.(\ref{eq:qsplitting}) over $x$ one gets the following
relations between  the first moments of the OFPD's and the elastic form factors 
(for one quark flavor)
\begin{eqnarray}
\label{eq:vecsumrule}
&&\int_{-1}^{+1} d x H^{q}(x,\xi,t) \,=\, F_1^{q}(t) \;, \hspace{0.5cm}
\int_{-1}^{+1} d x E^{q}(x,\xi,t) \,=\, F_2^{q}(t) \;, \\
&&\int_{-1}^{+1} d x \tilde H^{q}(x,\xi,t) \,=\, g_A^{q}(t) \;, \hspace{0.5cm}
\int_{-1}^{+1} d x \tilde E^{q}(x,\xi,t) \,=\, h_A^{q}(t) \;,
\label{eq:axvecsumrule}
\end{eqnarray} 
where $F_1^q$, $F_2^q$ are the Dirac and Pauli form factors respectively, 
$g_A^q$ is the axial form factor and $h_A^q$ is the induced pseudoscalar 
form factor. 
Note that in the sum rules of Eqs.(\ref{eq:vecsumrule}) and 
(\ref{eq:axvecsumrule}), the $\xi$-dependence drops out. 
\newline
\indent
According to the considered reaction, the proton OFPD's 
enter in different combinations due to the  charges and isospin factors. 
For the DVCS on the proton, the combination is  
\begin{equation}
H^p_{DVCS}(x,\xi,t) \,=\,{4 \over 9} H^{u/p} \,+\, {1 \over 9} H^{d/p} 
\,+\,{1 \over 9} H^{s/p} \;,
\end{equation}
and similarly for $\tilde H$, $E$ and $\tilde E$. For the 
electroproduction of $\rho^0$ and $\pi^0$ on the proton, 
the isospin structure yields the combination 
\begin{equation}
H^p_{\rho^0}(x,\xi,t) \,=\,{1 \over {\sqrt{2}}} \left\{ {2 \over 3} H^{u/p} 
\,+\, {1 \over 3} H^{d/p} \right\} \;,
\hspace{0.5cm}
\tilde H^p_{\pi^0}(x,\xi,t) \,=\,{1 \over {\sqrt{2}}} \left\{
{2 \over 3} \tilde H^{u/p} \,+\, {1 \over 3} \tilde H^{d/p} \right\} \;,
\end{equation}
and similar for $E$ and $\tilde E$.
The elastic form factors for one quark flavor 
have to be related to the physical ones. Restricting oneself to the 
$u, d$ and $s$ quark flavors, this yields 
\begin{equation}
F_1^{u/p} \,=\, 2 F_1^{p} + F_1^{n} + F_1^{s} \;,
\hspace{0.5cm}
F_1^{d/p} \,=\, 2 F_1^{n} + F_1^{p} + F_1^{s} \;,
\label{eq:vecff}
\end{equation}
where $F_1^{p}$ and $F_1^{n}$ are the proton and neutron electromagnetic 
form factors respectively. The strange form factor is given by 
$F_1^{s/p} \equiv F_1^{s}$, but since it is small and not so well known 
 we set it to zero in the following.
Relations similar to Eq.(\ref{eq:vecff}) hold also for the form factor $F_2^{q}$.
For the axial vector form factors one uses the isospin decomposition : 
\begin{equation}
g_A^{u/p} \,=\, {1 \over 2} g_A + {1 \over 2} g_A^0 \;,
\hspace{0.5cm}
g_A^{d/p} \,=\, -{1 \over 2} g_A + {1 \over 2} g_A^0 \;.
\label{eq:axff}
\end{equation}
The isovector axial form factor $g_A$ 
is known from experiment ($g_A(0) \approx 1.26$) 
and for the unknown isoscalar axial form factor 
$g_A^0$ we use the quark model relation : $g_A^0(t) = 3/5 \, g_A(t)$. 
For $h_A^{q}$ we have relations similar to Eq.(\ref{eq:axff}).
\newline
\indent
The second moment of the OFPD's is relevant for the nucleon spin structure. 
It was shown in Ref.\cite{Ji97a} that there exists a (color) 
gauge-invariant decomposition 
of the nucleon spin:
\begin{equation}
{1 \over 2} \,=\, J_q \,+\, J_g \;,
\end{equation}
where $J_q$ and $J_g$ are respectively the total quark and gluon spin
 contributions to the nucleon spin. 
The second moment of the  OFPD's gives 
\begin{equation}
J_q \,=\, {1 \over 2} \, \int_{-1}^{+1} d x \, x \, 
\left[ H^{q}(x,\xi,t = 0) + E^{q}(x,\xi,t = 0) \right] \;, 
\label{eq:dvcs_spin}
\end{equation}
and this relation is independent of $\xi$. 
The total quark spin contribution $J_q$ decomposes as 
\begin{equation}
J_q = {1 \over 2} \Delta \Sigma + L_q \;,
\end{equation}
where 1/2 $\Delta \Sigma$ and $L_q$ are respectively 
the quark spin and quark orbital contributions to the nucleon spin. 
We recall that $\Delta \Sigma$ is measured through polarized DIS experiments. 
So if one can measure the OFPD's through 
exclusive hard electroproduction reactions,  
 the sum rule of Eq.(\ref{eq:dvcs_spin}) will determine  the quark orbital 
contribution to the nucleon spin.

\subsection{DVCS amplitude}
\label{section-5.4a}

In the following, we assume as in Refs.\cite{Ji97a}\cite{Ji97b} that a 
factorization between hard and soft processes is valid for DVCS. 
A formal factorization proof has been given very recently
\cite{Col98}\cite{Ji98}. 
Using the parametrization of Eq.(\ref{eq:qsplitting}) 
for the bilocal quark operator, 
the leading order DVCS tensor $H^{\mu \nu}_{L.O. \, DVCS} $ 
(defined by Eq.(\ref{eq_2_18})) follows from the two handbag diagrams of 
Fig.(\ref{fig:diagrams}a) as 
\begin{eqnarray}
&&\hspace{-0.5cm}H^{\mu \nu}_{L.O. \, DVCS} \nonumber\\
&&\hspace{-0.5cm} = {1 \over 2} 
\left\{\left[ \tilde p^\mu n^\nu + \tilde p^\nu n^\mu - g^{\mu \nu} \right] 
\cdot \int_{-1}^{+1}d x \left[ {1 \over {x - \xi + i \epsilon}} 
+ {1 \over {x + \xi - i \epsilon}} \right] \right. \nonumber\\
&&\hspace{.5cm}\cdot \left[ H^p_{DVCS}(x,\xi,t) \;\bar u(p^{'}) \gamma .n u(p) 
+\; E^p_{DVCS}(x,\xi,t) \;\bar u(p^{'}) i \sigma^{\kappa \lambda} {{n_\kappa \Delta_{\lambda}} \over {2 m}} N(p) \right] \nonumber\\
&&\hspace{0.4cm}+\;\left[ -i \varepsilon^{\mu \nu \kappa \lambda} \tilde p_\kappa n_\lambda \right] 
\cdot \int_{-1}^{+1}d x \left[ {1 \over {x - \xi + i \epsilon}} 
- {1 \over {x + \xi - i \epsilon}} \right] \nonumber\\
&&\hspace{.5cm}\left. \cdot \left[ \tilde H^p_{DVCS}(x,\xi,t) \bar N(p^{'}) \gamma . n \gamma_5 N(p) 
+ \tilde E^p_{DVCS}(x,\xi,t) \bar N(p^{'}) \gamma_5 {{\Delta \cdot n} \over{2 m}} N(p)  
\right] \right\}, 
\label{eq:dvcs_ampl}
\end{eqnarray}
with $\varepsilon_{0123} = +1$. 
Tests of the handbag approximation to the DVCS amplitude have been proposed in 
Ref.\cite{Die97} and experiments are proposed to check the hypothesis
\cite{Ber98}. 
\newline
\indent
The leading order DVCS amplitude of Eq.(\ref{eq:dvcs_ampl}) 
as given first in Refs.\cite{Ji97a} \cite{Ji97b}, 
is exactly gauge invariant with respect to the virtual photon, 
i.e. $q_{\nu} \, H^{\mu \nu}_{L.O. \, DVCS} = 0$.   
However gauge invariance is violated by the 
real photon except  in the forward direction. 
In fact 
$q^{'}_{\mu} \, H^{\mu \nu}_{L.O. DVCS} \sim \Delta_{\perp}$. 
This violation of gauge invariance is a 
higher twist effect of order $1/Q^2$ compared to the leading order 
term $H^{\mu \nu}_{L.O. \, DVCS}$. So in the limit $Q^2\to\infty$ it is
innocuous but for actual experiments  it matters. Actually for any cross section
estimate one needs to choose a gauge and this explicit gauge dependence for
$\theta\ne0$ is unpleasant. 
In the absence of a dynamical gauge invariant higher twist calculation of 
the DVCS amplitude, we propose to restore  gauge invariance
 in a heuristic way based on physical considerations. We propose to write 
\begin{equation} 
H^{\mu \nu}_{DVCS} \,=\,H^{\mu \nu}_{L.O. \, DVCS} \,-\, 
{{a^\mu} \over {\left( a \cdot q^{'}\right)}} \;
\left( q^{'}_\lambda \, H^{\lambda \nu}_{L.O. \, DVCS} \right) \;,
\label{eq:dvcs_gaugeinv1}
\end{equation}
where $a^\mu$ is a four-vector  specified below. 
Obviously $H^{\mu \nu}_{DVCS}$ respects gauge invariance 
for both the virtual and the real photon :
\begin{equation}
q_{\nu} \, H^{\mu \nu}_{DVCS} = 0\;, \hspace{1cm} 
q^{'}_{\mu} \, H^{\mu \nu}_{DVCS} = 0 \;.
\end{equation}
Furthermore, as 
\begin{equation}
q^{'}_{\lambda} \, H^{\lambda \nu}_{L.O. \, DVCS} = 
- \left( \Delta_\perp \right)_\lambda H^{\lambda \nu}_{L.O. \, DVCS}\;,
\end{equation}
the gauge restoring term  gives zero in the forward 
direction ($ \Delta_\perp = 0 $) which is natural. 
We  choose 
$a^\mu=\tilde p^\mu$ because $\tilde p \cdot q^{'}$ is of order 
$Q^2$, which gives automatically a gauge restoring term of order 
$O\left( 1/Q^2 \right)$. 
This  choice for $a^\mu$ is furthermore motivated by the fact that  
in the derivation of the leading order amplitude of 
Eq.(\ref{eq:dvcs_ampl}), only the $\tilde p^\mu$ components at the 
electromagnetic vertices are retained \cite{Ji97a}\cite{Ji97b}. 
The above  arguments lead to the following fully gauge invariant 
DVCS amplitude:
\begin{equation}
H^{\mu \nu}_{DVCS} \,=\,H^{\mu \nu}_{L.O. \, DVCS} \,+\, 
{{\tilde p^\mu} \over {\left( \tilde p \cdot q^{'}\right)}} \;
\left( \Delta_\perp \right)_\lambda \, 
H^{\lambda \nu}_{L.O. \, DVCS}  \;.
\label{eq:dvcs_gaugeinv}
\end{equation}
We will illustrate the influence of this gauge invariance prescription, 
by comparing DVCS observables calculated with Eq.(\ref{eq:dvcs_gaugeinv}) 
and with the leading order formula of Eq.(\ref{eq:dvcs_ampl}). 

\subsection{Hard meson electroproduction amplitudes}
\label{section-5.4b}

\begin{minipage}[b]{13.9cm}
For the electroproduction of $\pi^0$ and $\rho^0_L$ mesons \footnotemark[7] 
at large values of $x_B$ (i.e. in the valence region), 
the leading order amplitude is given by the diagrams of Fig.(\ref{fig:diagrams}c).  
This amplitude was calculated in Ref.\cite{Vdh97c} where 
the following gauge invariant expressions were found 
for the current operators 
\begin{eqnarray}
&&\hspace{-.9cm} J^\mu_{\rho^0_L}
=\; (i e 4 \pi \alpha_s)\;{4 \over 9}\; 
{{x_B} \over {Q^2}} \; 
\left[ \; \int_0^1 d z {{\Phi_\rho(z)} \over z}\right] \cdot 
\left\{ \tilde p^\mu + {{Q^2} \over {2 x_B^2}}\, n^\mu\right\} \nonumber\\
&&\hspace{.1cm} \cdot {1 \over 2} \,  
\int_{-1}^{+1} d x \left[ {1 \over {x - \xi + i \epsilon}} 
+ {1 \over {x + \xi - i \epsilon}}\right] \nonumber\\
&&\hspace{1.6cm}\cdot \left\{ H^p_{\rho^0_L}(x,\xi,t) \bar u(p^{'}) \gamma . n u(p) 
+ E^p_{\rho^0_L}(x,\xi,t) \bar u(p^{'}) 
i \sigma^{\kappa \lambda} {{n_\kappa \Delta_{\lambda}} \over {2 m}} u(p) \right\},
\label{eq:rho_ampl}
\end{eqnarray}
\begin{eqnarray}
&&\hspace{-.9cm} J^\mu_{\pi^0}
=\; (i e 4 \pi \alpha_s)\; {4 \over 9} \; 
{{x_B} \over {Q^2}} \; 
\left[\int_0^1 d z {{\Phi_\pi(z)} \over z}\right] \cdot 
\left\{ \tilde p^\mu + {{Q^2} \over {2 x_B^2}}\, n^\mu\right\} \nonumber\\
&&\hspace{.1cm} \cdot {1 \over 2} \,  
\int_{-1}^{+1} d x \left[ {1 \over {x - \xi + i \epsilon}} 
+ {1 \over {x + \xi - i \epsilon}}\right] \nonumber\\
&&\hspace{1.6cm}\cdot \left\{ \tilde H^p_{\pi^0}(x,\xi,t) \bar u(p^{'}) \gamma . n \gamma_5 u(p) 
+ \tilde E^p_{\pi^0}(x,\xi,t) 
\bar u(p^{'}) \gamma_5 {{\Delta \cdot n} \over{2 m}} u(p) \right\},
\label{eq:pi_ampl}
\end{eqnarray}
where $\Phi_\rho(z)$ and $\Phi_\pi(z)$ are the $\rho$ and $\pi$ 
distribution amplitude (DA) respectively. 
For the pion, recent data \cite{Gro98} for the $\pi^0 \gamma^* \gamma$ 
transition form factor up to $Q^2$ = 9 GeV$^2$ support 
the asymptotic form: 
\begin{equation}
\Phi_\pi(z) = \sqrt{2} f_\pi 6 z (1 - z)\;,
\end{equation}
with $f_\pi$ = 0.093 GeV from the pion weak decay. 
For the rho meson, the DA is not as well  known. 
A recent theoretical analysis \cite{Bal96} updating the QCD sum rule 
analysis by 
\footnotetext[7]{Note that although we will only show results for the $\rho^0$ 
and $\pi^0$ in this work, the formalism also applies for other vector mesons 
such as the $\omega$ and $\phi$ or other pseudoscalar mesons such as
the $\eta$.} 
\end{minipage}

\noindent
including $O(\alpha_s)$ radiative corrections, 
favors a DA that is rather close to its asymptotic form. 
In the calculations shown below, we use the asymptotic DA for the rho : 
\begin{equation}
\Phi_\rho(z) = \sqrt{2} f_\rho 6 z (1 - z)\;,
\end{equation}
with $f_\rho$ = 0.153 GeV determined from the electromagnetic decay 
$\rho^0 \rightarrow e^+ e^-$. 
Note that the CZ DA for the $\rho$ \cite{Che84}, 
which is more concentrated near the end points, 
would yield an amplitude larger by a factor 5/3.
\newline
\indent
The ( $e, e^{'} M$ ) cross section for a meson $M$ is given 
by the formula of Eq.(\ref{eq_2_30}). The $T, L, TT, TL$ cross sections 
appearing in Eq.(\ref{eq_2_30}) are calculated from the helicity 
amplitudes for meson electroproduction. In the following, we give only 
predictions for $\sigma_L$ as, at large $Q^2$, only 
the meson production by a longitudinal virtual photon survives 
and because the factorization theorem \cite{Col97} only applies for  
a longitudinal virtual photon. 
The longitudinal $\gamma^* + p \rightarrow M + p$ two body cross section 
$d \sigma_L / d t$ of Eq.(\ref{eq_2_30}) is 
\begin{equation}
{{d \sigma_L} \over {d t}} = {1 \over {16 \pi \left( s - m^2\right)^2}} 
\; {1 \over 2} \sum_{h_N} \sum_{h^{'}_N} | 
{q_0 \over Q} \, \varepsilon_\mu (\lambda = 0) 
\cdot J^\mu (\lambda_M = 0, h^{'}_N; h_N) |^2 \;,
\label{eq:mescross}
\end{equation}
where $\varepsilon_\mu(\lambda = 0)$ is the polarization vector for a 
longitudinal photon (following the convention of App.A, Eq.(\ref{eq:app1pol})). 
It is expressed in terms of the lightlike vectors 
$\tilde p^\mu$ and $n^\mu$ as 
\begin{equation}
{q_0 \over Q} \, \varepsilon^\mu(\lambda = 0) \;=\;
- {1 \over Q} \left\{ x_B \, \tilde p^\mu \;+\; 
{Q^2 \over {2 x_B}} n^\mu \right\} \;.
\end{equation}
In Eq.(\ref{eq:mescross}), 
$J^\mu$ is the current operator for production of a $\rho^0_L$ or $\pi^0$ 
defined in Eqs(\ref{eq:rho_ampl}, \ref{eq:pi_ampl}).
One can see  from Eqs.(\ref{eq:rho_ampl}, \ref{eq:pi_ampl}) that 
the large $Q^2$ behaviour of the amplitude for $\rho^0_L$ and $\pi^0$ 
production is 
\begin{eqnarray}
{q_0 \over Q} \, \varepsilon_\mu (\lambda = 0) 
\cdot J^\mu (\lambda_M = 0, h^{'}_N; h_N) 
\sim {q_0 \over Q}\,\varepsilon_\mu (\lambda = 0) \cdot 
\left\{  \tilde p^\mu + {{Q^2} \over {2 x_B^2}} n^\mu  \right\} \; 
\left( {{x_B} \over {Q^2}} \right) = - {1 \over {Q}}.
\end{eqnarray}
As $ s =m^2 + Q^2 (1 - x_B)/x_B $, 
this leads to a $1/Q^6$ behavior of $d \sigma_L / d t$ at large $Q^2$.

\subsection{Models for the off-forward parton distributions}
\label{section-5.5}

Ultimately one wants to extract the OFPD's from data but in order 
to evaluate electroproduction observables, we need a first guess for 
the OFPD's. We shall restrict our considerations to the near forward 
direction because this kinematical domain is the closest to  
inclusive DIS, which we want to use as a guide. In this domain, the 
contribution of $E$ and $\tilde E$, which goes like $\Delta$, is suppressed. 
So in the following we shall keep only $H$ and $\tilde H$. Furthermore, we assume that, at small $-t$, 
$H^q$ is proportional to the unpolarized quark distribution for which 
we take the MRS ($S_0$) parametrization of Ref.\cite{MRS93}
\begin{eqnarray}
x \, d_V(x) &=& A_d x^{0.78} (1 - x)^{4.57} ( 1 - 0.87 \sqrt{x} + 0.82 x )\;, \nonumber\\
x \, \left( u_V(x) + d_V(x) \right) &=& A_{ud} x^{0.26} (1 - x)^{3.82} ( 1 + 14.4 \sqrt{x} + 16.99 x )\;, \nonumber\\
x \, S(x) &=& 1.87 (1 - x)^{10} ( 1 - 2.21 \sqrt{x} + 6.22 x )\;.
\label{eq:unpoldistr}
\end{eqnarray} 
The valence parts of the quark distributions are normalized as 
\begin{equation}
{1\over 2} \int_{0}^{+1}dx \, u_V(x) =  \int_{0}^{+1}dx \, d_V(x) = 1 \;,
\end{equation}
and the sea quark distributions are parametrized as 
$u_S = \bar u = d_S = \bar d = 2 s_s = 2 \bar s = 5/9 \, S$. 
Our ansatz for the OFPD's $H^q$ thus yields 
\begin{equation}
H^{u/p}(x,\xi,t) = {1\over 2} u(x) {F_1^{u/p}(t)}\;, \hspace{0.5cm}
H^{d/p}(x,\xi,t) = d(x) {F_1^{d/p}(t)}\;, \hspace{0.5cm}
H^{s/p}(x,\xi,t) = 0\;,
\label{eq:factunpol}
\end{equation}
and obviously the ansatz satisfies both Eq.(\ref{eq:qdistr}) and 
the sum rule Eq.(\ref{eq:vecsumrule}). 
\newline
\indent
Similarly, we assume that $\tilde H^q(x,\xi,t)$ is proportional to the 
polarized quark distribution. For the latter we neglect the not well known 
sea part and we adopt the valence parametrizations of Ref.\cite{Gos97}
\begin{eqnarray}
&&\Delta u_V(x) = \cos \Theta_D(x)\; \left[ u_V(x) - {2 \over 3} d_V(x) \right] \;,\hspace{0.5cm}
\Delta d_V(x) =  \cos \Theta_D(x)\; \left[ - {1 \over 3} d_V(x) \right] \;, \nonumber \\
&&\cos \Theta_D(x) = {1 \over { 1 + H_0 (1 - x)^2/\sqrt{x} } } \;,
\label{eq:poldistr}
\end{eqnarray}
with $H_0 = 0.06$ such that the Bjorken sum rule is satisfied. 
This yields the normalization
\begin{equation}
\int_{0}^{+1}dx \Delta u_V(x) = 0.98 \;, \hspace{0.5cm} 
\int_{0}^{+1}dx \Delta d_V(x) = -0.27 \;, 
\end{equation}
which leads to the correct proton 
to neutron magnetic moment ratio $\mu_p / \mu_n \approx -1.47$. 
Our ansatz for $\tilde H^q$ is thus
\begin{equation}
\tilde H^{u/p}(x,\xi,t) = \Delta u_V(x) {g_A^{u/p}(t)} / {g_A^{u/p}(0)} \;, \hspace{0.5cm}
\tilde H^{d/p}(x,\xi,t) = \Delta d_V(x) {g_A^{d/p}(t)} / {g_A^{d/p}(0)} \;, 
\label{eq:factpol}
\end{equation}
and one can check that both Eq.(\ref{eq:delqdistr}) 
and the sum rule of Eq.(\ref{eq:axvecsumrule}) are satisfied. 
Note that as suggested by the bag model estimate of Ref.\cite{Ji97c}, 
we have omitted any dependence of the OFPD's on $\xi$ which should be 
justified for moderate $Q^2$. At high values of $Q^2$, this approximation 
is expected to break down and the $Q^2$ evolution has to be considered. 
The evolution of the OFPD's has recently been studied by several authors 
\cite{Rad97}\cite{Bal97}\cite{Bel97}\cite{Fra97}. 

\newpage

\subsection{Results for DVCS and hard meson leptoproduction observables}
\label{section-5.6}

\begin{figure}[h]
\epsfxsize=10 cm
\epsfysize=11.6 cm
\centerline{\epsffile{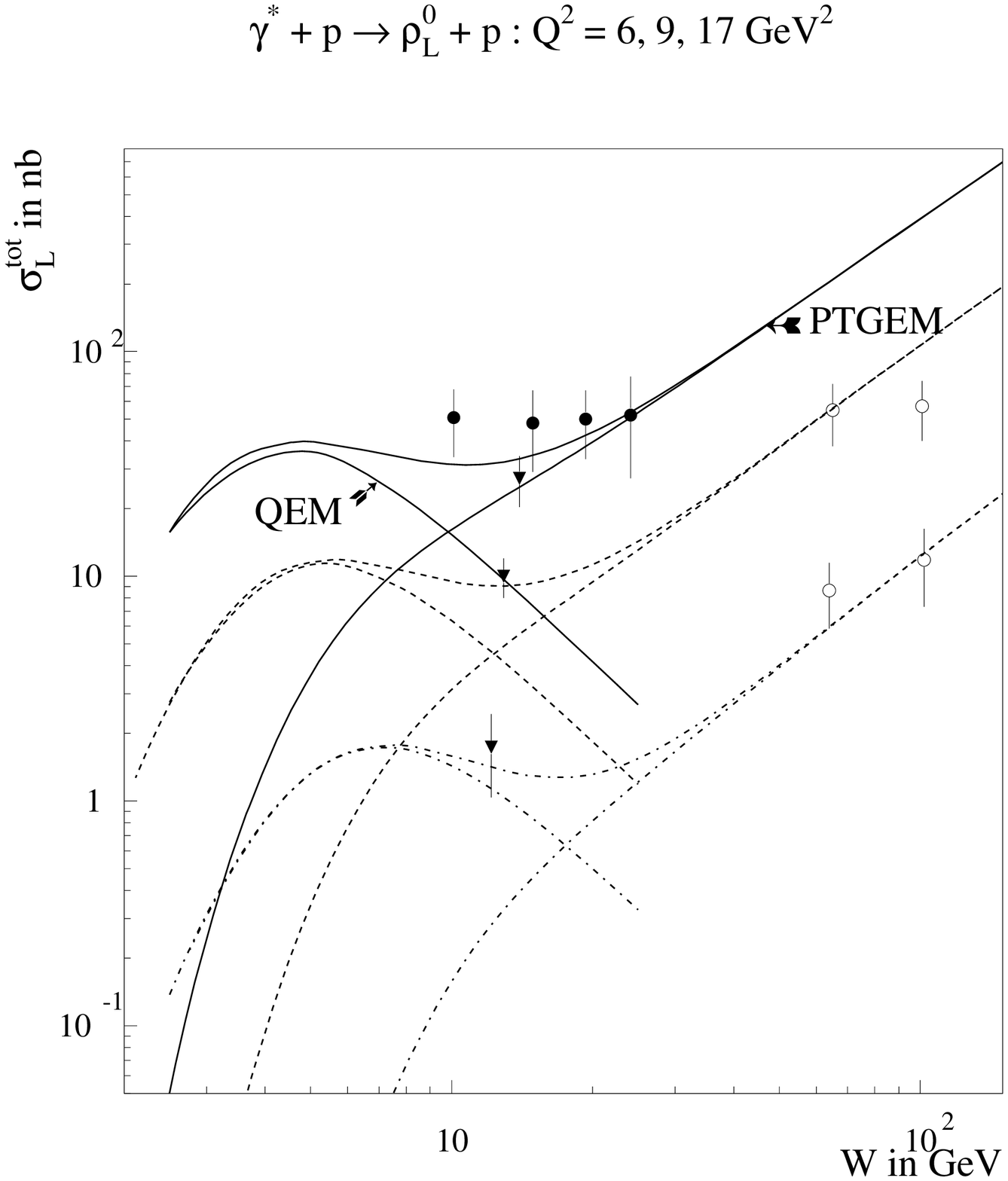}}
\vspace{-1cm}
\caption[]{\small Total longitudinal cross section for $\rho^0_L$ electroproduction. 
Data from NMC \cite{Arn94} (triangles) at $Q^2$ =  5.5 (highest point), 
8.8 and 16.9 (lowest point) GeV$^2$, 
E665 \cite{Ada97} (black circles) at $Q^2$ = 5.6 GeV$^2$
and ZEUS \cite{Der95} (open circles) at $Q^2$ = 8.8 (upper points) 
and 16.9 GeV$^2$ (lower points). 
Calculations are shown at $Q^2$ = 6 GeV$^2$ (full lines), 
$Q^2$ = 9 GeV$^2$ (dashed lines) and $Q^2$ = 17 GeV$^2$ (dashed-dotted lines). 
The curves which grow at high W correspond with gluon exchange whereas the curves 
which are peaked below W $\approx$ 10 GeV correspond with quark exchange. 
The incoherent sum of both mechanisms is also shown.}
\label{fig:rhotot}
\end{figure}

We first show results for the $\rho^0$ electroproduction cross sections. 
In Fig.(\ref{fig:rhotot}), the total longitudinal $\rho^0_L$ 
electroproduction cross section is shown as function of the C.M. energy 
$W$ for different values of $Q^2$. 
At high C.M. energies, it is well known that the Perturbative 
Two Gluon Exchange Mechanism (PTGEM)  
dominates as soon as  $Q^2\geq$ 6 GeV$^2$ and this mechanism implies a  
$1/Q^6$ behavior of $\sigma_L$
\cite{Bro94}\cite{Fra96}. The PTGEM  
cross section calculated with Eq.(5) of Ref.\cite{Fra96} is 
shown in Fig.(\ref{fig:rhotot}). We used the CTEQ3L parametrization 
\cite{Lai95} for the gluon distribution. 
The PTGEM explains well the fast increase at high energy of the cross section. 
Note that the data point shown on Fig.(\ref{fig:rhotot}) are  
consistent with the more recent ZEUS data \cite{Cri97}.

However, the PTGEM substantially underestimates the data 
at lower energies (around $W \approx 10~GeV$), 
where the quark exchange mechanism (QEM) of Fig.(\ref{fig:diagrams}b) 
is expected to contribute since $x_B$ is then in the valence region. 
In Fig.(\ref{fig:rhotot}) we show our predictions using the OFPD's proposed
 in the previous section.  The incoherent sum of both mechanisms is also 
indicated. Fig.(\ref{fig:rhotot}) 
provides a rather strong indication that the deviation 
from the PTGEM of the data at lower energies can be attributed 
to the onset of the QEM. The good agreement between the data in this region and
our prediction gives us some confidence in our factorized ansatz 
for the OFPD's.
\begin{figure}[h]
\epsfxsize=10 cm
\epsfysize=12 cm
\centerline{\epsffile{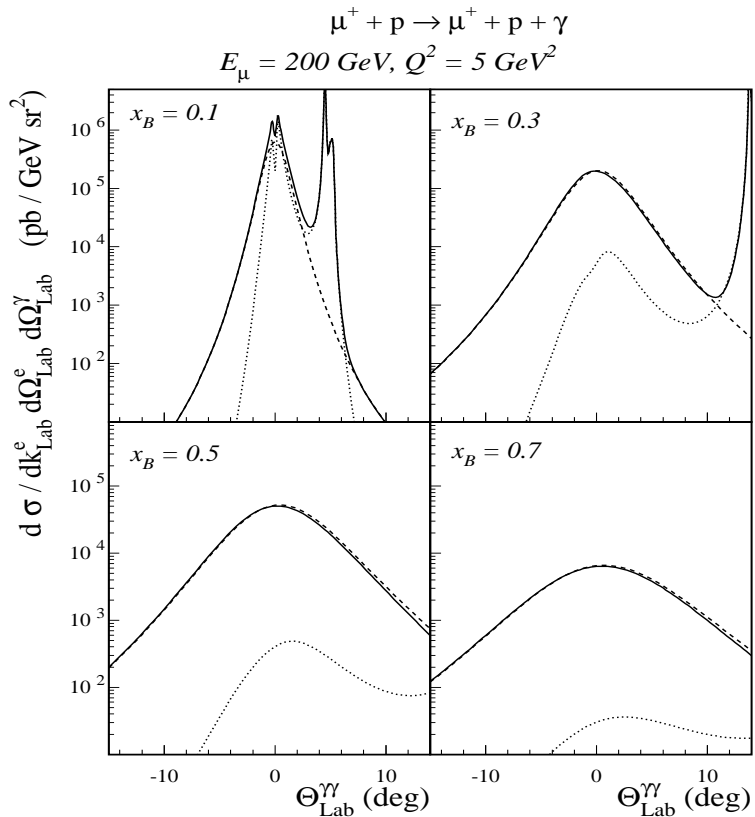}}
\vspace{-1cm}
\caption[]{\small Comparison between BH (dotted lines), 
DVCS (dashed lines) and total $\gamma$ (full lines) in-plane cross 
sections for COMPASS kinematics.}
\label{fig:dvcs200}
\end{figure}

\begin{figure}[h]
\vspace{-2.5cm}
\hspace{.5cm}
\epsfxsize=12.5 cm
\epsfysize=14.5 cm
\centerline{\epsffile{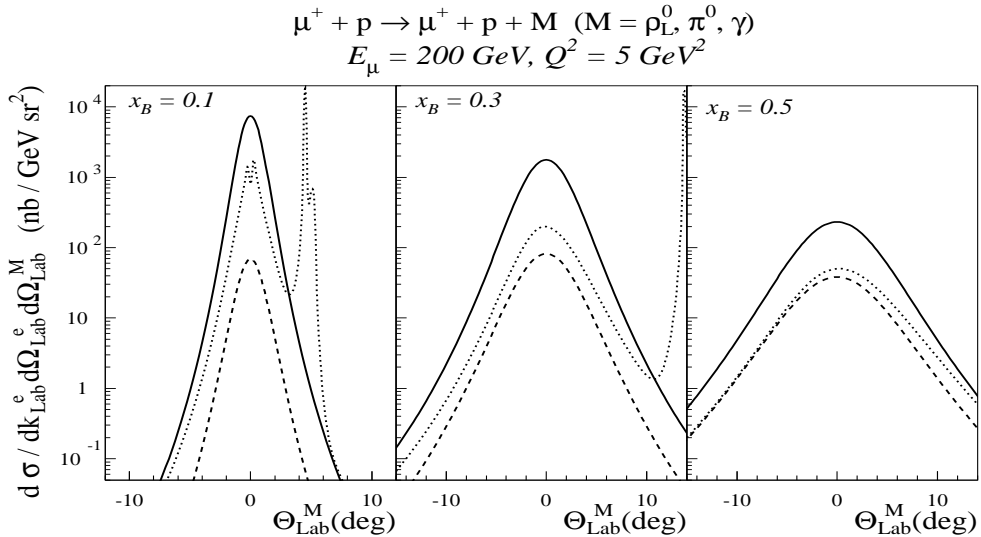}}
\vspace{-5cm}
\caption[]{\small Comparison between $\rho^0_L$ (full lines), 
$\pi^0$ (dashed lines) and total $\gamma$ (dotted lines) 
muoproduction in-plane cross sections for COMPASS kinematics.}
\label{fig:compass}
\end{figure}

\begin{figure}[h]
\epsfxsize=8 cm
\epsfysize=10 cm
\centerline{\epsffile{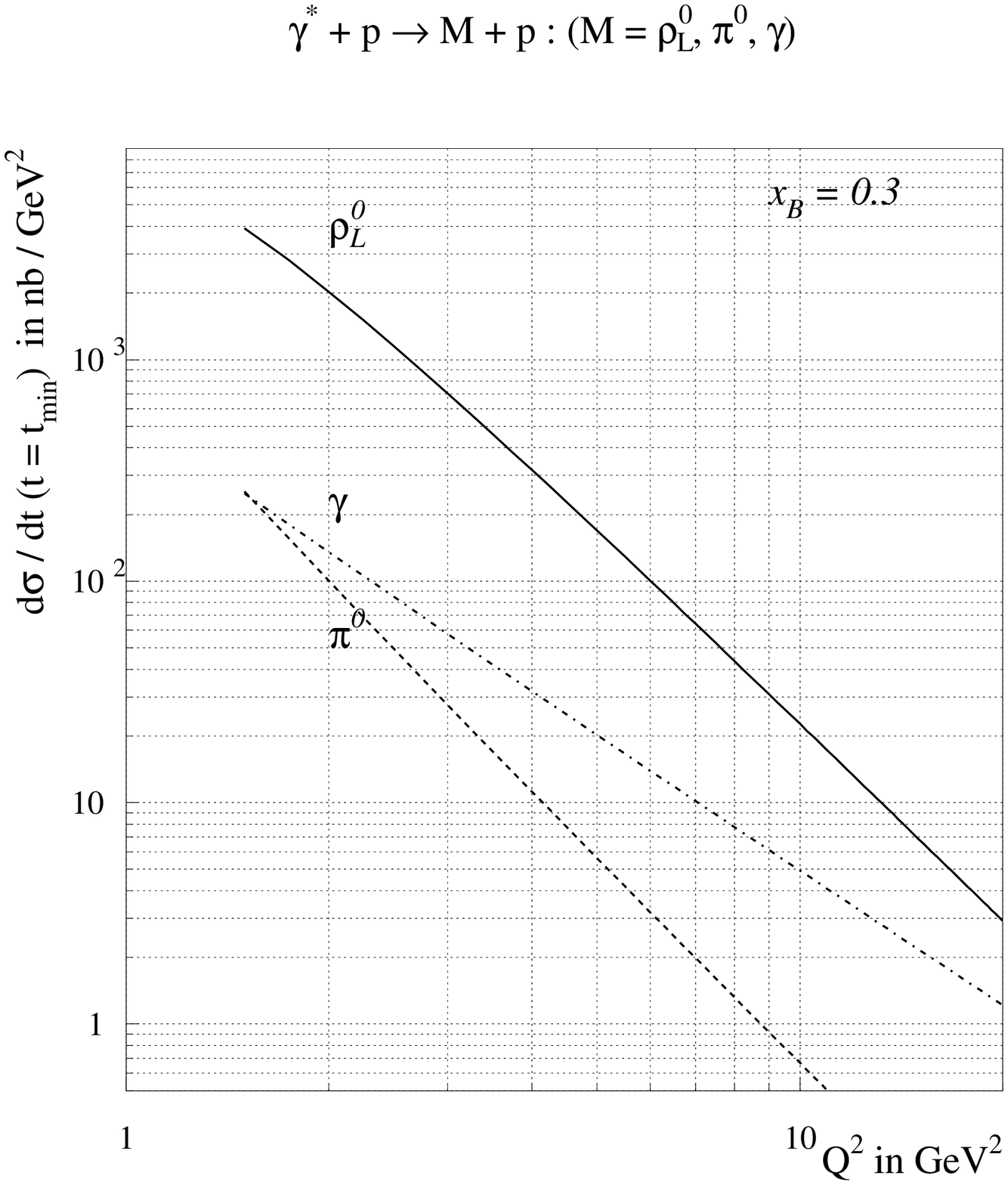}}
\vspace{-1cm}
\caption[]{\small Scaling behavior of the forward ( $t = t_{min}$ ) 
differential cross section $d \sigma_L/d t$ for 
$\rho^0_L$ and $\pi^0$ electroproduction 
and of $d \sigma_T/d t$ for the DVCS cross section. }
\label{fig:scaling}
\end{figure}

\noindent
\begin{minipage}[b]{13.9cm}

\hspace{0.45cm}
The QEM is studied further in
Figs.(\ref{fig:dvcs200}),(\ref{fig:compass}) 
where the angular dependence of the fivefold differential 
DVCS \footnotemark[8], 
$\pi^0$ and $\rho^0_L$ muoproduction cross sections are shown 
for kinematics accessible at COMPASS. 
In our calculations we add coherently the BH and DVCS amplitudes. 
From a phenomenological point of view, it is clear that the 
best situation occurs when the BH process is negligible. 
For fixed $Q^2$ and $s$, the only way to favor the DVCS 
\footnotetext[8]{The small differences between the results shown here and 
those of Ref.\cite{Gui97} are due to the use of different parton 
distributions.}
\end{minipage}

\noindent
over the BH is to 
increase the virtual photon flux and this amounts to increase the beam 
energy. According to our estimate, the unpolarized $(l, l^{'} \gamma)$  
cross section in the forward region is dominated by the BH process in the 
few GeV region. To get a clear dominance of the DVCS 
process one needs a beam energy in the 100 GeV range. 
This is illustrated in Fig.(\ref{fig:dvcs200}) 
where we show separately the BH, the DVCS and the coherent cross section 
at $E_\mu=200$ GeV and $Q^2$ = 5 GeV$^2$. 
In the valence region ($x_B \approx 0.3$), the DVCS cross section 
is more than one order of magnitude larger than the BH in the forward direction. 
The BH cross section increases if one goes to smaller $x_B$ and 
becomes comparable to the DVCS cross section around $x_B \approx 0.1$.
 This is due to the fact that in the BH process the exchanged photon has
 4-momentum $(q-q')$ which gives a $1/t$ behaviour to the amplitude.
  The value of $t$ in the forward direction ($t_{min}$) 
is very small for small $x_B$ values and it increases with  $x_B$. At 
$x_B = 0.7$, the value of $- t_{min}$ in Fig.(\ref{fig:dvcs200}) 
is around 1 GeV$^2$, which explains the sharp drop of the BH amplitude 
compared to the DVCS one. The sharp rise of the BH process 
in the forward direction at small $x_B$ puts a limit on the region where 
the DVCS can be studied experimentally. At COMPASS kinematics, 
$x_B \simeq 0.1$ seems to be the lower limit. 
Although the BH is not a limiting factor at  high $x_B$,
 one cannot go too close to $x_B = 1$ in order to stay well 
above the resonance region.   
\newline
\indent
In Fig.(\ref{fig:compass}), the $x_B$ dependence of 
the $\rho^0_L$, $\pi^0$ and $\gamma$ 
fivefold differential muoproduction cross sections are compared in the same 
kinematics as in Fig.(\ref{fig:dvcs200}). 
The $x_B$ dependence of the $\rho^0_L$, $\pi^0$ and DVCS cross sections 
is a direct reflection of the $x_B$ dependence of the OFPD's.
In the valence region ($x_B \approx 0.3$), the $\rho^0_L$ cross section, 
which is sensitive to the unpolarized OFPD's, is about one 
order of magnitude larger than the $\pi^0$ cross section 
which is sensitive to the polarized OFPD's. 
The DVCS cross section  is sensitive to both unpolarized and polarized OFPD's 
but, due to the additional electromagnetic coupling, it 
is also about one order of magnitude below the $\rho^0_L$ 
cross section. Although the $\rho^0_L$ cross section is the 
easiest to measure, the $\pi^0$ and DVCS cross sections seem 
large enough to encourage a study of the feasibility of the experiments. 
The three reactions are highly complementary due to their 
different dependence on the OFPD's. 
\newline
\indent
Before considering the extraction of  the OFPD's from the data,
 it is compulsory to demonstrate that the scaling regime has been reached.
  In Fig.(\ref{fig:scaling}) 
we show the forward $\gamma^* + p \rightarrow M + p$ 
 cross sections as a function of $Q^2$. This illustrates  
the lever arm one has to test the scaling behavior, 
as the maximum value of $Q^2$ will be given by the count rate
limit of an experiment. 
For the DVCS, the leading order amplitude is constant in $Q$ and is 
predominantly transverse. 
Therefore, the leading order DVCS transverse cross section $d \sigma_T/d t$ 
shows a $1/Q^4$ behavior. 
To test this scaling behavior, one needs of course a kinematical  situation 
where the DVCS dominates over the BH.
 
\begin{figure}[h]
\vspace{-2cm}
\hspace{.5cm}
\epsfxsize=12.5 cm
\epsfysize=13.5 cm
\centerline{\epsffile{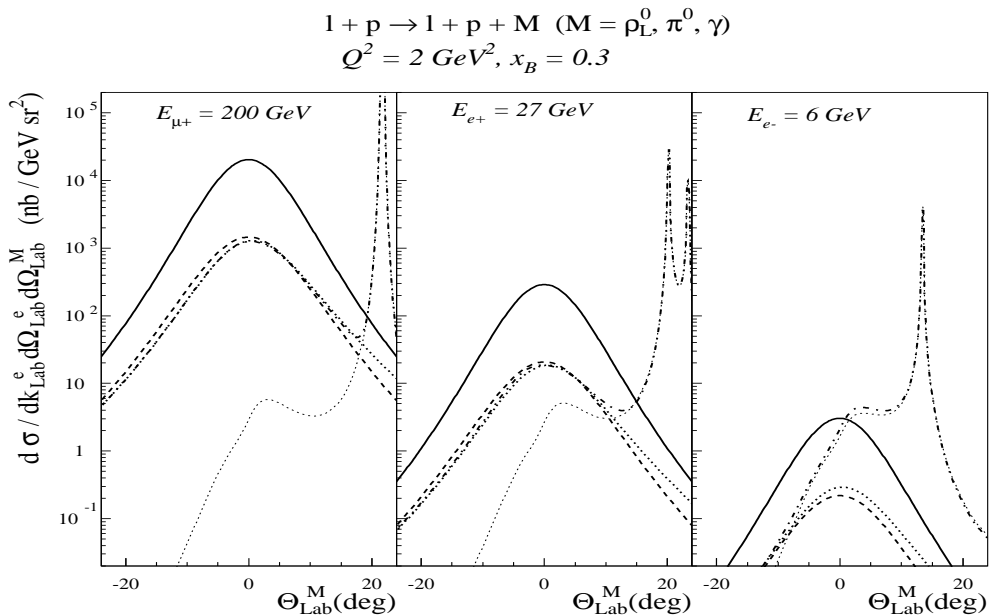}}
\vspace{-3.5cm}
\caption[]{\small Comparison between $\rho^0_L$ (full lines), 
$\pi^0$ (dashed lines), DVCS (dotted lines), BH (thin dotted lines) 
and total $\gamma$ (dashed-dotted lines) 
leptoproduction in-plane cross sections at $Q^2$ = 2 GeV$^2$, $x_B$ = 0.3 
and for different beam energies : 
$E_{\mu^+}$ = 200 GeV (COMPASS), $E_{e^+}$ = 27 GeV (HERMES), 
$E_{e^-}$ = 6 GeV (CEBAF).}
\label{fig:diffkin}
\end{figure}

\begin{figure}[h]
\vspace{1.5cm}
\epsfxsize=8 cm
\epsfysize=9 cm
\centerline{\epsffile{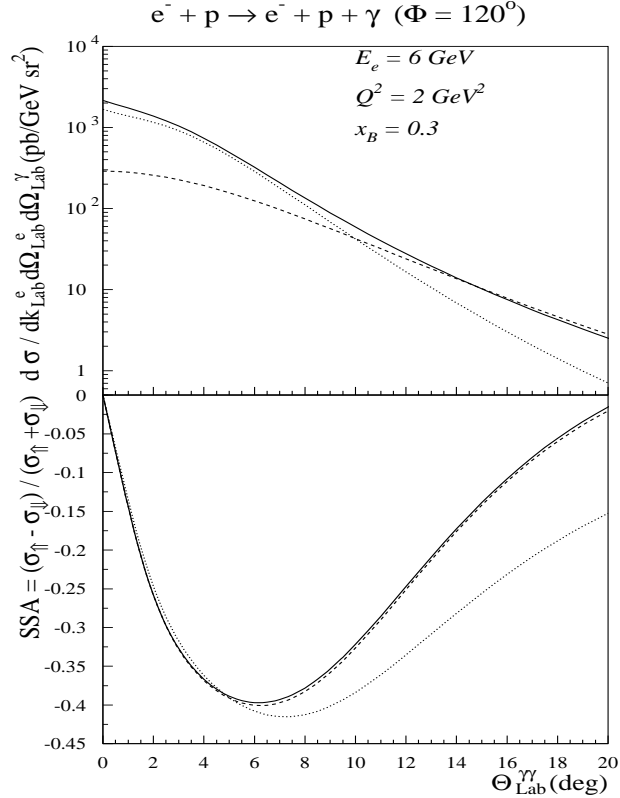}}
\vspace{.8cm}
\caption[]{\small DVCS at CEBAF for out-of-plane kinematics ($\phi = 120^o$). 
Upper part shows the differential cross section : 
DVCS (dashed line), BH (dotted line) and total $\gamma$ (full line). 
In the lower part, the electron single spin asymmetry is shown 
for the gaugeinvariant amplitude (full line) and for the leading order 
non-gaugeinvariant amplitude calculated in radiative gauge (dashed line) 
and in Feynman gauge (dotted line).}
\label{fig:dvcsasymm}
\end{figure}

The leading order 
meson electroproduction amplitude was seen to behave as $1/Q$ and is 
predominantly longitudinal. 
Therefore, the leading order longitudinal cross section $d \sigma_L/d t$ 
for meson electroproduction shows a $1/Q^6$ behavior. 
Note that eventhough  the $\rho^0_L$ electroproduction is not as 
clean  as the DVCS due to its  dependence  on  the $\rho^0$ wavefunction,  
a test of the $1/Q^6$ scaling behavior would nevertheless be 
 meaningful. The reason is that an uncertainty 
in the meson wavefunction would influence mostly the normalization 
of the cross section, but not its behavior in $Q$. 
\newline
\indent
\newline
\indent
Although a high energy such as planned at COMPASS is  preferable,  one 
can try  to undertake a preliminary study of the hard electroproduction 
reactions using the existing facilities such as HERMES or CEBAF, despite their
low energy. 
To this end, we compare in Fig.(\ref{fig:diffkin}), 
the $\rho^0_L$, $\pi^0$ and $\gamma$ cross sections as function of 
the beam energy at a fixed $Q^2$ = 2 GeV$^2$ and $x_B$ = 0.3. 
Going up in energy, the increasing virtual photon flux factor boosts the 
$\rho^0_L$, $\pi^0$ leptoproduction cross sections and the DVCS part of 
the $\gamma$ leptoproduction cross section.
For the $\gamma$ electroproduction cross section the BH process 
is hardly influenced by the beam 
energy and therefore overwhelms the DVCS cross section at low beam energies. 
\newline
\indent
To make a preliminary exploration of the OFPD's in the few GeV range  
 and to test the scaling, the measurement of 
the $\rho^0_L$ leptoproduction through its decay into charged pions seems the 
easiest from the experimental point of view as the count rates are the 
highest. In particular, in the $Q^2$ = 1 - 5 GeV$^2$ range, the 
$\rho^0_L$ cross section can be used to delineate 
the scaling region.   
\newline
\indent
For the $\gamma$ leptoproduction at low energies,  we suggest that an 
exploration of the DVCS part might be possible if the beam is polarized. 
The electron single spin asymmetry (SSA) 
does not vanish out of plane and is only due to 
the interference between the BH process 
and the imaginary part of the DVCS amplitude. 
Therefore, even if the cross section is dominated by the BH process, 
the SSA is {\it linear} in the OFPD's. To illustrate the point, we show on 
Fig.(\ref{fig:dvcsasymm}) the unpolarized cross section for a 6 GeV beam and 
the SSA at an azimuthal angle $\phi$ = 120$^0$. 
In the $0^0$ - $5^0$ region, 
a rather large asymmetry is predicted, even though the cross section is 
dominated by the BH process. Of course this does not exclude the risk 
that at this small  $Q^2$, higher twist effects might be important. 

\begin{figure}[h]
\vspace{1cm}
\epsfxsize=8 cm
\epsfysize=10 cm
\centerline{\epsffile{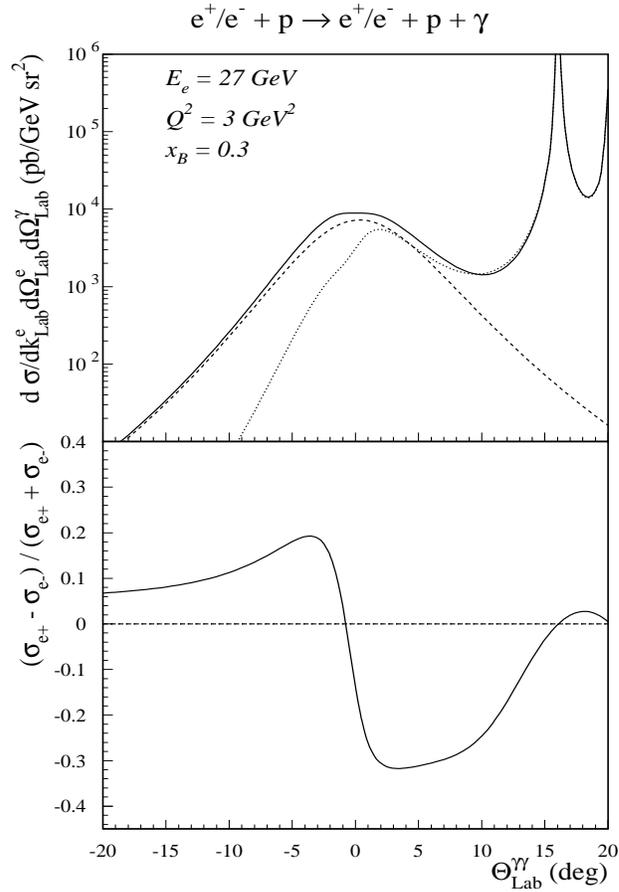}}
\vspace{1cm}
\caption[]{\small e$^+$e$^-$ asymmetry at HERMES. 
Upper part shows the in-plane e$^+$ differential cross section : 
BH + DVCS (full line), DVCS (dashed line), 
BH (dotted line). 
Lower part : e$^+$e$^-$ charge asymmetry with same curve conventions as 
in upper part of figure.}
\label{fig:chargeasymm}
\end{figure}

To illustrate the effect of the gauge restoring term in  Eq.(\ref{eq:dvcs_gaugeinv}), 
we have plotted in Fig.(\ref{fig:dvcsasymm}) the SSA both 
for the gauge invariant and non gauge invariant amplitudes. 
For the non gauge invariant amplitude of Eq.(\ref{eq:dvcs_ampl}), 
the SSA is shown both in the radiative gauge and in the Feynman gauge. 
As expected, all predictions are identical at small angle. 
At larger angles the  gauge dependence clearly shows up, especially in the
Feynman gauge. 
\newline
\indent
Finally, we recall  that the real part of the BH-VCS interference 
can be accessed by 
reversing the charge of the lepton beam since this 
 changes the relative sign the BH and FVCS amplitudes. This leads to the
 following charge asymmetry
\begin{equation}
\sigma_{e^+} - \sigma_{e^-} \sim 4 \Re e \left[ T^{BH} {T^{FVCS}}^* \right] \;.
\end{equation}
Our estimate for the e$^+$e$^-$ asymmetry is shown in 
Fig.(\ref{fig:chargeasymm}) at 27 GeV. The comfortable asymmetry 
that one can see in the small angle region may offer an 
interesting opportunity for HERMES.
\newline
\indent

\newpage

\section{OUTLOOK AND PERSPECTIVES}
\label{sect_6}

As we have seen in the previous sections, virtual Compton scattering is a very
versatile tool.

At threshold it gives us access to a new kind of electron scattering experiments
where the target is submitted to an external electric or magnetic field. The
outcome of the experiments are the form factors of the currents induced by the
applied field. This generalizes the notion of polarizability in the sense that
the spatial distribution of the induced current can be measured while in real
Compton scattering only the dipole moment is accessible. For the proton there
are six GP's which are functions of $Q^2$. Their experimental determination is now under
way but this is a delicate task. As explained in Section \ref{sec3}, the
response  at threshold is dominated by the global motion of the proton in the
external field. Scattering on this moving object teaches us nothing more 
than what we already know such as its mass, its charge and its elastic form factors. To
extract the true information one has to subtract this trivial but dominating
part and this supposes very accurate experiments. To obtain the GP's, with
an accuracy of 15\% one needs cross sections with an accuracy of 3\%.

The first experiment has been performed at MAMI \cite{d'Ho95}\cite{LHu97a} 
and at the time of writing, the analysis 
has not yet reached a stage where the polarizabilities can be
extracted from the data. 
On the other hand the low energy theorem is well satisfied by the very low
energy data \cite{d'Ho97}. To our knowledge it is the first time that this
theorem is tested.  The next experiments will
take place at CEBAF \cite{Ber93} in 1998 where the higher beam energy 
is expected to help by allowing
a better choice of the virtual photon polarization $\epsilon$. 
Finally an experiment at small
$Q^2$ has been accepted at Bates \cite{Sha97} with the aim of testing Chiral
perturbation theory predictions.

At present no experiments with polarization have been proposed. Though polarized
beams exist now at the high duty cycle electron accelerators, 
we have seen that one needs to measure the
polarization of the recoiling proton. Due to the low efficiency of the
polarimeters, below 10\%, one can foresee 
that this interesting physics will not start
before the unpolarized program has achieved its task.
We have demonstrated however that these recoil polarization observables 
provide three independent observables and complement the three unpolarized 
observables. Therefore, the measurement of these recoil polarization 
observables holds promise to disentangle the six independent GP's. 

The hard scattering regime of VCS discussed in Section \ref{sec4} gives access
to the most elementary part of the nucleon structure, its 3 quark valence wave
function. Thanks to the large energy-momentum transferred to the nucleon, the
complex configurations which build the full confinement are eliminated from the
amplitude. The price to pay is that the cross sections are extremely small and
without a dedicated accelerator like ELFE \cite{Arv93} the experiments look
impossible. However a preliminary attempt will be performed at CEBAF
\cite{Ber93}. The beam energy is definitively too low to access the hard
scattering regime but the use of the polarized beam to select the BH-VCS
interference will provide the first test of the diquark model prediction for the
phase of the VCS amplitude. Finally we recall that the physics case of 
real Compton scattering is similar to the one of VCS. Since
the experiments are much easier it may be a good strategy to try to develop
first the real Compton scattering in the hard scattering regime.
Such experiments might become possible with a high intensity, high energy 
($E_\gamma \sim$ 15 GeV) photon beam at the existing HERA ring 
\cite{d'Ho96}\cite{D'An97}.

VCS in the Bjorken regime or DVCS is the most recent development in the VCS
field. Because of its strong connection with the time honoured inclusive DIS, it
has become rapidely popular.  In this regime one can access the so called 
off-forward parton distributions which may shed a new light on the 
nucleon spin problem. We
have used a crude factorized model of the OFPD's to perform the first evaluation
of the cross sections including the BH amplitude. Gauge invariance 
with respect to the outgoing real photon has been
enforced in a minimal way in the sense that the corrective term is explicitely
a higher twist effect. Its effect with respect to a calculation in the 
radiative gauge is about 1\%. 

To devise a convincing test of scaling of the 
DVCS amplitude and to access the OFPD's from this reaction 
one needs a very high energy in order that the BH process be 
negligible. The 100-200 GeV muon beam of COMPASS seems adequate, 
at least for not too small values of $x_{B}$.

However it is tempting to make use of the existing machines to open the way. The
trick is to climb on the back of the BH amplitude using its interference with
the DVCS amplitude. Even though the cross section may be completely 
dominated by the BH,
the interference remains linear in the OFPD's that we are looking for. The
first possibility is the single spin asymmetry which we have found to be large
and which selects the imaginary part of the DVCS amplitude. A proposal of
experiment is now in preparation \cite{Ber98}. 
The energy might be too low to reach the
scaling regime but since the experiment does not imply a huge effort it is worth
an attempt. The second possiblity is the beam charge asymmetry which is also
large and selects the real part of the DVCS amplitude. In this case one does not
need an out of plane experiment. This could be a good case for HERMES or SLAC 
and their higher energy as compared to CEBAF   would be a serious advantage.

The high energy forward meson leptoproduction is closely related 
to DVCS because the same OFPD's enter the amplitudes. 
Moreover one can select the OFPD's by selecting the meson. In this way, 
the $\rho^0_L$ electroproduction gives access to the unpolarized OFPD's and 
the $\pi^0$ to the polarized OFPD's. 
One needs another input, the meson wave function,
and therefore the information one gets from this process is less pure than in
the DVCS case. However it is a useful complement and we have evaluated the
cross sections using the (conservative) asymptotic wave function, 
which for the pion is able to describe the recent 
data for the $\pi^0 \gamma^* \gamma$ transition form factor \cite{Gro98}. 
We found comfortable cross sections especially for $\rho^0_L$ leptoproduction. 
In the meson case the BH problem is absent 
and therefore the measurement of the $\rho^0_L$ 
electroproduction may provide a nice test to see if one can 
access the scaling region already at CEBAF.

An important point which has not much been discussed in the text is the problem of
radiative corrections, which is important in the case of experiments with
electrons. All the Feynman diagrams contributing to the cross section of 
order $\alpha^4_{QED}$ have been calculated \cite{Vdh98} using the dimensional
regularisation both in the infrared and in the ultraviolet regions. The only
approximation is that the radiative effects have been neglected for the
proton. This approximation is justified by its large mass and has been tested 
to be numerically negligible using a model \cite{LHu97b}. The difficulty of the
calculation is that it cannot be done analytically. After renormalization and
cancellation of the infrared divergences, there remain multidimensional
integrals over the Feynman parameters that one must evaluate numerically. The
hard point is that the functions to integrate have sometimes wild variations, of the
order $E_{\mathrm beam}/m_e$ in a range of the order $m_e/E_{\mathrm beam}$. It
has been necessary to develop very accurate integration methods to take care of
this \cite{LHu97b}. Another problem, which is particular to VCS, is that the
integrals may also have poles corresponding to the on shell propagation of the
intermediate photon. To cope with that it has been necessary to extend the
integration in the complex plane of the Feynman parameters.
 
As pointed out in the introduction of this review, three interesting kinematical 
domains of VCS have been identified and discussed in some detail in this review. 
The threshold regime, the hard scattering
regime and the DVCS have a well established physics case and are worth the
experimental effort. This does not mean that VCS experiments cannot be done in
other kinematical regimes. This is the case for the resonance region
($\sqrt{s} < 2$ GeV which will also be explored at CEBAF in 1998
\cite{Ber93}. In this region the advantage of VCS over other probes is not so
clear even if one forgets about the BH problem. 
The point is that in the
resonance region the photon channel is, by unitarity, strongly coupled to the
meson production channels. 
So the information is dominated by meson production and
rescattering. This may be interesting by itself but it is much easier to study
it  directly using meson electroproduction rather than with the very 
difficult ($e, e' \gamma$) experiments.
 
It seems now that, despite its experimental challenge, VCS has reached the
status of a powerful and promising probe of nucleon structure. Nevertheless 
some  work remains to be done to comfort this position. 
In particular, the higher twist effects need to be estimated 
and the model estimates of the OFPD's should
be developped. Finally, since the OFPD's enter the amplitude through a
convolution, one has to devise an efficient way to extract them from the data.
We hope that this review will stimulate such work and that the experiments in
the coming years will confirm our optimistic view of virtual Compton
scattering.

\section*{\center{Acknowledgements}}

This work was supported by the French Commissariat \`a 
l'Energie Atomique and in part by the EU/TMR contract ERB FMRX-CT96-0008.
The authors thank P.Y. Bertin, V. Breton, N. d'Hose, H. Fonvieille, 
C. Hyde-Wright, L. Van Hoorebeke and P. Vernin for their experimental efforts as well as their 
stimulating interest for this work. The
calculation of the radiative corrections would have been impossible without the
enormous work of D. Lhuillier, D. Marchand and J.Van de Wiele. This work has
been possible thanks to the collaboration with M. Guidal, P. Kroll, G.Q. Liu, 
M. Sch\"urmann and A.W. Thomas. We thank D. Drechsel, A. Metz, and V.
Brindejonc for many stimulating discussions.

\newpage

\begin{appendix}
\section{Polarization vectors and spinors}
\label{app1}

The unit vectors $\vec{e}(1),\vec{e}(2),\vec{e}(3)$ of the C.M. frame 
have been defined in the text. The polarization vectors of a photon with
momentum $\vec{k}$ are
\begin{equation}
\vec{\varepsilon}(\vec{k},1)=
\vec{\varepsilon}(\vec{k},2)\times\hat{k},\ 
\vec{\varepsilon}(\vec{k},2)=
\frac{\hat{k}\times\vec{e}(1)}{|\hat{k}\times\vec{e}(1)|},\ 
\vec{\varepsilon}(\vec{k},3)=\hat{k}.
\end{equation}
and the helicity states are

\begin{equation}
\vec{\varepsilon}(\vec{k},\pm 1)=\mp\frac{1}{\sqrt{2}}
\left[\vec{\varepsilon}(\vec{k},1)\pm\,i\,\vec{\varepsilon}(\vec{k},2) \right],\ 
\vec{\varepsilon}(\vec{k},0)=\vec{\varepsilon}(\vec{k},3).
\end{equation}
The 4-vector polarization states in the Lorentz gauge are then defined as
\begin{equation}
\varepsilon^\mu(\pm 1)=\left(
\begin{array}{c}
0  \\ \vec{\varepsilon}(\pm 1)  
\end{array}
\right),\ 
\varepsilon^\mu(0)=\left(
\begin{array}{c}
 {\mathrm k}/k_0 \\ \vec{\varepsilon}(0)  
\end{array}
\right).
\label{eq:app1pol}
\end{equation}
The helicity spinors are
\begin{equation}
u(\vec{k},h)=\left(
\begin{array}{c}
\sqrt{k_0+m}\ \chi_h(\hat{k})\\2h\sqrt{k_0-m}\ \chi_h(\hat{k})
\end{array}
\right)\;,
\end{equation}
with
\begin{equation}
\chi_{1/2}(\hat{k})=\left(
\begin{array}{c}
\cos(\alpha/2)\\{\mathrm e}^{i\phi}\sin(\alpha/2)
\end{array}
\right),\ \ 
\chi_{-1/2}(\hat{k})=\left(
\begin{array}{c}
-{\mathrm e}^{-i\phi}\sin(\alpha/2)\\ \cos(\alpha/2)
\end{array}
\right)\;,
\end{equation}
where ($\alpha,\phi$) are the polar and azimuthal angles of the direction
$\hat{k}$.
The rest frame spin projection spinors are
\begin{equation}
u(\vec{k},\sigma)=\left(
\begin{array}{c}
\sqrt{k_0+m}\ \chi_\sigma\\ \sqrt{k_0-m} \; \vec\sigma \cdot \hat k \ \chi_\sigma
\end{array}
\right),\ \ 
\chi_{1/2}=\left(
\begin{array}{c}
1\\ 0
\end{array}
\right),\ \ \chi_{-1/2}=\left(
\begin{array}{c}
0\\ 1
\end{array}
\right)\;.
\end{equation}
These spinors are the positive energy solutions of the Dirac equation
($\gamma .k-m)u(k)=0$ and the Dirac matrices are those of Ref.\cite{Bjo64}.

\newcommand{\omegal}{\omega_{L}^{\vphantom{a}}}
\newcommand{\omegals}{\omega_{L}^{\prime}}
\newcommand{\qlhat}{\hat{q}_{L}^{\vphantom{a}}}
\newcommand{\qlshat}{\hat{q}_{L}^{\prime}}
\newcommand{\qlvec}{\vec{q}_{L}^{\vphantom{a}}}
\newcommand{\qlsvec}{\vec{q}_{L}^{\, \prime}}
\newcommand{\elvec}{\vec{\epsilon}_{L}^{\vphantom{a}}}
\newcommand{\elsvec}{\vec{\epsilon}_{L}^{\, \prime \ast}}
\newcommand{\om}{\omega}
\newcommand{\omo}{\omega_{0}}
\newcommand{\oms}{\omega'}
\newcommand{\qbar}{\bar{q}}
\newcommand{\qbarqua}{\bar{q}\hspace{0.5mm}^{2}}
\newcommand{\qbardrei}{\bar{q}\hspace{0.5mm}^{3}}
\newcommand{\hilfa}{a'a\bar{q}^{2}}
\newcommand{\hilfb}{\omega'a'a\bar{q}}
\newcommand{\hilfc}{\omega'^{2}a'a}
\newcommand{\cosinus}{\cos\!\vartheta}
\newcommand{\cosqua}{\cos^{2}\!\vartheta}
\newcommand{\cosdrei}{\cos^{3}\!\vartheta}
\newcommand{\sinus}{\sin\!\vartheta}
\newcommand{\sinqua}{\sin^{2}\!\vartheta}

\section{Vector basis}
\label{app2}

The vector spherical harmonics are defined according to
\begin{equation}
\vec{\mathcal{Y}}_{LM}^{l} (\hat{k}) = 
 \sum_{\lambda,\mu} 
 \langle l \lambda , 1 \mu \mid L M \rangle
 Y_{l \lambda} (\hat{k}) \vec{e}(\mu) \,, 
 \quad \textrm{with} \; \mu = 0,\pm1 \,.
\end{equation}
The magnetic, (transverse) electric and (electric) longitudinal vectors of
the multipole expansion are, respectively 
\begin{eqnarray} 
\vec{\mathcal{M}}_{LM} (\hat{k}) & = &
 \vec{\mathcal{Y}}_{LM}^{L} (\hat{k}) \vphantom{\frac{1}{1}} \,,
 \nonumber \\
\vec{\mathcal{E}}_{LM} (\hat{k}) & = &
 \sqrt{\frac{L+1}{2L+1}} \vec{\mathcal{Y}}_{LM}^{L-1} (\hat{k})
 + \sqrt{\frac{L}{2L+1}} \vec{\mathcal{Y}}_{LM}^{L+1} (\hat{k}) \,,
 \nonumber \\
\vec{\mathcal{L}}_{LM} (\hat{k}) & = &
 \sqrt{\frac{L}{2L+1}} \vec{\mathcal{Y}}_{LM}^{L-1} (\hat{k})
 - \sqrt{\frac{L+1}{2L+1}} \vec{\mathcal{Y}}_{LM}^{L+1} (\hat{k}) \,.
\end{eqnarray}
They have the following useful properties
\begin{equation}
\vec{\mathcal{M}}_{LM} (\hat{k}) =
 \frac{\vec{L}_{k} Y_{LM}(\hat{k})}{\sqrt{L(L+1)}} \, , \quad
\vec{\mathcal{E}}_{LM} (\hat{k}) =
 - i \hat{k} \times \vec{\mathcal{M}}_{LM} (\hat{k}) \, , \quad
\vec{\mathcal{L}}_{LM} (\hat{k}) =
 \hat{k} Y_{LM} (\hat{k}) \, ,
\end{equation}
where $\vec{L}_{k}$ is the angular momentum with respect to $\vec{k}$. The
4-dimensional basis $V^{\mu} (\rho L M , \hat{k} )$ is defined as  
\begin{eqnarray} \label{a_basis}
V^{\mu} (0 L M , \hat{k} ) & = &
 \bigl( Y_{LM}(\hat{k}) , \vec{0} \, \bigr) \, , 
 \vphantom{\frac{1}{1}} \nonumber \\
V^{\mu} (1 L M , \hat{k} ) & = &
 \bigl( 0 , \vec{\mathcal{M}}_{LM} (\hat{k}) \bigr) \, , 
 \vphantom{\frac{1}{1}} \nonumber \\
V^{\mu} (2 L M , \hat{k} ) & = &
 \bigl( 0 , \vec{\mathcal{E}}_{LM} (\hat{k}) \bigr) \, , 
 \vphantom{\frac{1}{1}} \nonumber \\
V^{\mu} (3 L M , \hat{k} ) & = &
 \bigl( 0 , \vec{\mathcal{L}}_{LM} (\hat{k}) \bigr) \, .
 \vphantom{\frac{1}{1}}
\end{eqnarray}
and one has the closure relation
\begin{equation}
\sum_{L,M} \sum_{\rho = 0}^{3} g_{\rho \rho}
 V^{\mu} (\rho L M , \hat{k} ) V^{\nu\ast} (\rho L M , \hat{k}' )
 = g^{\mu\nu} \delta ( \hat{k} - \hat{k}' ) \, ,
\end{equation}
and the orthogonality relation
\begin{equation} 
 \int d\hat{k} V^{\mu\ast} (\rho L M , \hat{k} ) 
 V_{\mu} (\rho' L' M' , \hat{k} )
 = g_{\rho \rho'} \delta_{L L'} \delta_{M M'} \,.
\end{equation}

\end{appendix}

\end{document}